\documentclass[]{elsarticle}

\usepackage{lineno,hyperref}
\modulolinenumbers[5]
\usepackage[inline,shortlabels]{enumitem}
\usepackage{amssymb}
\usepackage{amsmath}
\usepackage{amsthm}
\usepackage{enumitem}
\usepackage{accents}
\usepackage{amsbsy}
\usepackage[flushleft]{threeparttable}
\usepackage{graphicx}
\usepackage{color}
\usepackage{float}
\usepackage{mwe}
\usepackage[labelformat=simple]{subcaption}

\usepackage{caption}
\usepackage{algorithm}
\usepackage{algpseudocode}
\usepackage{fullpage}
\usepackage[czech,english]{babel}
\makeatletter
\algnewcommand{\LineComment}[1]{\Statex \hskip\ALG@thistlm \(\triangleright\) #1}
\makeatother
\newtheorem*{remark}{Remark}

\makeatletter
\def\ps@pprintTitle{%
\let\@oddhead\@empty
\let\@evenhead\@empty
\def\@oddfoot{\reset@font\hfil\thepage\hfil}
\let\@evenfoot\@oddfoot
}
\makeatother

\biboptions{numbers,sort&compress}
\biboptions{numbers,sort&compress}

\begin{document}

\begin{frontmatter}

\title{Bayesian inference for the stochastic identification of elastoplastic material parameters: Introduction, misconceptions and insights}


\author[University of Luxembourg,University of Liege]{H.~Rappel}

\author[University of Luxembourg]{L.A.A.~Beex}

\author[University of Luxembourg]{J.S.~Hale}

\author[University of Luxembourg,Cardiff University,The University of Western Australia]{S.P.A.~Bordas\corref{mycorrespondingauthor}}
\cortext[mycorrespondingauthor]{Corresponding author}
\ead{stephane.bordas@uni.lu}

\address[University of Luxembourg]{Faculty of Science, Technology and Communication, University of Luxembourg,
Campus~Kirchberg, 6, Rue Coudenhove-Kalergi, L-1359 Luxembourg, Luxembourg.}
\address[University of Liege]{Computational \& Multiscale Mechanics of Materials (CM3), Department of Aerospace and Mechanical Engineering, University of Li\`ege, Quartier Polytech 1, All\'ee de la D\'ecouverte 9, B-4000 Li\`ege, Belgium.}
\address[Cardiff University]{School of Engineering, Cardiff University,
Queens Buildings, The Parade, Cardiff~CF243AA, Wales, UK.}
\address[The University of Western Australia]{Adjunct Professor, Intelligent Systems for Medicine Laboratory School of Mechanical and Chemical Engineering, The University of Western Australia, 35 Stirling Highway, Crawley/Perth WA 6009, Australia.}

\begin{abstract}
We discuss Bayesian inference (BI) for the probabilistic identification of material parameters. This contribution aims to shed light on the use of BI for the identification of elastoplastic material parameters. For this purpose a single spring is considered, for which the stress-strain curves are artificially created. Besides offering a didactic introduction to BI, this paper proposes an approach to incorporate statistical errors both in the measured stresses, and in the measured strains. It is assumed that the uncertainty is only due to measurement errors and the material is homogeneous. Furthermore, a number of possible misconceptions on BI are highlighted based on the purely elastic case.
\end{abstract}


\end{frontmatter}


\section{Introduction}

\label{section:Introduction}

The most commonly used approach to identify material parameters in mechanics is to formulate an error function that measures the difference between the response of a model and experimental data \cite{C_Gogu_2010}. This error functional is then minimised with respect to the material parameters of interest to calculate them. Such an approach provides a deterministic estimate of the unknown material parameters, unable to account for the unavoidable uncertainties associated with any experimental observation. Accounting for such uncertainties is of critical importance to all engineering disciplines. An alternative, and rather different approach is to use Bayesian inference (BI). Using Bayes' theorem a probability density function (PDF), the so-called posterior distribution (or simply posterior), can be formulated as a function of the material parameters of interest. Subsequently, the PDF is analysed to determine relevant values, such as the mean of the material parameters, the material properties at which the PDF is maximum and the covariance. Only for simplistic cases can the PDF be explored analytically. Hence, numerical methods must generally be employed, e.g.~Markov chain Monte Carlo (MCMC) techniques \cite{D_Higdon_2002,J_Wang_2004,P_Risholm_2013, S_Lan_2016}. An alternative is to first approximate the PDF (e.g.~by a Laplace approximation) and then analyse it\cite{J_Beck_1998,C_oh_2008}. 

In contrast to conventional error minimisation approaches, BI incorporates the statistical noises of experimental devices \cite{C_Gogu_2010}. Consequently, the identified material parameters come with a level of uncertainty. Additionally, BI, also provides an intrinsic statistical regularisation which makes inverse problems with limited observations solvable \cite{J_Kaipio_2006}. On the other hand, applying Bayes' theorem for material parameter identification does require the measurement noise to be known, i.e.~the noise distributions and the parameters of these distributions must be established. The numerical techniques to analyse the PDF may furthermore need careful attention. 

The developments of BI in the field of parameter identification for mechanical models started with the identification of elastic constants. Isenberg \cite{J_Isenberg_1979} proposed a Bayesian approach for the identification of elastic parameters in 1979. This framework was subsequently used by various researchers to identify elastic material parameters based on dynamic responses \cite{K_Alvin_1997,J_Beck_1998, T_Marwala_2005}. Lai and Ip \cite{T_Lai_1996} used BI to identify the elastic properties of a thin composite plate. Daghia et al.~\cite{F_Daghia_2007} used the Bayesian framework for the dynamic identification of the elastic constants of thick laminated composite plates. Koutsourelakis \cite{P_Koutsourelakis_2012} used Bayesian inference to identify spatially varying elastic material parameters. In 2010 Gogu et al.~\cite{C_Gogu_2010} presented an introduction in the Bayesian approach for the identification of elastic constants, and compared the results with those of error minimisation. The influence of the prior distributions was however not systematically studied. In another study, Gogu et al.~\cite{C_Gogu_2013} used the Bayesian framework to identify elastic constants in multi-directional laminates.

BI is also used for parameter identification of nonlinear constitutive models. Muto and Beck \cite{M_Muto_2008} and Liu and Au \cite{P_Liu_2013} used the approach for hysteretic models, whereas Fitzenz et al.~\cite{D_Fitzenz_2007} used BI for a creep model of quartz. Most \cite{T_most_2010} used the Bayesian updating procedure for the parameter identification of an elastoplastic model without hardening (perfect plasticity). Rosi\'{c} et al.~\cite{B_Rosic_2013} used linear Bayesian updating via polynomial chaos expansion for an elastoplastic system. Hernandez et al.~\cite{W_Hernandez_2015} used BI for viscoelastic materials.

Another study that uses Bayes' theorem to identify material parameters is the work of Nichols et al.~\cite {J_Nichols_2010}. They employed the Bayesian approach to identify the nonlinear stiffness of a dynamic system. Furthermore, Nichols et al.~\cite{J_Nichols_2010} used the method to find the location, size and depth of delamination in a composite beam. Abhinav and Manohar \cite{S_Abhinav_2015} used BI to characterise the dynamic parameters of a structural system with geometrical nonlinearities. The approach is also employed as a framework to assess the quality of different models with respect to measured data (i.e.~model selection):~e.g.~hyperelastic constitutive models for soft tissue \cite{S_Madireddy_2015}, phenomenological models for tumour growth \cite{J_Oden_2013}, models for damage progression in composites due to fatigue \cite{J_Chiachio_2015} and fatigue models for metals \cite{I_Babuska_2015}. Sarkar et al.~\cite{S_Sarkar_2012} used the Bayesian method to identify thermodynamical parameters of cementitious materials. BI is also used in the field of mechanics for inverse problems employing differential equations \cite{J_Wang_2004,S_Cotter_2009}.

Bayesian inversion relies on concept that are not generally intuitive to those who have focused upon deterministic inversion methods. This is particularly true for nonlinear material models. Understanding the role of prior knowledge on the results is essential.

We have not encountered any Bayesian inversion study within the field of mechanics where errors on strain and stress measures are simultaneously taken into account. In other fields, a few studies can be found that deal with two error sources \cite{S_Gull_1989,D_Stephen_1995,J_Caroll_1999,R_Scheines_1999,B_Kelly_2007,D_Hogg_2010}. Yet, the methods described in these papers are only investigated in the context of linear regression, which is insufficient to tackle the problems we are interested in, in particular nonlinear hardening.

\begin{remark}
In practice displacement and force are usually the only quantities which can be directly measured (strains and stresses are deduced from those displacement and force measurements). In this work, we focus on a single spring. In this case, the strain is proportional to the displacement and the force to the stress. Consequently, in the remainder of this paper, we will abusively assume that the stress and strains are the measured quantities, which will facilitate writing the constitutive relations directly in terms of measured data.
\end{remark}

This contribution aims to show how Bayesian inference can be applied for the stochastic identification of elastoplastic material parameters. The important issues focused on are

\begin{enumerate}[(1)]
\item a formulation to incorporate the split in the purely elastic part of the response and the elastoplastic part,
\item a formulation to incorporate the uncertainty in the measured stresses as well as the uncertainty in the measured strains for elastic(-plastic) models and
\item the presentation of possible misconceptions on BI. This is important to help newcomers to the field. The elastic case is used as an illustration for this, because it is the simplest repetitive case to grasp.
\end{enumerate}

Our contribution focuses on one dimensional stress-strain results of a single spring in order to be as straightforward as possible. This single spring can be seen as an isotropic material model with no influence of the Poisson's ratio. We also propose a Bayesian framework to not only take the statistical errors in the stress measurements into account, but also the statistical error in the strain measurements. The latter part shows similarities with the study of Kelly \cite{B_Kelly_2007} in 2007, by computing the likelihood function considering all measured data, and the marginalisation integral.

The structure of the paper is as follows. Section \ref{section:Concepts} discusses the theoretical fundamentals behind Bayes' theorem, the elastoplastic material models and MCMC methods as the numerical techniques to analyse the posterior distribution. Section \ref{section:Bayesian inference for tensile tests with noise in the stress} describes a Bayesian approach for the stochastic identification of elastoplastic material parameters, when only the stress measurements include stochastic errors. Section \ref{section:Bayesian inference for tensile tests with noise in the stress} presents how to split the experimental data in a purely elastic domain and an elastoplastic domain. In section \ref{section:Bayesian inference for tensile tests with noise in both stress and strain} the Bayesian method to incorporate both the statistical errors in the stresses and strains is presented. In section \ref{section:Examples} a considerable number of relevant examples for all the aforementioned approaches is presented. Furthermore, in this section some misconceptions about the Bayesian method are indicated. In section \ref{section:Conclusion} conclusions are finally presented.
 
\section{Concepts}

\label{section:Concepts}

A conventional procedure to identify material parameters in mechanics relies on error minimisation. Often, a least-squared error functional is formulated that measures the squared difference between the experimental data and the predicted model response. This error functional is subsequently minimised with respect to the material parameters of interest. A measure for the quality of the model response with respect to the measured data is then formed by the residual (i.e.~the final least-squared difference between the measured data and the model response). This formulation does not take into account the statistical information. However there exist some advanced least-squares formulations that use the statistical information \cite{L_Le_Magorou_2002, K_Genovese_2005}.

Identification approaches based on BI provide an alternative that aims to construct a probability distribution function (PDF) based on the measured data and expert's prior knowledge, as a function of the material parameters of interest. In the case of one parameter, the PDF measures the probability that a particular value of the parameter occurs. Hence, the PDF must be analysed to investigate which values the parameter will most probably be. Important quantities of interest are the value of parameter at which the PDF is maximum (i.e.~`modes' in general statistical terminology or `maximum-a-posteriori-probability', abbreviated to `MAP', in Bayesian terminology), the mean and the variance (i.e.~the average of the squared differences from the mean).

BI may be considered as an approach that `accounts' for the fact that only a limited number of observations are made. This is achieved by incorporating some assumed prior knowledge about the parameters of interest. For inverse modelling, the prior knowledge (prior distribution, or simply `prior' in Bayesian terminology) regularises the inverse problem. This ensures that the systems to be solved are not ill-posed.
 
Note that conventional error minimisation approaches based on least-squares, are the same as the frequentist approach based on the maximum likelihood (ML) i.e.~maximising the likelihood function see sub section \ref{subsection:Bayesian inference} with respect to the parameter of interest, if the statistical error is additive, uncorrelated and normally distributed with a zero standard deviation \cite{J_Kaipio_2006}. Note that in the limit of an infinite number of measurements, the frequentist approach based on maximum likelihood and the Bayesian approach tend to the same results because the likelihood function will become dominant in the Bayesian formulation \cite{A_W_vandervaart_1998}. 

The remainder of this section presents the essential concepts of necessary to apply BI to the identification of elastoplastic material parameters in the case of a single spring. The concepts involve Bayes' theorem, the material models of interest and the adaptive MCMC method to analyse the posterior distributions.

Note that in this section capital letters denote random variables and bold ones denote vectors and matrices.

\subsection{Bayesian inference}

\label{subsection:Bayesian inference}

We start by considering events $A$ and $B$, and the discrete probabilities of each event: $P(A)$ and $P(B)$. The probability that events $A$ and $B$ both occur, is given by the joint probability, $P(A,B)$, which can be expanded as:

\begin{equation}
\label{eq:1}
P(A,B)=P(A|B)P(B)=P(B|A)P(A),
\end{equation}

\noindent where $P(A|B)$ and $P(B|A)$ are conditional probabilities. Conditional probability $P(A|B)$ expresses the probability that event $A$ occurs, if it is certain that event $B$ occurs. Using Eq.~(\ref{eq:1}), the simplest form of Bayes' theorem can be written as: 

\begin{equation}
\label{eq:2}
P(A|B)=\frac{P(A)P(B|A)}{P(B)}.
\end{equation}

If one regards two continuous random variables $ \textbf{X} \in \mathbb{R}^{n}$ and $\textbf{Y} \in \mathbb{R}^{k}$, instead of discrete variables, where $\textbf{X}$ is a random vector with $n$ unknown parameters and $\textbf{Y}$ a random vector with $k$ measurements, Eq.~(\ref{eq:2}) can be rewritten in terms of the following probability distribution functions (where $\pi$ denotes a PDF):

\begin{equation}
\label{eq:3}
\pi(\textbf{x}|\textbf{y})=\frac{\pi(\textbf{x})\pi(\textbf{y}|\textbf{x})}{\pi(\textbf{y})},
\end{equation} 

\noindent where $\pi(\textbf{x})$, $\pi(\textbf{y}|\textbf{x})$ and $\pi(\textbf{x}|\textbf{y})$ are referred to as the prior distribution (i.e.~the PDF that includes the prior knowledge), the likelihood function ( i.e.~the PDF of the observed data $\mathbf{y}$ as a function of unknown parameters $\mathbf{x}$) and the posterior distribution (i.e.~the PDF of the unknown parameters $\mathbf{x}$ given the observations $\mathbf{y}$), respectively.

Using the law of total probabilities \cite{J_Tadeusz_2001} which relates the marginal probabilities ($\pi(\textbf{x})$ and $\pi(\textbf{y})$) to the conditional probabilities ($\pi(\textbf{y}|\textbf{x})$), the denominator in Eq.~(\ref{eq:3}) can be written as:

\begin{equation}
\label{eq:4}
\pi(\textbf{y})=\int_{\mathbb{R}^n} \pi(\textbf{x})\pi(\textbf{y}|\textbf{x})d\textbf{x}.
\end{equation}

\noindent Since the data ($\textbf{y}$) is already measured, the denominator in Eq.~(\ref{eq:3}) is a constant number, $C \in \mathbb{R}^{+}$. This constant number can be regarded as a normalisation factor that ensures that the integral of the posterior ($\pi(\textbf{x}|\textbf{y})$) equals 1:

\begin{equation}
\label{eq:5}
\pi(\textbf{x}|\textbf{y})=\frac{1}{C}\pi(\textbf{x})\pi(\textbf{y}|\textbf{x}).
\end{equation}

\noindent In the case of parameter identification, one is generally interested in the maximum-a-posteriori-probability point (the value of the unknown parameter at which the PDF has the maximum value), the mean and the covariance matrix, which are independent of $C$. Hence, it suffices to rewrite Eq.~(\ref{eq:5}) as:

\begin{equation}
\label{eq:6}
\pi(\textbf{x}|\textbf{y})\propto\pi(\textbf{x})\pi(\textbf{y}|\textbf{x}).
\end{equation}

In order to obtain the posterior (i.e.~the PDF of the vector of unknown parameters given the observations $\pi(\textbf{x}|\textbf{y})$), the likelihood function ($\pi(\textbf{y}|\textbf{x})$) and the prior ($\pi(\textbf{x})$) need to be formulated. First, the likelihood function is considered.

In order to construct the likelihood function, a noise model has to be formulated and a noise distribution ($\pi_\text{noise}$) has to be determined and calibrated. For the moment, we assume that the noise distribution is known (including its parameters). The noise model used in this study is additive, which is frequently employed, amongst others in several inverse studies \cite{J_Kaipio_2006,F_Daghia_2007}. The additive noise model can be written as follows:

\begin{equation}
\label{eq:7}
\textbf{Y}=\textbf{f}(\textbf{X})+\boldsymbol{\Omega},
\end{equation}

\noindent where $\textbf{X} \in \mathbb{R}^{n}$ is again the vector with unknown material parameters, $\textbf{Y} \in \mathbb{R}^{k}$ the vector with the measured data and $\mathbf{\Omega} \in \mathbb{R}^{k}$ the noise vector. $\textbf{f}: \mathbb{R}^n\rightarrow \mathbb{R}^k$ denotes the model and is a function of the unknown material parameters ($\textbf{X}$). Given the realisations $\textbf{X}=\textbf{x}$ and $\boldsymbol{\Omega}=\boldsymbol{\omega}$, and assuming that the parameters ($\textbf{X}$) and the error ($\boldsymbol{\Omega}$) are statistically independent, the likelihood function reads:

\begin{equation}
\label{eq:8}
\pi(\textbf{y}|\textbf{x})=\pi_\text{noise}(\textbf{y}-\textbf{f}(\textbf{x})),
\end{equation}

\noindent where $\pi_\text{noise}(\boldsymbol{\omega})$ is the PDF of the noise (which is assumed to be identified based on separate calibration experiments, see subsections \ref{subsection:noise distribution} and \ref{subsection:BI double error Noise distribution}). Substitution of Eq.~(\ref{eq:8}) in Eq.~(\ref{eq:6}) yields:

\begin{equation}
\label{eq:9}
\pi(\textbf{x}|\textbf{y})\propto\pi(\textbf{x})\pi_\text{noise}(\textbf{y}-\textbf{f}(\textbf{x})).
\end{equation}

A critical aspect of the Bayesian framework is the selection of the prior distribution $(\pi(\textbf{x}))$ \cite{J_Kaipio_2006} in which \emph{a-priori} knowledge about the parameters is translated in terms of a PDF. If enough data is observed the prior hardly has an influence \cite{S_Madireddy_2015}. The influence of the prior distribution will be considered in more detail in section \ref{section:Examples}.

Once the posterior is formulated (Eq.~(\ref{eq:9})), the mean parameter values, MAP parameter values and the covariance matrix have to be extracted from it in an analytical or numerical manner (e.g.~using Markov chain Monte Carlo methods). The adaptive Markov chain Monte Carlo is considered in subsection \ref{subsection:Markov chain Monte Carlo method}, as it is the numerical approach employed in this contribution to evaluate the posteriors.

\subsection{Material models}

\label{subsection:Material models}

In this contribution, BI is developed to identify the parameters of four one-dimensional material models: linear elasticity, linear elasticity with perfect plasticity, linear elasticity with linear hardening and linear elasticity with nonlinear hardening. Hardening is considered to be kinematic and associative. For each model, the identification is based on the results of monotonic uniaxial tensile tests. Below, the governing equations of the models are given, as well as those for when monotonic tensile loading takes place.

\subsubsection{Linear elastic}

\label{subsubsection:Linear elastic}

The linear elastic model assumes a linear relationship between the stress and the strains. In the one dimensional case, this writes:

\begin{equation}
\label{eq:10}
\sigma(\epsilon,\textbf{x})=E\epsilon,
\end{equation}

\noindent where $\sigma$ is the stress, $\epsilon$ is the strain, $\textbf{x}$ is the material parameter vector (here $\textbf{x}=E$) and $E$ is Young's modulus and assumed to be constant in the domain.

\subsubsection{Linear elastic-perfectly plastic}

\label{subsubsection:Linear elastic-perfectly plastic}

The linear elastic-perfectly plastic model neglects the effect of work hardening, assuming that purely plastic deformation occurs when the stress reaches its yield value. The total strain ($\epsilon$) in this contribution is additively split into  an elastic part $\epsilon_{e}$ and a plastic part $\epsilon_{p}$:

\begin{equation}
\label{eq:11}
\epsilon=\epsilon_{e}+\epsilon_{p},
\end{equation}

\noindent and the stress is defined as a function of the elastic strain, $\epsilon_{e}$:

\begin{equation}
\label{eq:12}
\sigma(\epsilon,\textbf{x})=E\epsilon_{e}=E(\epsilon-\epsilon_{p}).
\end{equation}

The yield condition at which plastic yielding occurs, is written as:

\begin{equation}
\label{eq:13}
f(\sigma)=\left |\sigma \right |-\sigma_{y0}\leq0,
\end{equation}

\noindent where $\sigma_{y0}$ is the initial yield stress and $f$ is the yield function. Consequently, $\textbf{x}=\begin{bmatrix}E& \sigma_{y0}\end{bmatrix}^{T}$. 

Furthermore, the flow rule for the plastic strain can be written as: 

\begin{equation}
\label{eq:14}
\dot{\epsilon}_{p}=\dot{\alpha}\frac{\partial{f}}{\partial{\sigma}}=\dot{\alpha}\,\textrm{sgn}({\sigma}),
\end{equation}

\noindent where $\textrm{sgn}(\cdot)$ is the sign function and $\alpha$ is the cumulative plastic strain. Finally, the Kuhn-Tucker conditions \cite{J_C_Simo_2000} must hold:

\begin{equation}
\label{eq:15}
\, \begin{matrix}
 \dot{\alpha}\geq0,& f(\sigma)\leq0, & \dot{\alpha} f(\sigma)=0.
\end{matrix}
\end{equation}

The stress-strain response of the linear elastic-perfectly plastic model during monotonic tension can be written as:

\begin{equation}
\label{eq:16}
\sigma=\begin{cases}
E\epsilon& \text{ if } \epsilon\leq\frac{\sigma_{y0}}{E} \\ 
\sigma_{y0}& \text{ if } \epsilon >\frac{\sigma_{y0}}{E} 
\end{cases}.
\end{equation}

\noindent Using the Heaviside step function ($h(\cdot)$), Eq.~(\ref{eq:16}) can be conveniently expressed as:

\begin{equation}
\label{eq:17}
\sigma(\epsilon,\textbf{x})=E\epsilon\bigg(1-h\Big(\epsilon-\frac{\sigma_{y0}}{E}\Big)\bigg)+\sigma_{y0}h\Big(\epsilon-\frac{\sigma_{y0}}{E}\Big).
\end{equation}

\subsubsection{Linear elastic-linear hardening}

\label{subsubsection:Linear elastic-linear hardening}

The linear elastic-linear hardening model is identical to the linear elastic-perfectly plastic model, except for the yield function, which writes:

\begin{equation}
\label{eq:18}
 f(\sigma)=\left |\sigma \right |-\sigma_{y0}-H\alpha\leq0,
\end{equation}

\noindent where $H$ is the plastic modulus. Here, $\textbf{x}=\begin{bmatrix} E&\sigma_{y0}&H\end{bmatrix}^{T}$.

Consequently, the stress-strain response of the model during monotonic tension writes:

\begin{equation}
\label{eq:19}
\sigma=\begin{cases}
E\epsilon& \text{ if } \epsilon\leq\frac{\sigma_{y0}}{E} \\ 
\sigma_{y0}+H\epsilon_{p}& \text{ if } \epsilon >\frac{\sigma_{y0}}{E}  
\end{cases},
\end{equation}

\noindent which can again be expressed using the Heaviside step function:

\begin{equation}
\label{eq:20}
\sigma(\epsilon,\textbf{x})=E\epsilon\bigg(1-h\Big(\epsilon-\frac{\sigma_{y0}}{E}\Big)\bigg)+\bigg(\sigma_{y0}+\frac{HE}{H+E}\Big(\epsilon-\frac{\sigma_{y0}}{E}\Big)\bigg)h\Big(\epsilon-\frac{\sigma_{y0}}{E}\Big).
\end{equation}

\subsubsection{Linear elastic-nonlinear hardening}

\label{subsubsection:Linear elastic-nonlinear hardening}

The linear elastic-nonlinear hardening model only differs from the linear elastic-perfectly plastic model through the yield function which writes:

\begin{equation}
\label{eq:21}
f(\sigma)=\left |\sigma \right |-\sigma_{y0}-H\alpha^{n}\leq0,
\end{equation}

\noindent where $n$ is an additional plastic material parameter and hence, $\textbf{x}=\begin{bmatrix}E&\sigma_{y0}&H&n\end{bmatrix}^{T}$.

For the monotonic loading case, the stress-strain response can be written as:

\begin{equation}
\label{eq:22}
\sigma=\begin{cases}
E\epsilon& \text{ if } \epsilon\leq\frac{\sigma_{y0}}{E} \\ 
\sigma_{y0}+H\epsilon^{n}_{p}& \text{ if } \epsilon >\frac{\sigma_{y0}}{E}  
\end{cases},
\end{equation}

\noindent or using the Heaviside step function:

\begin{equation}
\label{eq:23}
\sigma(\epsilon,\textbf{x})=E\epsilon\bigg(1-h\Big(\epsilon-\frac{\sigma_{y0}}{E}\Big)\bigg)+\bigg(\sigma_{y0}+H\Big(\epsilon-\frac{\sigma(\epsilon,\textbf{x})}{E}\Big)^{n}\bigg)h\Big(\epsilon-\frac{\sigma_{y0}}{E}\Big).
\end{equation} 

It is important to note that Eq.~(\ref{eq:23}) is an implicit function of the stress ($\sigma(\epsilon,\textbf{x})$ appears both on the left hand side and right hand side of Eq.~(\ref{eq:23}) and cannot immediately be identified from Eq.~(\ref{eq:23})). This is in contrast to the stress-strain expressions of the previous material models for monotonic tension (Eqs.~(\ref{eq:10}), (\ref{eq:17}) and (\ref{eq:20})), which are all explicit functions (i.e. $\sigma (\epsilon, \textbf{x})$ can be directly computed when one has $\epsilon$). This changes the likelihood function construction procedure for the linear elastic-nonlinear hardening case which is discussed in subsection \ref{subsection:BI single error Linear elastic-linear hardening}.

\subsection{Markov chain Monte Carlo method (MCMC)}

\label{subsection:Markov chain Monte Carlo method}

Once a posterior has been constructed, it needs to be analysed to determine the unknown parameters. Only in a few straightforward cases, the analysis can be performed analytically. A numerical approach needs to be employed in the majority of the cases, e.g.~when the model is nonlinear or when the model is an implicit function. If the posterior is continuous one may use approximation approaches (e.g.~a Laplace approximation \cite{J_Beck_1998}) that require the derivatives of the posterior. If the posterior is $C_{0}$-continuous however, numerical approaches which do not require any derivatives need to be employed. The posteriors for for the aforementioned elastoplastic models are all $C_{0}$-continuous due to the abrupt transition between the purely elastic part and the elastoplastic part of the response (see e.g.~Eq.~(\ref{eq:22})). 

Markov chain Monte Carlo (MCMC) techniques are frequently employed numerical approaches to investigate posteriors, that do not require any derivatives \cite{J_Beck_2002,Y_Marzouk_2007,J_Kristensen_2014}. MCMC approaches are based on drawing samples from the posterior to approximate the parameters' statistical properties (e.g.~the mean value and variance). Below, the fundamental concepts of the Monte Carlo method are discussed as well as the adaptive Metropolis algorithm to perform the sampling.

\subsubsection{Monte Carlo method}

\label{subsubsection:Monte Carlo method}

The main purpose of the Monte Carlo method is to approximate integrals of the following form:

\begin{equation}
\label{eq:24}
\textbf{I}=\int_{\mathbb{R}^n}\textbf{g}(\textbf{x})\pi(\textbf{x})d\textbf{x},
\end{equation} 

\noindent where $\pi$ is the PDF of interest (in our case the posterior) and $\textbf{g}: \mathbb{R}^n\rightarrow \mathbb{R}^k$ is an integrable function over $\mathbb{R}^{n}$. This integral can be approximated using the following quadrature:

\begin{equation}
\label{eq:25}
\hat{\textbf{I}}=\frac{1}{N}\sum_{i=1}^{N}\textbf{g}(\textbf{x}_{i}),
\end{equation}

\noindent where $\left \{ \textbf{x}_{i}\right \}_{i}^{N}$ is the set of samples drawn from the PDF of interest ($\pi$) and the hat on $\hat{\textbf{I}}$ represents the numerically approximated equivalent of $\textbf{I}$. The drawing of samples from $\pi$ implies that most of the samples are in the domain in which numerical evaluations of $\pi$ do not equal zero. By the law of large numbers (LLN) \cite{C_Andrieu_2003}, $\hat{\textbf{I}}$ converges as follows:

\begin{equation}
\label{eq:26}
\lim_{N \to +\infty}\frac{1}{N}\sum_{i=1}^{N}\textbf{g}(\textbf{x}_{i})=\textbf{I}.
\end{equation}

\noindent The numerical approximation of the components of the covariance matrix for $\textbf{g}(\textbf{x})$ ($\widehat{\textbf{cov}}_{\textbf{g}}$) is \cite{S_Brooks_2011}: 

\begin{equation}
\label{eq:27}
(\widehat{\mathrm{cov}}_{\textbf{g}})_{jm}=\frac{1}{N}\sum_{i=1}^{N}\Big(g_{j}(\textbf{x}_{i})-I_{j}\Big)\Big((g_{m}(\textbf{x}_{i})-I_{m}\Big),\ j=1,2,\cdots,k,\ m=1,2,\cdots,k.
\end{equation}

The mean of the posterior ($\boldsymbol{\mu}_{\text{post}}$) can be computed by substituting $\textbf{g}(\textbf{x})=\textbf{x}$ and $\pi=\pi_{\textrm{post}}$ in Eq.~(\ref{eq:24}), which yields:

\begin{equation}
\label{eq:28}
\boldsymbol{\mu}_\text{post}=\int_{\mathbb{R}^n}\textbf{x}\,\pi_{\textrm{post}}(\textbf{x})d\textbf{x}=\lim_{N \to +\infty}\frac{1}{N}\sum_{i=1}^{N}\textbf{x}_{i}.
\end{equation}

\noindent Furthermore, the components of the posterior's covariance matrix are approximated as follows:

\begin{equation}
\label{eq:29}
(\widehat{\mathrm{cov}}_{\textrm{post}})_{jm}=\frac{1}{N}\sum_{i=1}^{N}\Big((x_{i})_{j}-(\mu_{\textrm{post}})_{j}\Big)\Big((x_{i})_{m}-(\mu_{\textrm{post}})_{m}\Big),\ j=1,2,\cdots,k,\ m=1,2,\cdots,k.
\end{equation}

If we again assume that a large number of samples is taken (i.e.~$N$ is large), the MAP estimate can furthermore be approximated as \cite{C_Andrieu_2003}:

\begin{equation}
\label{eq:30}
\widehat{\textbf{MAP}}=\underset{\textbf{x}_{i;i=1,...,N}}{\textrm{argmax}}\pi(\textbf{x}_{i}).
\end{equation}

The essential part of a Monte Carlo procedure is the drawing of admissible samples ($\textbf{x}_{i}$). Below, the standard and adaptive Metropolis algorithms are discussed. The adaptive one is the algorithm used in section \ref{section:Examples}.

\subsubsection{The standard and the adaptive Metropolis algorithm}

\label{subsubsection:The standard and the adaptive Metropolis algorithm}

The standard Metropolis-Hastings approach is a frequently employed MCMC algorithm \cite{C_Andrieu_2003}. The basic idea of the Metropolis-Hastings algorithm is to explore the PDF of interest by making a random walk across the parameter space $\textbf{x}$. Considering sample $\mathbf{x}_{i}$ and an evaluation of the PDF $\pi(\mathbf{x}_{i})$, a new sample $\mathbf{x}_{p}$ is proposed by drawing from a proposal distribution ($q$ in Algorithm \ref{algorithm:The Metropolis-Hastings algorithm in practice}). If the PDF evaluated at the proposed sample ($\pi(\mathbf{x}_{p})$) is larger than at the current sample ($\pi(\mathbf{x}_{i})$), the proposed sample is \emph{always} accepted. However, if the PDF at the proposed sample is smaller than at the current sample, the proposed sample \emph{may} be accepted based on the ratio of the PDF evaluated at the proposed and current sample ($r$ in Algorithm \ref{algorithm:The Metropolis-Hastings algorithm in practice}). The ratio is compared to a random number generated from a uniform distribution. If the ratio is greater than the random number, the proposed sample is accepted. If the ratio is smaller than the random number, the proposed sample is rejected, and we remain at the current sample. In the case the proposed sample was accepted, the proposed sample becomes the current sample. The algorithm repeats until convergence is achieved.

Critically, as the number of samples increases, it can be shown that the distribution of values approximates the target distribution. For more information, see \cite{S_Brooks_2011}. The standard Metropolis-Hastings algorithm is presented in Algorithm \ref{algorithm:The Metropolis-Hastings algorithm in practice} in more detail.

\begin{algorithm}
\caption{The standard Metropolis-Hastings algorithm}
\label{algorithm:The Metropolis-Hastings algorithm in practice}
\begin{algorithmic}[1]
\State select the initial sample $\textbf{x}_{0}\in \mathbb{R}^{n}$ and $\gamma$
\For{$i=0,1,2,...,N-1$ }
\State draw $\textbf{x}_{p} \in \mathbb{R}^{n}$ from the proposal distribution $q(\textbf{x}_{p}|\textbf{x}_{i})$ in Eq.~(\ref{eq:33})
\State calculate the ratio $r(\textbf{x}_{i},\textbf{x}_{p})=\textrm{min}\Big(1,\frac{\pi(\textbf{x}_{p})q(\textbf{x}_{i}|\textbf{x}_{p})}{\pi(\textbf{x}_{i})q(\textbf{x}_{p}|\textbf{x}_{i})}\Big)$\label{line:4}\LineComment{$\pi(\cdot)$ denotes the target distribution (i.e.~posterior).}
\State draw $u \in [0,1]$ from uniform probability density
\If {$r(\textbf{x}_{i},\textbf{x}_{p})\geq u$}
\State $\textbf{x}_{i+1}=\textbf{x}_{p}$
\Else
\State $\textbf{x}_{i+1}=\textbf{x}_{i}$ 
\EndIf
\EndFor
\end{algorithmic}
\end{algorithm}

In the case that the transition kernel is symmetric (as in this contribution), the following relation holds:

\begin{equation}
\label{eq:31}
q(\textbf{x}_{i}|\textbf{x}_{p})=q(\textbf{x}_{p}|\textbf{x}_{i}).
\end{equation}

\noindent Consequently, step \ref{line:4} in the algorithm simplifies to: 

\begin{equation}
\label{eq:32}
r(\textbf{x}_{i},\textbf{x}_{p})=\textrm{min}\Big(1,\frac{\pi(\textbf{x}_{p})}{\pi(\textbf{x}_{i})}\Big).
\end{equation}

The stability and convergence of the algorithm can be checked by tracing the generated samples and analysing their characteristics. The evolution of the mean value and the standard deviation can for instance be checked for convergence \cite{S_Sarkar_2012}. The distribution approximated by MCMC converges to the target distribution if the samples show a stationary statistical behaviour, i.e.~if the mean and standard deviation remain constant \cite{S_Sarkar_2012}.

The efficiency of the algorithm is influenced by the initial sample ($\textbf{x}_{0}$) and the proposal distribution ($q$) \cite{J_Kaipio_2006}. The most common proposal distribution for the Metropolis-Hastings algorithm (as employed here) is of the following Gaussian form:

\begin{equation}
\label{eq:33}
q(\textbf{x}_{i}|\textbf{x}_{p})=q(\textbf{x}_{p}|\textbf{x}_{i})\propto \textrm{exp}\Big(-\frac{1}{2\gamma^{2}}\left \| \textbf{x}_{i}-\textbf{x}_{p} \right \|^{2}\Big),
\end{equation}

\noindent where $\gamma$ is the parameter that determines the width of the proposal distribution and must be tuned to obtain an efficient and converging algorithm. An efficient starting value is $\gamma=\frac{2.38}{\sqrt{n}}$ \cite{A_Gelman_1996}, where $n$ is the number of unknown parameters and hence, the dimension of the posterior.

To overcome the tuning of $\gamma$, Haario et al.~\cite{H_Haario_1999} introduced the Adaptive Proposal (AP). The AP method updates the width of the proposal distribution, based on the existing knowledge of the posterior. The existing knowledge is based on the previous samples. For sample $i+1$, the update employs the following formulation:

\begin{equation}
\label{eq:34}
q(\textbf{x}_{p}|\textbf{x}_{i})\sim N(\textbf{x}_{i},\gamma^{2}\textbf{R}_{i}), 
\end{equation}

\noindent where $N(\textbf{x}_{i},\gamma^2\textbf{R}_{i})$ denotes a normal distribution with mean $\textbf{x}_{i}$ and covariance matrix $\gamma^{2}\textbf{R}_{i}$, of size $n \times n$. To establish $\textbf{R}_{i}$, all $i$ previous samples are first stored in matrix $\textbf{K}$ of size $i \times n$. $\textbf{R}_{i}$, is then computed as:

\begin{equation}
\label{eq:35}
\textbf{R}_{i}=\frac{1}{i-1}\widetilde{\textbf{K}}^{T}\widetilde{\textbf{K}},
\end{equation}

\noindent where $\widetilde{\textbf{K}}=\textbf{K}-\textbf{K}_\text{mean}$ and $\textbf{K}_\text{mean}$ reads:

\begin{equation}
\label{eq:36}
\textbf{K}_\text{mean}=\begin{bmatrix}
\textbf{k}_\text{mean} \\ \textbf{k}_\text{mean} \\ \vdots \\ \textbf{k}_\text{mean}
\end{bmatrix}_{i \times n},
\end{equation}

\noindent and $\textbf{k}_\text{mean}$ is a row matrix of length $n$ which is determined as follows:

\begin{equation}
\label{eq:37}
\textbf{k}_\text{mean}=\frac{1}{i} \begin{bmatrix}
\sum\limits_{j=1}^{i} (K)_{j1} & \sum\limits_{j=1}^{i} (K)_{j2} & \cdots & \sum\limits_{j=1}^{i} (K)_{jn}
\end{bmatrix}.
\end{equation}

The following relation is used for $N(\textbf{x}_{i},\gamma^{2}\textbf{R}_{i})$ in this contribution:

\begin{equation}
\label{eq:38}
N(\textbf{x}_{i},\gamma^{2}\textbf{R}_{i})=\textbf{x}_{i}+\frac{\gamma}{\sqrt{i-1}}\widetilde{\textbf{K}}^{T}N(0,\textbf{I}_{i}),
\end{equation}

\noindent where $\textbf{I}_{i}$ is the identity matrix of size $i \times i$ and $N(0,\textbf{I}_{i})$ is the $i$-dimensional normal distribution.
 
Note that it is computationally inefficient to update the proposal distribution after each new sample is generated. In the numerical examples in this study therefore, updating takes place once per 1000 sample generations. Algorithm \ref{algorithm:The Metropolis-Hastings algorithm with symmetric AP proposal} shows the Metropolis-Hastings algorithm with the symmetric AP proposal (Eq.~(\ref{eq:38})) that is employed here.

\begin{algorithm}
\caption{The Metropolis-Hastings algorithm with symmetric AP proposal}
\label{algorithm:The Metropolis-Hastings algorithm with symmetric AP proposal}
\begin{algorithmic}[1]
\State select the initial sample $\textbf{x}_{0}\in \mathbb{R}^{n}$ and set $\gamma=\frac{2.38}{\sqrt{n}}$
\For{$i=0,1,2,...,N-1$ }
\State draw $\textbf{x}_{p} \in \mathbb{R}^{n}$ from the proposal distribution $q(\textbf{x}_{p}|\textbf{x}_{i})$ in Eq.~(\ref{eq:38})
\State calculate the ratio $r(\textbf{x}_{i},\textbf{x}_{p})=\textrm{min}\Big(1,\frac{\pi(\textbf{x}_{p})}{\pi(\textbf{x}_{i})}\Big)$ \LineComment{$\pi(\cdot)$ denotes the target distribution (i.e.~posterior).}
\State draw $u \in [0,1]$ from uniform probability density
\If {$r(\textbf{x}_{i},\textbf{x}_{p})\geq u$}
\State $\textbf{x}_{i+1}=\textbf{x}_{p}$
\Else
\State $\textbf{x}_{i+1}=\textbf{x}_{i}$
\EndIf
\State \textbf{per 1000 samples}
\State \indent update matrix $\widetilde{\textbf{K}}$
\EndFor
\end{algorithmic}
\end{algorithm}

\section{Bayesian inference for uniaxil tensile tests with noise in the stress measurements}

\label{section:Bayesian inference for tensile tests with noise in the stress}

The current section focuses on the formulation of Bayesian identification approaches for the material models of subsection \ref{subsection:Material models}, when only the stress measurements are polluted with statistical errors. The measured strains are thus considered to be exact. In section \ref{section:Bayesian inference for tensile tests with noise in both stress and strain}, Bayesian updating is considered when also the measured strains are also polluted with a statistical noise.

\subsection{Noise distribution}

\label{subsection:noise distribution}

\begin{figure}
\begin{minipage}[t]{0.5\linewidth}
\centering
\subcaptionbox{Fitted noise distribution\label{fig:1_a}}{\frame{\def\svgwidth{0.8\columnwidth}
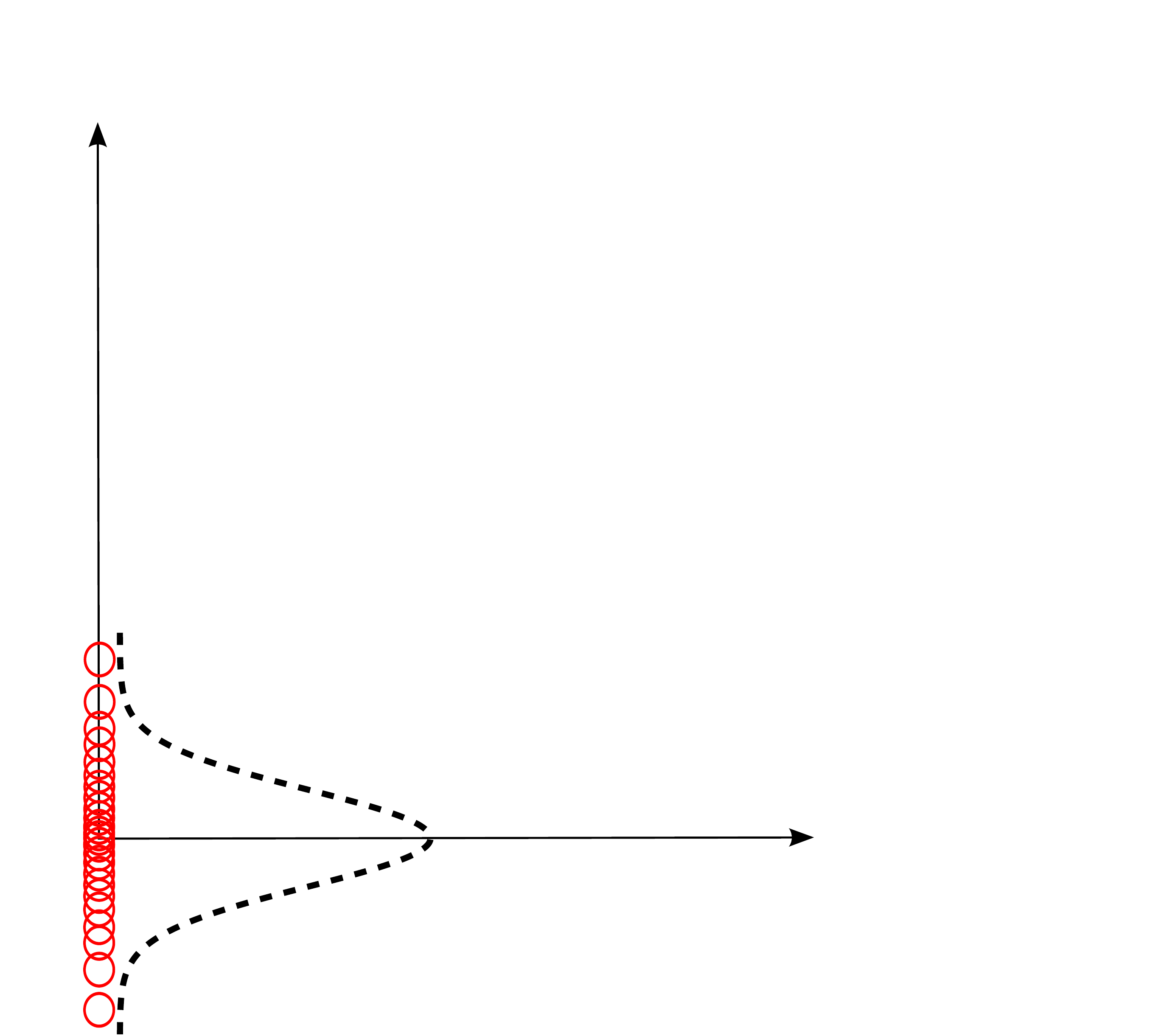}}
\end{minipage}
\begin{minipage}[t]{0.5\linewidth}
\centering
\subcaptionbox{The shifted noise distribution\label{fig:1_b}}{\frame{\def\svgwidth{0.8\columnwidth}
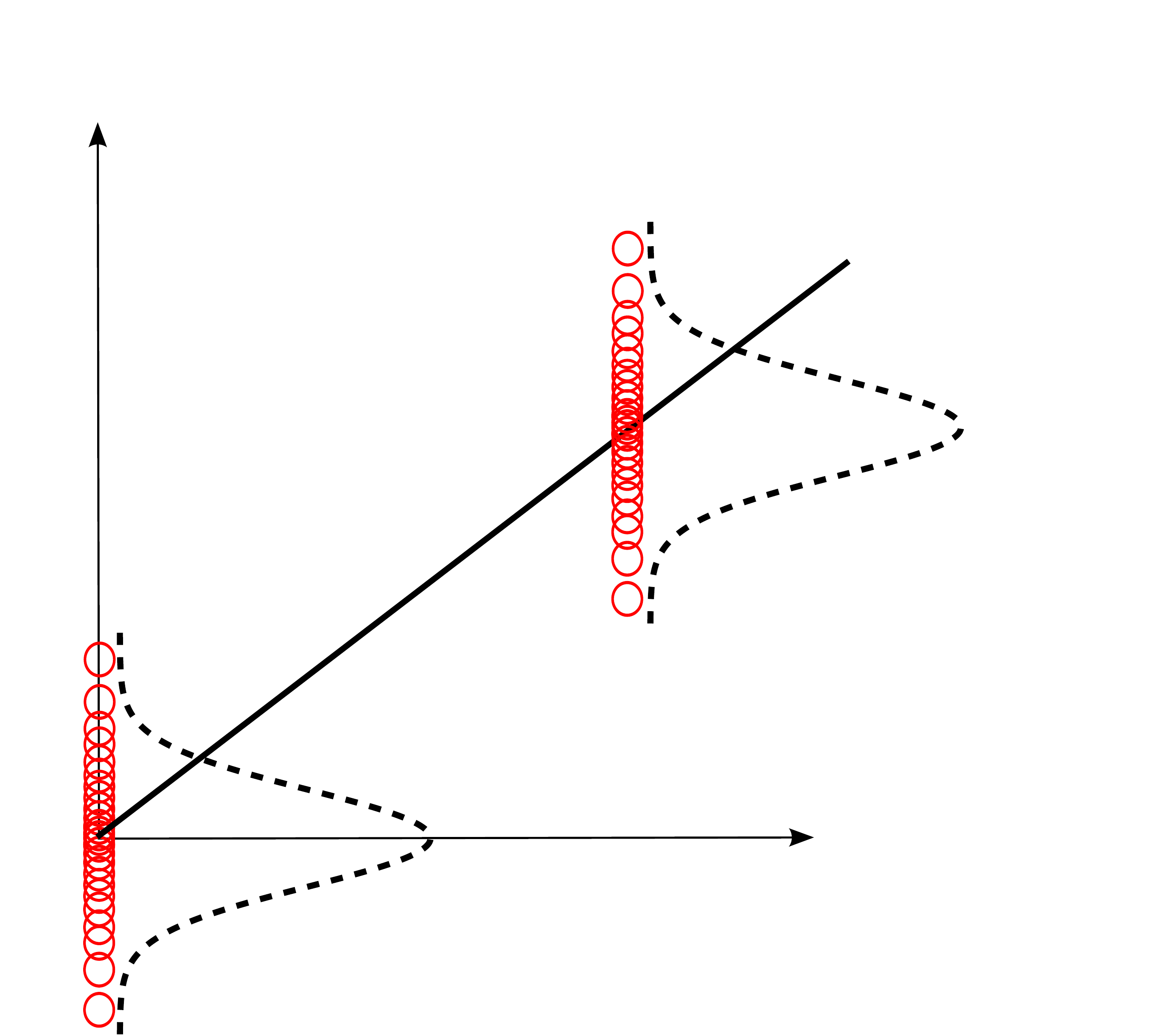}}
\end{minipage}
\caption{Schematic of the stress-strain measurements (red circles) of the `calibration experiments', including the noise distributions (dashed). The theoretical stress-strain relation (which is exact for the calibration experiments) is presented as the bold straight line on the diagram on the right.}
\label{fig:1}
\end{figure}

To determine the noise distribution and its parameters, two sets of `calibration experiments' can be performed. First, a test is performed without any specimen. The stress-strain measurements of this test are shown in Fig.~\ref{fig:1_a}. It shows that the PDF of the noise in the `stress measurements' is a normal distribution with a zero mean and a standard deviation of $S_\text{noise}$.

Second, the evolution of the noise distribution (including its parameters) must be determined. To this purpose, a tensile test is performed on a calibration specimen (of which the Young's modulus is known). The assumed results are presented in Fig.~\ref{fig:1_b}. The mean stress value varies linearly with the strain. The standard deviation $S_\text{noise}$ however remains the same.

As the `calibration measurements' indicate that an additive noise model can be used and the stresses are polluted by a normal noise distribution with standard deviation $S_\text{noise}$, BI is first employed to identify the Young's modulus of the linear elastic model.

\subsection{Linear elastic}

\label{subsection:BI single error Linear elastic}

The only unknown material parameter in the linear elastic model is the Young's modulus ($E$). Based on section \ref{section:Concepts}, the additive noise model for a single stress measurement can be written as follows:

\begin{equation}
\label{eq:39}
\sigma^{m}=E\epsilon+\Omega,
\end{equation}

\noindent where $\sigma^{m}$ is the measured stress and $\Omega$ is the random variable representing the noise in the stress measurement. Based on the artificially generated calibration results, the noise distribution is normal and can be written as:

\begin{equation}
\label{eq:40}
\pi_\text{noise}(\omega)=\frac{1}{\sqrt{2\pi}S_\text{noise}}\textrm{exp}\Big(-\frac{\omega^2}{2S_\text{noise}^{2}}\Big).
\end{equation} 

\noindent Using Eq.~(\ref{eq:8}), the likelihood function is expanded as:

\begin{equation}
\label{eq:41}
\pi(\sigma^{m}|E)=\pi_\text{noise}(\sigma^{m}-E\epsilon)=\frac{1}{\sqrt{2\pi}S_\text{noise}}\textrm{exp}\Big(-\frac{(\sigma^{m}-E\epsilon)^2}{2S_\text{noise}^{2}}\Big),
\end{equation}

\noindent Substitution of Eq.~(\ref{eq:41}) in Eq.~(\ref{eq:9}) then yields the following expression for the posterior:

\begin{equation}
\label{eq:42}
\pi(E|\sigma^{m})\propto \pi(E)\textrm{exp}\Big(-\frac{(\sigma^{m}-E\epsilon)^{2}}{2S_\text{noise}^{2}}\Big).
\end{equation}

If we use a prior in the form of a modified normal distribution as follows:

\begin{equation}
\label{eq:43}
\pi(E)\propto \begin{cases}
\textrm{exp}\Big(-\frac{(E-\overline{E})^{2}}{2S_{E}^{2}}\Big)& \text{ if } E\ge0  \\ 
0& \text{ otherwise }
\end{cases},
\end{equation}

\noindent the posterior distribution reads:

\begin{equation}
\label{eq:44}
\pi(E|\sigma^{m})\propto \begin{cases}
\textrm{exp}\Big(-\Big[\frac{(E-\overline{E})^{2}}{2S_{E}^{2}}+\frac{(\sigma^{m}-E\epsilon)^{2}}{2S_\text{noise}^2}\Big]\Big)& \text{ if } E\ge0  \\ 
0& \text{ otherwise } 
\end{cases},
\end{equation} 

\noindent where $\overline{E}$ and $S_{E}$ are the mean and standard deviation of the prior distribution, respectively. Note that the Young's modulus cannot be negative for an actual material, which is taken into account in the prior distribution (Eq.~(\ref{eq:43})). 

If we now consider the posterior distribution of the previous measurement to be the prior distribution of the current measurement, the posterior for all $k$ measurements can be expressed as:

\begin{equation}
\label{eq:45}
\begin{matrix}
 \pi(E|\boldsymbol{\sigma}^{m})\propto \textrm{exp}\Big(-\Big[\frac{(E-\overline{E})^{2}}{2S_{E}^{2}}+\frac{\sum\limits_{i=1}^{k} (\sigma_{i}^{m}-E\epsilon_{i})^{2}}{2S_\text{noise}^{2}}\Big]\Big),& E\geq0
\end{matrix}
\end{equation}

\noindent where $\pi(E|\boldsymbol{\sigma}^{m})=\pi(E|\sigma_{1}^{m},\sigma_{2}^{m},\cdots,\sigma_{k}^{m})$. Eq.~(\ref{eq:45}) can now be written in the following form:

\begin{equation}
\label{eq:46}
\begin{matrix}
\pi(E|\boldsymbol{\sigma}^{m})\propto \textrm{exp}\Big(-\frac{(E-\mu_\text{post})^{2}}{2S_\text{post}^{2}}\Big),& E\geq0
\end{matrix}
\end{equation}

\noindent where $\mu_\text{post}$ and $S_\text{post}$ are the mean and standard deviation of the posterior distribution, which is again a normal distribution (with the condition $E\geq0$). Both can be expressed as:

\begin{equation}
\label{eq:47}
\begin{matrix}
\mu_\text{post}=\frac{S_\text{noise}^{2}\overline{E}+S_{E}^{2}\sum\limits_{i=1}^{k}\epsilon_{i}\sigma_{i}^{m}}{S_\text{noise}^{2}+S_{E}^{2}\sum\limits_{i=1}^{k}\epsilon_{i}^{2}},& S_\text{post}=\sqrt{\frac{S_\text{noise}^{2}S_{E}^{2}}{S_\text{noise}^{2}+S_{E}^{2}\sum\limits_{i=1}^{k} \epsilon_{i}^{2}}}
\end{matrix}.
\end{equation}

Hence, it is possible to analytically examine the posterior distribution for the linear elastic model, in case the noise model is additive and the noise distribution as well as the prior distribution are (modified) normal distributions. For the other cases below, MCMC approaches are required and the adaptive MCMC approach used for those is verified based on the analytical expressions in Eq.~(\ref{eq:47}) for the linear elastic model.

\subsection{Linear elastic-perfectly plastic}

\label{subsection:BI single error Linear elastic-perfectly plastic}

The parameters to be identified for the linear elastic-perfectly plastic model are the Young's modulus and the initial yield stress which are stored in parameter vector $\textbf{x}=\begin{bmatrix}
E&
\sigma_{y0}\end{bmatrix}^{T}$. Since the experimental equipment and experimental condition both remain the same (i.e.~the measured stresses are still polluted by noise stemming from the same normal distribution and the measured strains are still exact), the same additive noise model applies:

\begin{equation}
\label{eq:48}
\sigma^{m}=\sigma(\epsilon,\textbf{x})+\Omega,
\end{equation}

\noindent where $\sigma(\epsilon,\textbf{x})$ is referred to as the theoretical stress and takes the form of the stress-strain relation for monotonic tension in Eq.~(\ref{eq:16}). Using Eq.~(\ref{eq:40}) for the noise distribution, the likelihood function for a single stress measurement reads:

\begin{equation}
\label{eq:49}
\pi(\sigma^{m}|\textbf{x})=\pi_\text{noise}(\sigma^{m}-\sigma(\epsilon,\textbf{x}))=\frac{1}{\sqrt{2\pi}S_\text{noise}}\textrm{exp}\Big(-\frac{(\sigma^{m}-\sigma(\epsilon,\textbf{x}))^2}{2S_\text{noise}^{2}}\Big),
\end{equation}

\noindent or:

\begin{equation}
\label{eq:50}
\pi(\sigma^{m}|\textbf{x})=\frac{1}{\sqrt{2\pi}S_\text{noise}}\textrm{exp}\Bigg(-\frac{\bigg(\sigma^{m}-E\epsilon\bigg(1-h\Big(\epsilon-\frac{\sigma_{y0}}{E}\Big)\bigg)-\sigma_{y0}h\Big(\epsilon-\frac{\sigma_{y0}}{E}\Big)\bigg)^2}{2S_\text{noise}^{2}}\Bigg).
\end{equation}

Taking the physical constraints into account that the Young's modulus and the initial yield stress must be nonnegative, the following prior distribution is selected:

\begin{equation}
\label{eq:51}
\pi(\textbf{x})\propto \begin{cases}
\textrm{exp}\Big(-\frac{(\textbf{x}-\overline{\textbf{x}})^{T}\boldsymbol{\Gamma}^{-1}_{\textbf{x}}(\textbf{x}-\overline{\textbf{x}})}{2}\Big)&  \text{ if } E\ge0 \textrm{ and } \sigma_{y0}\geq0  \\ 
0& \text{ otherwise }
\end{cases},
\end{equation}

\noindent where $\overline{\textbf{x}}$ is the mean value vector of the prior distribution and $\boldsymbol{\Gamma}_{\textbf{x}}$ is the covariance matrix of the prior. Substitution of Eq.~(\ref{eq:50}) and Eq.~(\ref{eq:51}) in the reduced variant of Bayes' formula of Eq.~(\ref{eq:9}), yields the following posterior distribution for $k$ measurements:

\begin{equation}
\label{eq:52}
\pi(\textbf{x}|\boldsymbol{\sigma}^{m})\propto\textrm{exp}\Bigg(-\bigg[\frac{(\textbf{x}-\overline{\textbf{x}})^{T}\boldsymbol{\Gamma}^{-1}_{\textbf{x}}(\textbf{x}-\overline{\textbf{x}})}{2}+\frac{\sum\limits_{i=1}^{k}\bigg(\sigma_{i}^{m}-E\epsilon_{i}\bigg(1-h\Big(\epsilon_{i}-\frac{\sigma_{y0}}{E}\Big)\bigg)-\sigma_{y0}h\Big(\epsilon_{i}-\frac{\sigma_{y0}}{E}\Big)\bigg)^2}{2S_{noise}^{2}}\bigg]\Bigg),
\end{equation}

\noindent where the probability of obtaining a negative Young's modulus and initial  yield stress is zero thanks to the selected prior distribution.

 It is important to realise that the posterior distribution in this case is not of a form that allows an analytical evaluation due to the presence of the Heaviside function. Hence, the adaptive MCMC approach is employed to analyse the posterior for the linear elastic-perfectly plastic model in section \ref{section:Examples}.

\subsection{Linear elastic-linear hardening}

\label{subsection:BI single error Linear elastic-linear hardening}

The parameter vector for the linear elastic-linear hardening model reads $\textbf{x}=\begin{bmatrix}
E&\sigma_{y0}&H\end{bmatrix}^{T}$. Assuming the same experimental equipment and condition (and hence, the same noise model and noise distribution), the likelihood function for a single measurement reads:

\begin{equation}
\label{eq:53}
\pi(\sigma^{m}|\textbf{x})\propto\textrm{exp}\Bigg(-\frac{\bigg(\sigma^{m}-E\epsilon\bigg(1-h\Big(\epsilon-\frac{\sigma_{y0}}{E}\Big)\bigg)-\bigg(\sigma_{y0}+\frac{HE}{H+E}\Big(\epsilon-\frac{\sigma_{y0}}{E}\Big)\bigg)h\Big(\epsilon-\frac{\sigma_{y0}}{E}\Big)\bigg)^{2}}{2S_\text{noise}^{2}}\Bigg).
\end{equation}

In addition to the physical constraints for the Young's modulus and the initial yield stress, the plastic modulus ($H$) must also be nonnegative. The following prior distribution is therefore selected:

\begin{equation}
\label{eq:54}
\pi(\textbf{x})\propto \begin{cases}
\textrm{exp}\Big(-\frac{(\textbf{x}-\overline{\textbf{x}})^{T}\boldsymbol{\Gamma}^{-1}_{\textbf{x}}(\textbf{x}-\overline{\textbf{x}})}{2}\Big)& \text{ if } E\ge0 \textrm{ and } \sigma_{y0}\geq0 \textrm{ and } H\geq0 \\ 
0& \text{ otherwise } 
\end{cases}.
\end{equation}

\noindent Using Bayes' formula, the posterior distribution for $k$ observations reads:

\begin{multline}
\label{eq:55}
\pi(\textbf{x}|\boldsymbol{\sigma}^{m})\propto\textrm{exp}\Bigg(-\bigg[\frac{(\textbf{x}-\overline{\textbf{x}})^{T}\boldsymbol{\Gamma}^{-1}_{\textbf{x}}(\textbf{x}-\overline{\textbf{x}})}{2}+\\\frac{\sum\limits_{i=1}^{k}\bigg(\sigma^{m}_{i}-E\epsilon_{i}\bigg(1-h\Big(\epsilon_{i}-\frac{\sigma_{y0}}{E}\Big)\bigg)-\bigg(\sigma_{y0}+\frac{HE}{H+E}\Big(\epsilon_{i}-\frac{\sigma_{y0}}{E}\Big)\bigg)h\Big(\epsilon_{i}-\frac{\sigma_{y0}}{E}\Big)\bigg)^{2}}{2S_\text{noise}^{2}}\bigg]\Bigg).
\end{multline}

\subsection{Linear elastic-nonlinear hardening}

\label{subsection:BI single error Linear elastic-nonlinear hardening}

The parameter vector for the linear elastic-nonlinear hardening model is $\textbf{x}=\begin{bmatrix}
E&\sigma_{y0}&H&n\end{bmatrix}^{T}$. Considering no change of experimental equipment (and hence, the same noise model and noise distribution), the expression for the measured stress reads again:

\begin{equation}
\label{eq:56}
\sigma^{m}=\sigma(\epsilon,\textbf{x})+\Omega,
\end{equation}

\noindent which, together with Eq.~(\ref{eq:23}), results in the following expression for the measured stress: 

\begin{equation}
\label{eq:57}
\sigma^{m}=E\epsilon\bigg(1-h\Big(\epsilon-\frac{\sigma_{y0}}{E}\Big)\bigg)+\bigg(\sigma_{y0}+H\Big(\epsilon-\frac{\sigma(\epsilon, \textbf{x})}{E}\Big)^{n}\bigg)h\Big(\epsilon-\frac{\sigma_{y0}}{E}\Big)+\Omega.
\end{equation}

\noindent It is important to note that the expression for the measured stress is now a function of the theoretical stress ($\sigma(\epsilon,\textbf{x})$). This is in contrast to the expressions of the measured stresses of the other material models. Consequently, the construction of the likelihood function changes. First, one needs to determine the probability that the measured stress occurs, for a given theoretical stress and a set of given material parameters:

\begin{equation}
\label{eq:58}
\pi(\sigma^{m}|\textbf{x},\sigma(\epsilon,\textbf{x}))=\pi_\text{noise}(\sigma^{m}-\sigma(\epsilon,\textbf{x})).
\end{equation}

\noindent The likelihood function required for Bayes' theorem however, expresses the likelihood that measured stresses occur for given parameters. This can be obtained by integrating Eq.~(\ref{eq:58}), together with the theoretical stress likelihood function, over the theoretical stress \cite{A_Gelman_2003}:

\begin{equation}
\label{eq:59}
\pi(\sigma^{m}|\textbf{x})=\int_{0}^{+\infty}\pi(\sigma^{m}|\textbf{x},\sigma(\epsilon,\textbf{x}))\pi(\sigma(\epsilon,\textbf{x})|\textbf{x})d\sigma.
\end{equation}

\noindent In Eq.~(\ref{eq:59}), the theoretical stress is integrated from 0 to infinity, because it cannot be smaller than zero in monotonic tension (otherwise compression would occur). As the theoretical stress can in principle be infinitely large, no true upper bound is included. 

In the case that the parameters are given (as in MCMC approaches in which each sample is a realisation of the parameters), the theoretical stress likelihood function can be expressed as follows:

\begin{equation}
\label{eq:60}
\pi(\sigma(\epsilon,\textbf{x})|\textbf{x})=\delta(\sigma(\epsilon,\textbf{x})-w),
\end{equation}

\noindent where $\delta$ is the Dirac delta function and $w$ is used here to represent the right hand side of Eq.~(\ref{eq:23}). Note that in the case that all the parameters are given, the only possible value for the theoretical stress is the same expression of the theoretical stress, but only as a function of the parameters. In Eq.~(\ref{eq:60}) this is imposed by the Dirac delta function. Substitution of Eq.~(\ref{eq:58}) and Eq.~(\ref{eq:60}) in Eq.~(\ref{eq:59}) yields:

\begin{equation}
\label{eq:61}
\pi(\sigma^{m}|\textbf{x})=\int_{0}^{+\infty}\pi_\text{noise}(\sigma^{m}-\sigma(\epsilon,\textbf{x}))\delta(\sigma(\epsilon,\textbf{x})-w)d\sigma.
\end{equation}

\noindent The integral in Eq.~(\ref{eq:61}) can be computed using \cite{S_Hassani_2008}:

\begin{equation}
\label{eq:62}
\int_{a}^{b}f(z)\delta(g(z))dz=\begin{cases}
\sum\limits_{i=1}^{n}\frac{f(c_{i})}{\left | {g}'(c_{i}) \right |}& \text{ if } a<c_{i}<b\\ 
0& \textrm{ otherwise }
\end{cases},
\end{equation}

\noindent where $c_i$ are the $i$ roots of $g(z)$, for which holds $\frac{dg(c_{i})}{dz}\neq0$. Combining Eqs.~(\ref{eq:61}) and (\ref{eq:62}) yields the final likelihood function for a single measurement as:

\begin{equation}
\label{eq:63}
\pi(\sigma^{m}|\textbf{x})\propto\begin{cases}
\textrm{exp}\Big(-\frac{(\sigma^{m}-E\epsilon)^{2}}{2S_\text{noise}^{2}}\Big)& \text{ if } \epsilon\leq\frac{\sigma_{y0}}{E} \\ 
\frac{\textrm{exp}\Big(-\frac{(\sigma^{m}-c)^{2}}{2S_\text{noise}^{2}}\Big)}{\left | 1+\frac{Hn}{E}(\epsilon-{\frac{c}{E}})^{n-1} \right |}& \text{ if }\epsilon>\frac{\sigma_{y0}}{E}
\end{cases},
\end{equation} 

\noindent where $c$ is the value of the theoretical stress when $\sigma(\epsilon,\textbf{x})-w=0$ is solved for $\sigma(\epsilon,\textbf{x})$. Choosing the prior distribution in the form of a modified normal distribution as:

\begin{equation}
\label{eq:64}
\pi(\textbf{x})\propto \begin{cases}
\textrm{exp}\Big(-\frac{(\textbf{x}-\overline{\textbf{x}})^{T}\boldsymbol{\Gamma}^{-1}_{\textbf{x}}(\textbf{x}-\overline{\textbf{x}})}{2}\Big)& \text{ if } E\ge0 \textrm{ and } \sigma_{y0}\geq0 \textrm{ and } H\geq0 \textrm{ and } n\geq0 \\ 
0& \text{ otherwise }
\end{cases}.
\end{equation}

\noindent The final form of the posterior distribution reads:

\begin{equation}
\label{eq:65}
\pi(\textbf{x}|\boldsymbol{\sigma}^{m})\propto \textrm{exp}\Big(-\frac{(\textbf{x}-\overline{\textbf{x}})^{T}\boldsymbol{\Gamma}^{-1}_{\textbf{x}}(\textbf{x}-\overline{\textbf{x}})}{2}\Big)\prod_{i=1}^{k}\pi(\sigma_{i}^{m}|\textbf{x}),
\end{equation}

\noindent where $\pi(\sigma_{i}^{m}|\textbf{x})$ must be calculated according to Eq.~(\ref{eq:63}) for each measurement. To investigate the posterior, first $\sigma(\epsilon,\textbf{x})-w=0$ is numerically solved for $\sigma(\epsilon,\textbf{x})$ and then the adaptive MCMC is employed to find the statistical characteristics of the posterior.

\section{Bayesian inference for uniaxial  tensile tests with noise in both stress and strain measurements}

\label{section:Bayesian inference for tensile tests with noise in both stress and strain}

All the cases studied in section \ref{section:Bayesian inference for tensile tests with noise in the stress} consider an error in the stress, but not in the strain. Both the measured stresses and the measured strains may be polluted by statistical errors however. Both statistical errors are not related to each other, as the devices to measure the forces and strains are independent of each other (e.g.~when a load cell and digital image correlation are used). This section deals with the same four material models as in the previous sections but considers that the stresses as well as the strains are influenced by noise.

\subsection{Noise distribution}

\label{subsection:BI double error Noise distribution}

\begin{figure}
\begin{minipage}[t]{0.5\linewidth}
\centering
\subcaptionbox{Fitted noise distribution\label{fig:2_a}}{\frame{\def\svgwidth{0.8\columnwidth}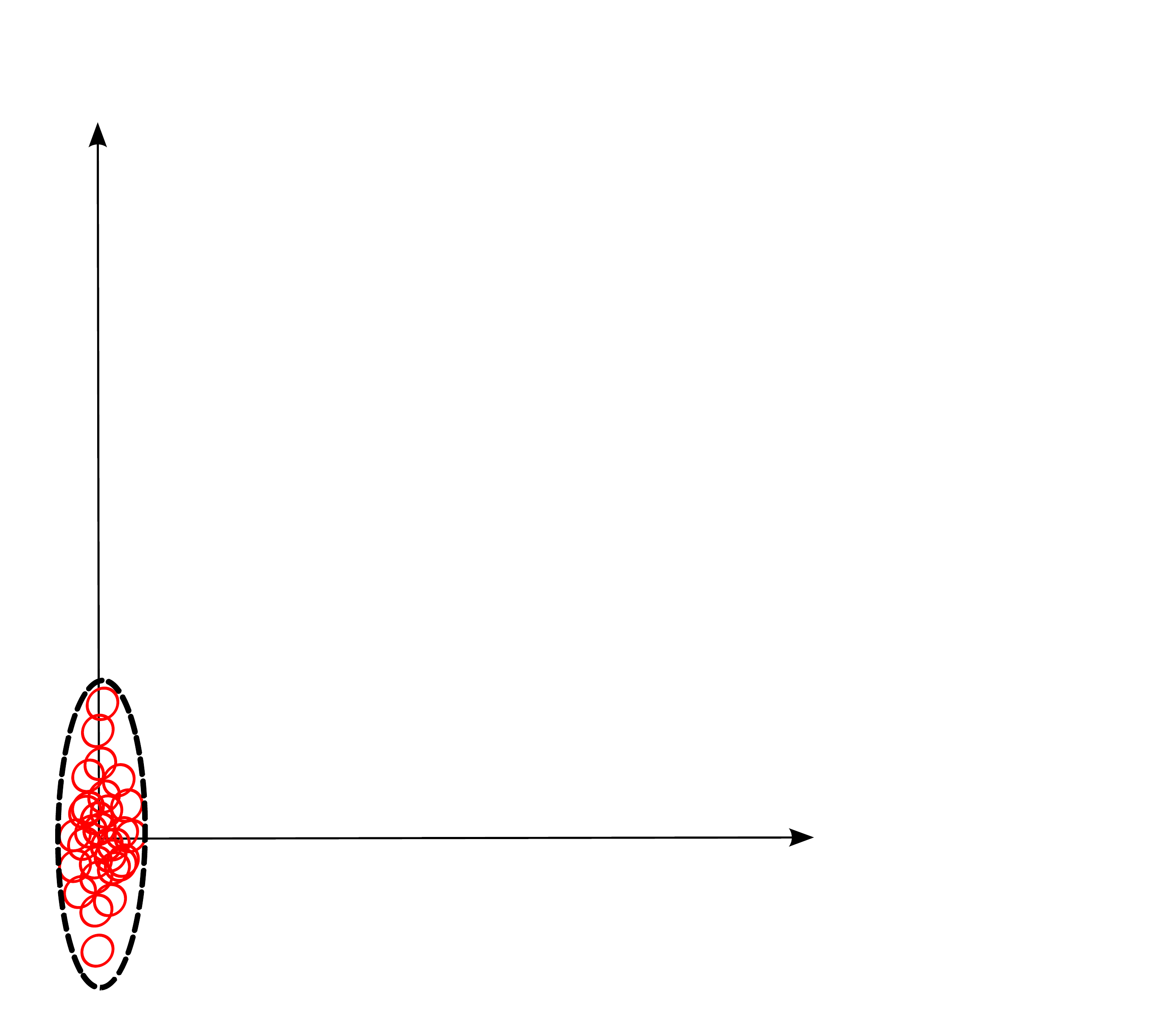}}
\end{minipage}
\begin{minipage}[t]{0.5\linewidth}
\centering
\subcaptionbox{The shifted noise distribution\label{fig:2_b}}{\frame{\def\svgwidth{0.8\columnwidth}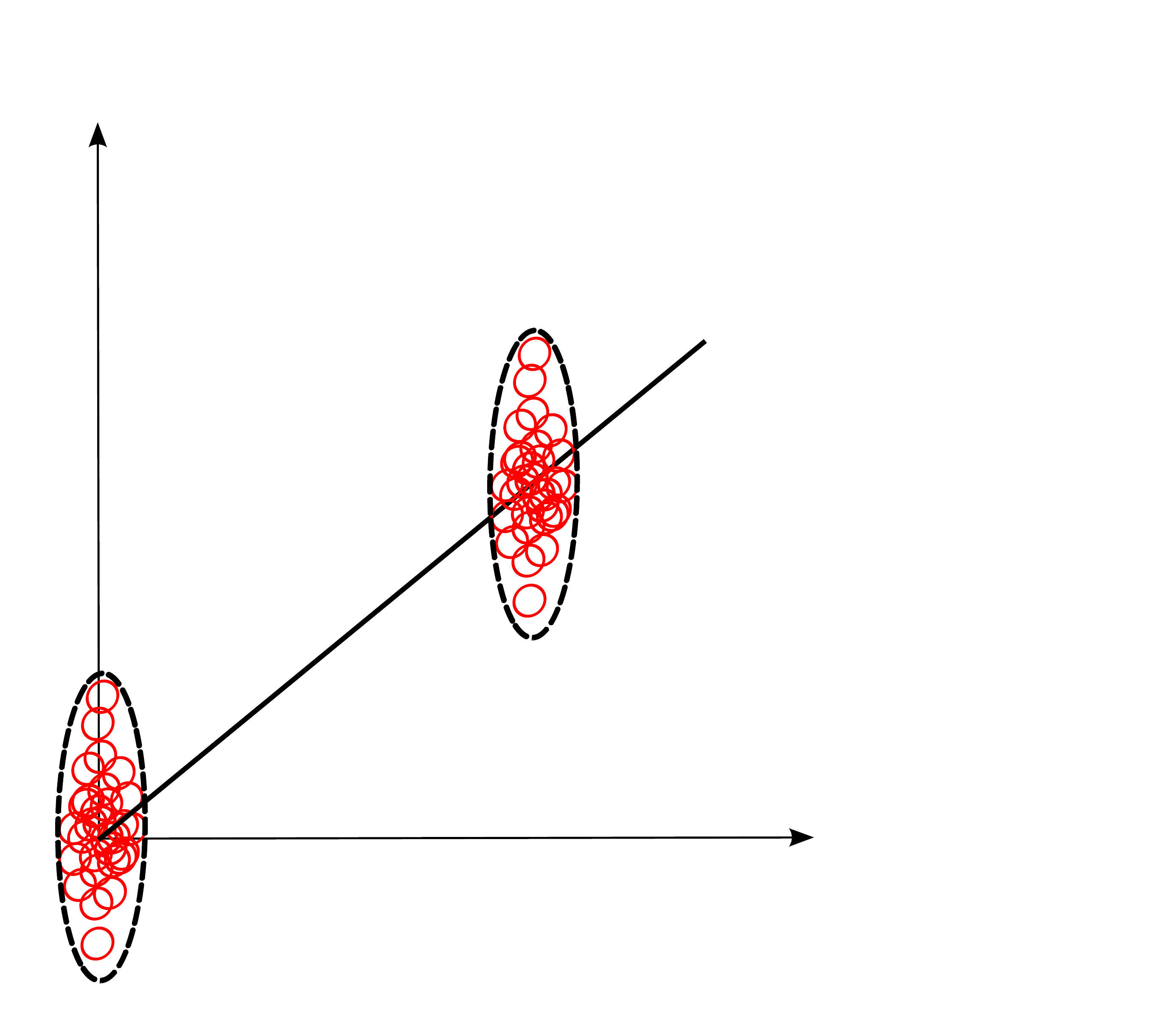}}
\end{minipage}
\caption{Schematic of the stress-strain measurements (red circles) of the `calibration experiments' for the case with uncertainty in both stress and strain, including some isolines of the noise distributions (dashed). The theoretical stress-strain relation (which is exact for the calibration experiments) is presented as the bold straight line on the diagram on the right.}
\label{fig:2}
\end{figure}

The first calibration experiment is again performed without the use of a specimen. The stress-strain measurements of this test are shown in Fig.~\ref{fig:2_a}. The measurements in Fig.~\ref{fig:2_a} indicate a normal distribution with a zero mean and a diagonal covariance $\boldsymbol{\Gamma}_\text{noise}$. The noise model and the evolution of the noise distribution are again investigated by performing a tensile test on a calibration specimen with an exactly known Young's modulus. The schematic results are shown in Fig.~\ref{fig:2_b}. It is clearly visible that the mean stress and strain values follow a linear, theoretical stress-strain relationship and hence, the covariance does not change. The calibration results show that the additive noise model can also be used in this section.

\subsection{Linear elastic}

\label{subsection:BI double error Linear elastic}

The additive noise model when both the stresses and strains are contaminated by stochastic noise, can be expressed as follows for the linear elastic model:

\begin{equation}
\label{eq:66}
\begin{cases}
\sigma^{m}=E\epsilon+\Omega_{\sigma}\\
\epsilon^{m}=\epsilon+\Omega_{\epsilon}
\end{cases},
\end{equation}

\noindent where $\sigma^{m}$ is the measured stress, $\epsilon^{m}$ is the measured strain, $\Omega_{\sigma}$ is the stochastic error of the stress measurement and $\Omega_{\epsilon}$ is the stochastic error of the strain measurement. Because the information from both the measured stress and the measured strain is used here, Bayes' formula for multiple variables must be employed \cite{S_Prince_2012}:

\begin{equation}
\label{eq:67}
\pi(E|\sigma^{m},\epsilon^{m})=\frac{\pi(E)\pi(\epsilon^{m}|E)\pi(\sigma^{m}|E,\epsilon^{m})}{\pi(\epsilon^{m})\pi(\sigma^{m}|\epsilon^{m})}.
\end{equation} 

\noindent Since the measured strain and the Young's modulus are statistically independent (measuring a specific strain does not interfere with the probability that a certain Young's modulus occurs, i.e.~$\pi(\epsilon^{m} | E) = \pi(\epsilon^{m})$), Eq.~(\ref{eq:67}) can be written as follows:

\begin{equation}
\label{eq:68}
\pi(E|\sigma^{m},\epsilon^{m})=\frac{\pi(E)\pi(\sigma^{m}|E,\epsilon^{m})}{\pi(\sigma^{m}|\epsilon^{m})},
\end{equation}

\noindent or, when one is only interested in relative probabilities, as:

\begin{equation}
\label{eq:69}
\pi(E|\sigma^{m},\epsilon^{m})\propto\pi(E)\pi(\sigma^{m}|E,\epsilon^{m}).
\end{equation}

\noindent The issue with Eq.~(\ref{eq:69}) is that the the likelihood function, $\pi(\sigma^{m}|E,\epsilon^{m})$, must again be determined by integration (over $\epsilon$ here \cite{A_Gelman_2003}), because $\pi(\sigma^{m}|E,\epsilon)$ can be determined directly, but $\pi(\sigma^{m}|E,\epsilon^{m})$ not. To this end, we write:

\begin{equation}
\label{eq:70}
\pi(\sigma^{m}|E,\epsilon^{m})=\int_{0}^{a}\pi(\sigma^{m}|E,\epsilon)\pi(\epsilon|\epsilon^{m})d\epsilon,
\end{equation}

\noindent where $a$ is defined by the physical upper bound of the tensile tester (i.e.~the ratio of the original length of the specimen and the maximum distance that the clamps can move). Based on Eq.~(\ref{eq:66}) one can express the conditional probabilities $\pi(\sigma^{m}|E,\epsilon)$ and $\pi(\epsilon|\epsilon^{m})$ as follows:

\begin{equation}
\label{eq:71}
\begin{cases}
\pi(\sigma^{m}|E,\epsilon)=\pi_{\Omega_{\sigma}}(\sigma^{m}-E\epsilon)\\
\pi(\epsilon|\epsilon^{m})=\pi_{\Omega_{\epsilon}}(\epsilon^{m}-\epsilon)
\end{cases},
\end{equation}

\noindent where $\pi_{\Omega_{\sigma}(\omega_{\sigma})}$ and $\pi_{\Omega_{\epsilon}(\omega_{\epsilon})}$ are the noise distributions of the errors in the stress measurements and the strain measurements, respectively. Based on the calibration results in the previous subsection, the noise distribution is a two-dimensional normal distribution with no correlation between the noise in the stress measurements and the noise in the strain measurements (diagonal covariance matrix) and zero mean value as follows: 

\begin{equation}
\label{eq:72}
\pi_\text{noise}(\boldsymbol{\omega})=\frac{1}{2\pi\sqrt{\left |\boldsymbol{\Gamma}_\text{noise} \right |}}\textrm{exp}\bigg(-\frac{\boldsymbol{\omega}^{T}\boldsymbol{\Gamma}_\text{noise}^{-1}\boldsymbol{\omega}}{2}\bigg),
\end{equation}

\noindent where $\boldsymbol{\omega}=\begin{bmatrix} \omega_{\sigma}& \omega_{\epsilon}\end{bmatrix}^{T}$ is the noise vector and $\boldsymbol{\Gamma}_\text{noise}$ is the covariance matrix of the noise distribution in the following form:

\begin{equation}
\label{eq:73}
\begin{bmatrix}
S_{\sigma}^{2}&0\\
0&S_{\epsilon}^{2}
\end{bmatrix}, 
\end{equation}

Combining Eqs.~(\ref{eq:71}) and (\ref{eq:72}), the following two conditional probabilities are required to construct the likelihood function (Eq.~(\ref{eq:70})):

\begin{equation}
\label{eq:74}
\begin{cases}
\pi(\sigma^{m}|E,\epsilon)=\frac{1}{\sqrt{2\pi}S_{\sigma}}\textrm{exp}\Big(-\frac{(\sigma^{m}-E\epsilon)^{2}}{2S_{\sigma}^{2}}\Big)\\
\pi(\epsilon|\epsilon^{m})=\frac{1}{\sqrt{2\pi}S_{\epsilon}}\textrm{exp}\Big(-\frac{(\epsilon^{m}-\epsilon)^{2}}{2S_{\epsilon}^{2}}\Big)
\end{cases},
\end{equation}

\noindent Substitution of Eq.~(\ref{eq:74}) in Eq.~(\ref{eq:70}) then results in the following expression for the likelihood function:

\begin{equation}
\label{eq:75}
\pi(\sigma^{m}|E,\epsilon^{m})=\frac{1}{2\pi S_{\sigma}S_{\epsilon}}\int_{0}^{a}\textrm{exp}\bigg(-\Big[\frac{(\sigma^{m}-E\epsilon)^{2}}{2S_{\sigma}^{2}}+\frac{(\epsilon^{m}-\epsilon)^{2}}{2S_{\epsilon}^2}\Big]\bigg)d\epsilon.
\end{equation}

\noindent The result of this integral can be expressed analytically as:

\begin{equation}
\label{eq:76}
\pi(\sigma^{m}|E,\epsilon^{m})=\frac{\sqrt{p_{3}}}{2\sqrt{2\pi}S_{\sigma}S_{\epsilon}}\textrm{exp}\Big(-\frac{p_{2}-p_{1}^{2}}{2p_{3}}\Big)\bigg[\textrm{erf}\Big(\frac{a-p_{1}}{\sqrt{2p_{3}}}\Big)-\textrm{erf}\Big(\frac{-p_{1}}{\sqrt{2p_{3}}}\Big)\bigg],
\end{equation}

\noindent where $\textrm{erf}(\cdot)$ is the error function \cite{S_Hassani_2008} and $p_{1}$, $p_{2}$ and $p_{3}$ are formulated as follows:

\begin{equation}
\label{eq:77}
p_{1}=\frac{E\sigma^{m}S_{\epsilon}^{2}+S_{\sigma}^{2}\epsilon^{m}}{E^{2}S_{\epsilon}^{2}+S_{\sigma}^{2}},\ p_{2}=\frac{S_{\sigma}^{2}(\epsilon^{m})^{2}+S_{\epsilon}^{2}(\sigma^{m})^{2}}{E^{2}S_{\epsilon}^{2}+S_{\sigma}^{2}},\ p_{3}=\frac{(S_{\sigma}S_{\epsilon})^{2}}{E^{2}S_{\epsilon}^{2}+S_{\sigma}^{2}}.
\end{equation}

Consequently, choosing the same prior as in Eq.~(\ref{eq:43}), the posterior distribution for $k$ measurements reads:

\begin{equation}
\label{eq:78}
\pi(E|\boldsymbol{\sigma}^{m},\boldsymbol{\epsilon}^{m})\propto\textrm{exp}\Big(\frac{(E-\overline{E})^{2}}{2S_{E}^{2}}\Big)\bigg(\frac{\sqrt{p_{3}}}{2\sqrt{2\pi}S_{\sigma}S_{\epsilon}}\bigg)^{k}\prod_{i=1}^{k}\textrm{exp}\Big(-\frac{(p_{2i}-p_{1i}^{2})}{2p_{3}}\Big)\bigg[\textrm{erf}\Big(\frac{a-p_{1i}}{\sqrt{2p_{3}}}\Big)-\textrm{erf}\Big(\frac{-p_{1i}}{\sqrt{2p_{3}}}\Big)\bigg],
\end{equation}

\noindent where $\pi(E|\boldsymbol{\sigma}^{m},\boldsymbol{\epsilon}^{m})=\pi(E|(\sigma_{1}^{m},\epsilon_{1}^{m}),(\sigma_{2}^{m},\epsilon_{2}^{m}),\cdots,(\sigma_{k}^{m},\epsilon_{k}^{m}))$.

\subsection{Linear elastic-perfectly plastic}

\label{subsection:BI double error Linear elastic-perfectly plastic}

In case the experimental devices remain the same, but the linear elastic-perfectly plastic model is used, the additive noise model reads:

\begin{equation}
\label{eq:79}
\begin{cases}
\sigma^{m}=\sigma(\epsilon,\textbf{x})+\Omega_{\sigma}\\
\epsilon^{m}=\epsilon+\Omega_{\epsilon}
\end{cases}
\end{equation}

\noindent where $\sigma(\epsilon,\textbf{x})$ is given by Eq.~(\ref{eq:17}) and $\textbf{x}=\begin{bmatrix}
E& \sigma_{y0}\end{bmatrix}^{T}$. For the same noise distribution as in Eq.~(\ref{eq:72}), the conditional probabilities required to construct the likelihood function (Eq.~(\ref{eq:70})) are:

\begin{equation}
\label{eq:80}
\begin{cases}
\pi(\sigma^{m}|E,\epsilon)=\frac{1}{\sqrt{2\pi}S_{\sigma}}\textrm{exp}\Bigg(-\frac{\bigg(\sigma^{m}-E\epsilon \bigg(1-h\Big(\epsilon-\frac{\sigma_{y0}}{E}\Big)\bigg)-\sigma_{y0}h\Big(\epsilon-\frac{\sigma_{y0}}{E}\Big)\bigg)^{2}}{2S_{\sigma}^{2}}\Bigg)\\
\pi(\epsilon|\epsilon^{m})=\frac{1}{\sqrt{2\pi}S_{\epsilon}}\textrm{exp}\Big(-\frac{(\epsilon^{m}-\epsilon)^{2}}{2S_{\epsilon}^{2}}\Big)
\end{cases}.
\end{equation}

\noindent Combining Eqs.~(\ref{eq:70}) and (\ref{eq:80}) the likelihood function can be written as:
 
\begin{multline}
\label{eq:81}
\pi(\sigma^{m}|\textbf{x},\epsilon^{m})=\frac{1}{2\sqrt{2\pi}S_{\epsilon}S_{\sigma}}\bigg(\sqrt{p_{3}}\,\textrm{exp}\Big(-\frac{p_{2}-p_{1}^{2}}{2p_{3}}\Big)\Big[\textrm{erf}\Big(\frac{\frac{\sigma_{y0}}{E}-p_{1}}{\sqrt{2p_{3}}}\Big)-\textrm{erf}\Big(\frac{-p_{1}}{\sqrt{2p_{3}}}\Big)\Big]+\\
S_{\epsilon}\textrm{exp}\Big(-\frac{(\sigma^{m}-\sigma_{y0})^{2}}{2S_{\sigma}^{2}}\Big)\Big[\textrm{erf}\Big(\frac{\epsilon^{m}-\frac{\sigma_{y0}}{E}}{\sqrt{2}S_{\epsilon}}\Big)-\textrm{erf}\Big(\frac{\epsilon^{m}-a}{\sqrt{2}S_{\sigma}}\Big)\Big].
\end{multline}

\noindent where $p_{1}$, $p_{2}$ and $p_{3}$ again given by Eq.~(\ref{eq:77}). Finally, selecting the prior distribution as in Eq.~(\ref{eq:51}), the posterior distribution for $k$ measurements reads:

\begin{equation}
\label{eq:82}
\pi(\textbf{x}|\boldsymbol{\sigma}^{m},\boldsymbol{\epsilon}^{m})\propto\textrm{exp}\Big(\frac{(\textbf{x}-\overline{\textbf{x}})^{T}\boldsymbol{\Gamma}_{\textbf{x}}^{-1}(\textbf{x}-\overline{\textbf{x}})}{2}\Big)\prod_{i=1}^{k}\pi(\sigma_{i}^{m}|\textbf{x},\epsilon_{i}^{m}).
\end{equation}

\subsection{Linear elastic-linear hardening}

\label{subsection:BI double error Linear elastic-linear hardening}

For the linear elastic-linear hardening model, the unknown parameters are $\textbf{x}=\begin{bmatrix} E&\sigma_{y0}&H\end{bmatrix}^{T}$. Assuming the same experimental equipment as in subsections \ref{subsection:BI double error Linear elastic} and \ref{subsection:BI double error Linear elastic-perfectly plastic} and employing Eqs.~(\ref{eq:20}) and (\ref{eq:70}), the likelihood function can be expressed as:

\begin{multline}
\label{eq:83}
\pi(\sigma^{m}|\textbf{x},\epsilon^{m})=\frac{1}{2\sqrt{2\pi}S_{\sigma}S_{\epsilon}}\bigg(\sqrt{p_{3}}\,\textrm{exp}\Big(-\frac{p_{2}-p_{1}^{2}}{2p_{3}}\Big)\Big[\textrm{erf}\Big(\frac{\frac{\sigma_{y0}}{E}-p_{1}}{\sqrt{2p_{3}}}\Big)-\textrm{erf}\Big(\frac{-p_{1}}{\sqrt{2p_{3}}}\Big)\Big]+\\
\frac{1}{\sqrt{\beta_{1}}}\textrm{exp}\Big(-\frac{\beta_{1}\beta_{3}-\beta_{2}^{2}}{2\beta_{1}}\Big)\Big[\textrm{erf}\Big(\frac{\sqrt{\beta_{1}}a-\frac{\beta_{2}}{\sqrt{\beta_{1}}}}{\sqrt{2}}\Big)-\textrm{erf}\Big(\frac{\frac{\sigma_{y0}\sqrt{\beta_{1}}}{E}-\frac{\beta_{2}}{\sqrt{\beta_{1}}}}{\sqrt{2}}\Big)\Big],
\end{multline}

\noindent where $p_{1}$, $p_{2}$ and $p_{3}$ are again given by Eq.~(\ref{eq:77}) and $\beta_{1}$, $\beta_{2}$ and $\beta_{3}$ are determined as follows:

\begin{equation}
\label{eq:84}
\begin{aligned}
\beta_{1}=\frac{\Big(\frac{HE}{H+E}\Big)^{2}}{S_{\sigma}^{2}}+\frac{1}{S_{\epsilon}^{2}}, \ \beta_{2}=\frac{(\sigma^{m}-\sigma_{y0})\frac{HE}{H+E}+\Big(\frac{HE}{H+E}\Big)^{2}\frac{\sigma_{y0}}{E}}{S_{\sigma}^{2}}+\frac{\epsilon^{m}}{S_{\epsilon}^{2}},\\
\beta_{3}=\frac{(\sigma^{m}-\sigma_{y0})^{2}+2(\sigma^{m}-\sigma_{y0})\frac{HE}{H+E}\frac{\sigma_{y0}}{E}+\Big(\frac{HE}{H+E}\Big)^{2}\Big(\frac{\sigma_{y0}}{E}\Big)^{2}}{S_{\sigma}^{2}}+\frac{(\epsilon^{m})^{2}}{S_{\epsilon}^{2}}.
\end{aligned}
\end{equation}

\noindent Selecting the same posterior distribution as in Eq.~(\ref{eq:54}), the posterior distribution is again of the following form:

\begin{equation}
\label{eq:85}
\pi(\textbf{x}|\boldsymbol{\sigma}^{m},\boldsymbol{\epsilon}^{m})\propto\textrm{exp}\Big(\frac{(\textbf{x}-\overline{\textbf{x}})^{T}\boldsymbol{\Gamma}_{\textbf{x}}^{-1}(\textbf{x}-\overline{\textbf{x}})}{2}\Big)\prod_{i=1}^{k}\pi(\sigma_{i}^{m}|\textbf{x},\epsilon_{i}^{m}).
\end{equation}

\noindent where $\pi(\sigma_{i}^{m}|\textbf{x},\epsilon_{i}^{m})$ is calculated using Eq.~(\ref{eq:83}) for each pair of measurements ($\sigma_{i}^{m},\epsilon_{i}^{m}$).

\subsection{Linear elastic-nonlinear hardening}

\label{subsection:BI double error Linear elastic-nonlinear hardening}

Like in subsection \ref{subsection:BI single error Linear elastic-nonlinear hardening} the parameter vector for the current case reads $\textbf{x}=\begin{bmatrix} E&\sigma_{y0}&H&n\end{bmatrix}^{T}$. As the experimental equipment remains the same, the additive noise model is still valid and hence, the following expressions for the measured stresses and strains are employed:

\begin{equation}
\label{eq:86}
\begin{cases}
\sigma^{m}=\sigma(\textbf{x},\epsilon)+\Omega_{\sigma}\\
\epsilon^{m}=\epsilon+\Omega_{\epsilon}
\end{cases}.
\end{equation}

\noindent In contrast to the previous models in this subsection, the theoretical stress ($\sigma(\textbf{x},\epsilon)$ in the first equation of Eq.~(\ref{eq:86})) is an implicit function (Eq.~(\ref{eq:23})) and as a result the construction of the likelihood function differs. First, Eq.~(\ref{eq:86}) is rewritten using Eq.~(\ref{eq:23}):

\begin{equation}
\label{eq:87}
\begin{cases}
\sigma^{m}=E\epsilon\bigg(1-h\Big(\epsilon-\frac{\sigma_{y0}}{E}\Big)\bigg)+\bigg(\sigma_{y0}+H\Big(\epsilon-\frac{\sigma(\epsilon, \textbf{x})}{E}\Big)^{n}\bigg)h\Big(\epsilon-\frac{\sigma_{y0}}{E}\Big)+\Omega_{\sigma}\\
\epsilon^{m}=\epsilon+\Omega_{\epsilon}
\end{cases},
\end{equation}

\noindent from which it is clear that the measured stress is not only a function of the material parameters ($\textbf{x}$) and the theoretical strain ($\epsilon$) but also of the theoretical stress ($\sigma(\textbf{x},\epsilon)$).

The following conditional probabilities for the measured stress and measured strain are established:

\begin{equation}
\label{eq:88}
\begin{cases}
\pi(\sigma^{m}|\textbf{x},\epsilon,\sigma(\textbf{x},\epsilon))=\pi_{\Omega_{\sigma}}(\sigma^{m}-\sigma(\textbf{x},\epsilon))\\
\pi(\epsilon|\epsilon^{m})=\pi_{\Omega_{\epsilon}}(\epsilon^{m}-\epsilon)
\end{cases}.
\end{equation}

\noindent The required likelihood function then reads:

\begin{equation}
\label{eq:89}
\pi(\sigma^{m}|\textbf{x},\epsilon^{m})=\int_{0}^{a}\int_{0}^{+\infty}\pi(\sigma^{m}|\textbf{x},\epsilon,\sigma(\textbf{x},\epsilon))\pi(\sigma(\textbf{x},\epsilon)|\textbf{x},\epsilon)\pi(\epsilon|\epsilon^{m})d\sigma d\epsilon,
\end{equation}

\noindent where $\pi(\sigma^{m}|\textbf{x},\epsilon,\sigma(\textbf{x},\epsilon))$ and $\pi(\epsilon|\epsilon^{m})$ are given by Eq.~(\ref{eq:88}). If the material parameters and the theoretical strain are given, the conditional probability $\pi(\sigma(\textbf{x},\epsilon)|\textbf{x},\epsilon)$ reads:

\begin{equation}
\label{eq:90}
\pi(\sigma(\textbf{x},\epsilon)|\textbf{x},\epsilon)=\delta(\sigma(\textbf{x},\epsilon)-w),
\end{equation}

\noindent where $w$ is again used to abbreviate the right hand side of Eq.~(\ref{eq:23}). \mbox{Consequently}, the resulting likelihood function reads:

\begin{multline}
\label{eq:91}
\pi(\sigma^{m}|\textbf{x},\epsilon^{m})=\frac{1}{2\pi S_{\sigma}S_{\epsilon}}\int_{0}^{a}\int_{0}^{+\infty}\delta(\sigma(\textbf{x},\epsilon)-w)\textrm{exp}\bigg(-\frac{(\epsilon^{m}-\epsilon)^{2}}{2S_{\epsilon}^{2}}\\-\frac{\Big(\sigma^{m}-E\epsilon\Big(1-h\big(\epsilon-\frac{\sigma_{y0}}{E}\big)\Big)-\Big(\sigma_{y0}+H\big(\epsilon-\frac{\sigma(\textbf{x},\epsilon)}{E}\big)^{n}\Big)h\big(\epsilon-\frac{\sigma_{y0}}{E}\big)\Big)^{2}}{2S_{\sigma^{2}}}\bigg)d\sigma d\epsilon.
\end{multline}

\noindent This can furthermore be expressed as:

\begin{equation}
\label{eq:92}
\pi(\sigma^{m}|\textbf{x},\epsilon^{m})=\frac{1}{2\pi S_{\sigma}S_{\epsilon}}(I_{1}+I_{2}),
\end{equation}

\noindent where

\begin{equation}
\label{eq:93}
I_{1}=\int_{0}^{\frac{\sigma_{y0}}{E}}\int_{0}^{+\infty}\delta(\sigma(\textbf{x},\epsilon)-w)\textrm{exp}\bigg(-\frac{(\epsilon^{m}-\epsilon)^{2}}{2S_{\epsilon}^{2}}-\frac{\Big(\sigma^{m}-E\epsilon\Big)^{2}}{2S_{\sigma^{2}}}\bigg)d\sigma d\epsilon,
\end{equation}

\noindent and

\begin{equation}
\label{eq:94}
I_{2}=\int_{\frac{\sigma_{y0}}{E}}^{a}\int_{0}^{+\infty}\delta(\sigma(\textbf{x},\epsilon)-w)\textrm{exp}\bigg(-\frac{(\epsilon^{m}-\epsilon)^{2}}{2S_{\epsilon}^{2}}-\frac{\Big(\sigma^{m}-\Big(\sigma_{y0}+H\big(\epsilon-\frac{\sigma(\textbf{x},\epsilon)}{E}\big)^{n}\Big)h\big(\epsilon-\frac{\sigma_{y0}}{E}\big)\Big)^{2}}{2S_{\sigma^{2}}}\bigg)d\sigma d\epsilon.
\end{equation}

Using the Dirac delta function properties and Eq.~(\ref{eq:62}), $I_{1}$ equals to:

\begin{equation}
\label{eq:95}
I_{1}=\frac{\sqrt{\pi p_{3}}}{\sqrt{2}}\textrm{exp}\Big(-\frac{p_{2}-p_{1}^{2}}{2p_{3}}\Big)\Big[\textrm{erf}\Big(\frac{\frac{\sigma_{y0}}{E}-p_{1}}{\sqrt{2p_{3}}}\Big)-\textrm{erf}\Big(\frac{-p_{1}}{\sqrt{2p_{3}}}\Big)\Big],
\end{equation}

\noindent where $p_{1}$, $p_{2}$ and $p_{3}$ are given by Eq.~(\ref{eq:77}). To calculate $I_{2}$, one first needs to find the solution for $g(z)$ in Eq.~(\ref{eq:62}), for which it must be substituted by $w$. Again using the Dirac delta function properties and integrating Eq.~(\ref{eq:94}) over the theoretical stress, $I_{2}$ reads: 

\begin{equation}
\label{eq:96}
I_{2}=\int_{\frac{\sigma_{y0}}{E}}^{a}\textrm{exp}\bigg(-\frac{(\epsilon^{m}-\epsilon)^{2}}{2S_{\epsilon}^{2}}-\frac{\Big(\sigma^{m}-\big(\sigma_{y0}+H\big(\epsilon-\frac{w}{E}\big)^{n}\big)\Big)^{2}}{2S_{\sigma}^{2}}\bigg)\frac{1}{\left|1+\frac{Hn}{E}(\epsilon-\frac{w}{E})^{n-1}\right|} d\epsilon,
\end{equation}

\noindent which can be calculated by numerical approaches (e.g.~Simpson's rule). Finally for the same prior distribution as in Eq.~(\ref{eq:64}), the posterior distribution for $k$ observations becomes:

\begin{equation}
\label{eq:97}
\pi(\textbf{x}|\boldsymbol{\sigma}^{m},\boldsymbol{\epsilon}^{m})\propto\textrm{exp}\Big(\frac{(\textbf{x}-\overline{\textbf{x}})^{T}\boldsymbol{\Gamma}_{\textbf{x}}^{-1}(\textbf{x}-\overline{\textbf{x}})}{2}\Big)\prod_{i=1}^{k}\pi(\sigma_{i}^{m}|\textbf{x},\epsilon_{i}^{m}).
\end{equation}

\noindent Note that in practice once a sample is drawn by the adaptive MCMC approach, the integral in Eq.~(\ref{eq:97}) and subsequently the likelihood function can be calculated.

\section{Examples}

\label{section:Examples}

All formulations derived in the previous two sections are investigated below. The effect of the prior distribution on the posterior distribution is studied, as well as BI's ability to recover a material parameter distribution when they are taken from a specific distribution. Also, BI's ability to recover correlations between different material parameters is exposed.

\subsection{Bayesian inference with noise in stress}

\label{subsection: Examples Bayesian inference with noise in stress}

This subsection presents several examples for the BI formulations of section \ref{section:Bayesian inference for tensile tests with noise in the stress}, in which only a statistical noise in the stress measurements is considered.

\subsubsection{Linear elastic (LE)}

\label{subsubsection:Examples linear elastic single noise}

\paragraph{Identification of the Young's modulus} In the first example a specimen with a Young's modulus of $210\ \textrm{GPa}$ is considered, which is to be identified. `Calibration experiments' were performed and the noise in the stress follows the normal distribution of Eq.~(\ref{eq:40}) with $S_\text{noise}=0.01\ \textrm{GPa}$. For only one stress measurement of $\sigma^{m}=0.1576\ \textrm{GPa}$ with corresponding strain $\epsilon=7.25\times10^{-4}$, the posterior distribution is calculated using Eq.~(\ref{eq:46}). Selecting the prior distribution as in Eq.~(\ref{eq:43}) with mean $\overline{E}=150\ \textrm{GPa}$ and a relatively large standard deviation $S_{E}= 50\ \textrm{GPa}$, the posterior reads:

\begin{equation}
\label{eq:98}
\begin{matrix}
\pi(E|\sigma^{m})\propto \textrm{exp}\Big(-\frac{(E-\mu_\text{post})^{2}}{2S_\text{post}^{2}}\Big),& E\geq0,
\end{matrix}
\end{equation}

\noindent where $\mu_\text{post}=212.6486\ \textrm{GPa}$ and $S_\text{post}=13.2964\ \textrm{GPa}$. 

\begin{figure}[H]
\begin{minipage}[t]{0.5\linewidth}
\centering
\subcaptionbox{One observation\label{fig:3_a}}{\includegraphics[width=\textwidth,trim={1cm 0.5cm 1cm 1cm},clip]{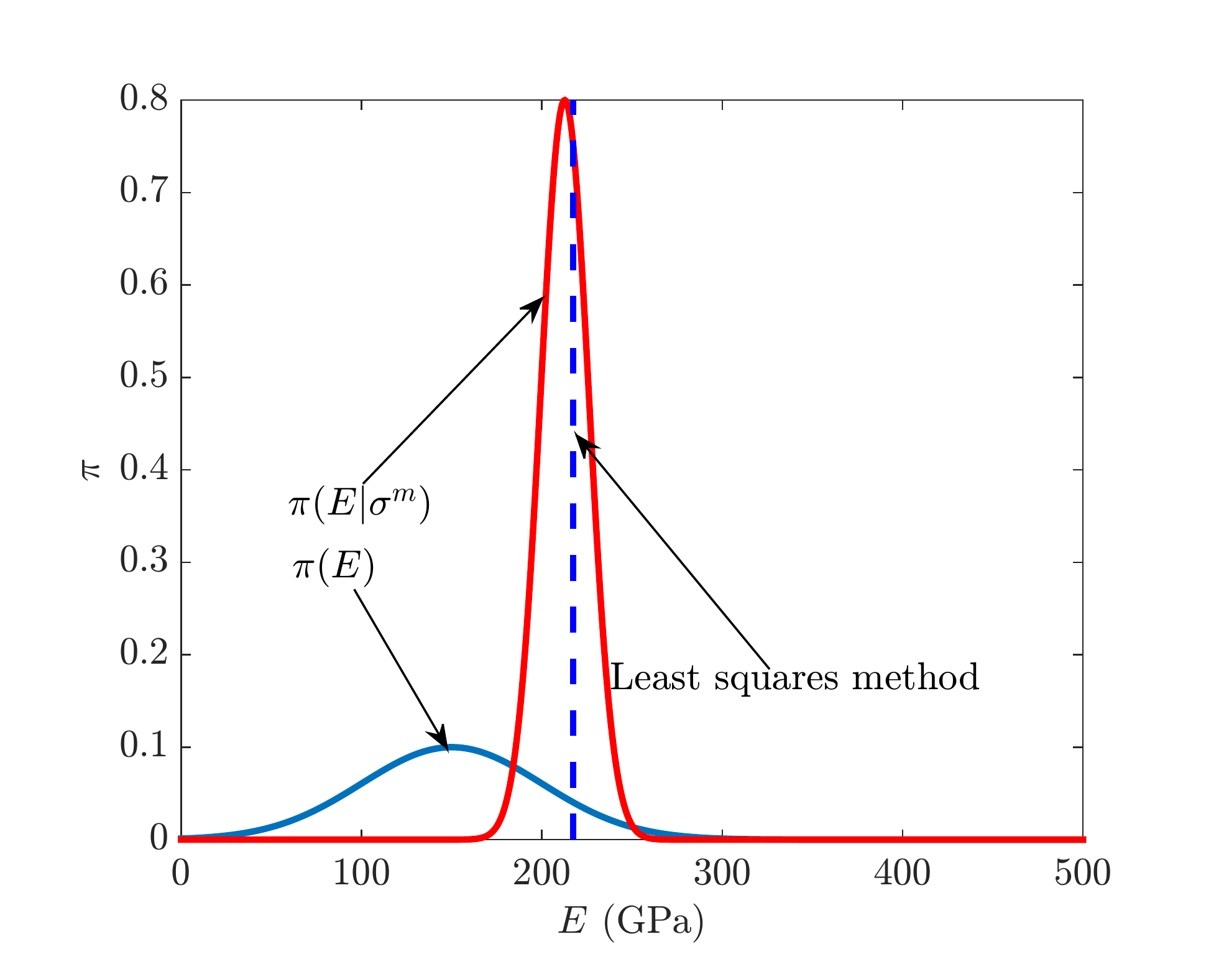}}
\end{minipage}
\begin{minipage}[t]{0.5\linewidth}
\centering
\subcaptionbox{Five successive observations\label{fig:3_b}}{\includegraphics[width=\textwidth,trim={1cm 0.5cm 1cm 1cm},clip]{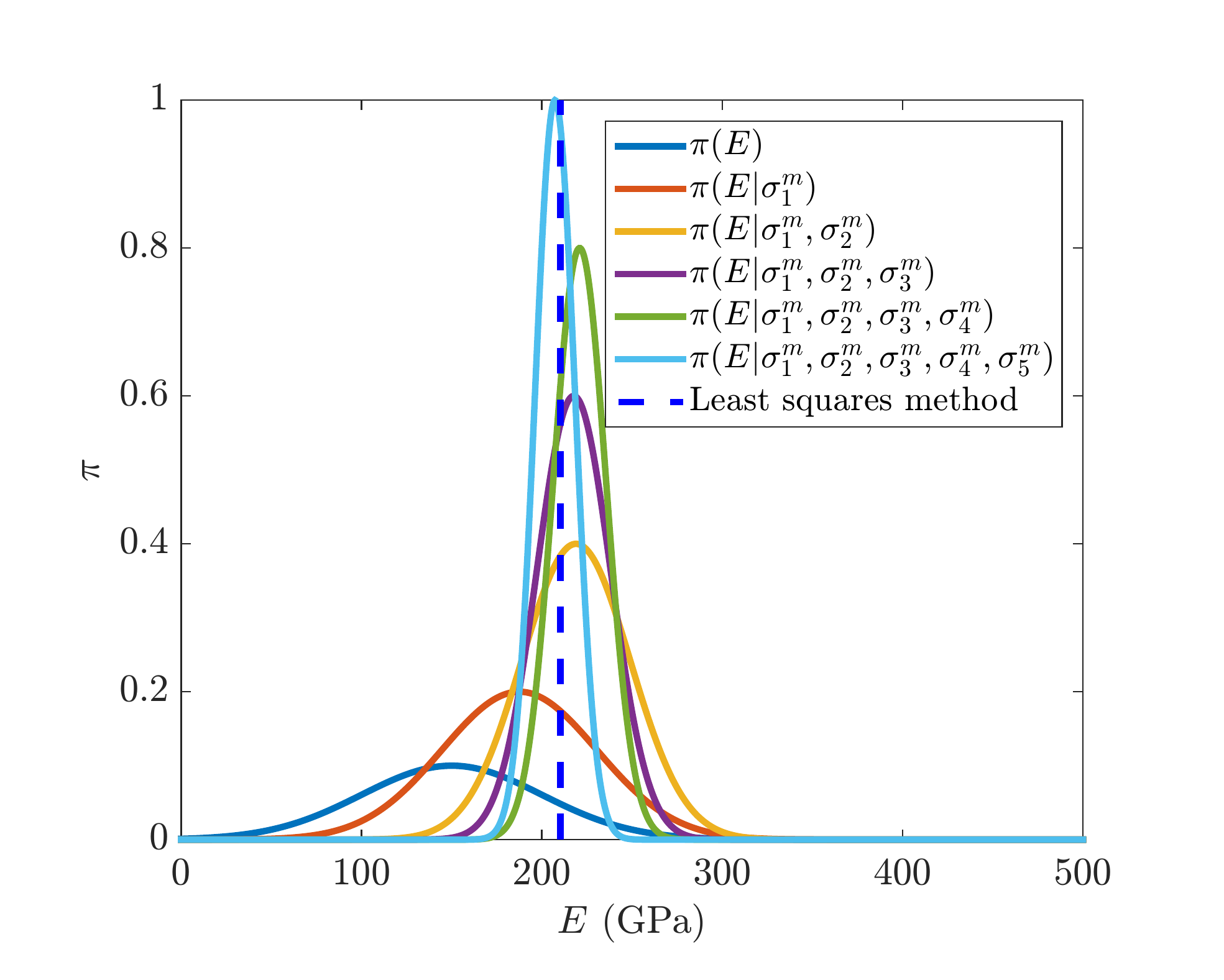}}
\end{minipage}
\caption{Linear elastic: The prior, the posterior and the value predicted by least squares method for one measurement \subref{fig:3_a} and five measurements \subref{fig:3_b}. The distributions are not normalised. The strain at which a measurement is made has a considerable influence on the posterior. This can be observed when the posterior of \subref{fig:3_a} ($\pi(E|\sigma^{m})$, red line) is compared to the posterior of \subref{fig:3_b} when only the first measurement is incorporated ($\pi(E|\sigma_{1}^{m})$, red line). An increase of the number of measurements leads to narrow posteriors with their MAP estimates closer to the specimen's Young's modulus.}
\label{fig:3}
\end{figure}

\begin{figure}[H]
\begin{minipage}[t]{0.5\linewidth}
\centering
\subcaptionbox{One observation\label{fig:4_a}}{\includegraphics[width=\textwidth,trim={0.75cm 0.5cm 1cm 1cm},clip]{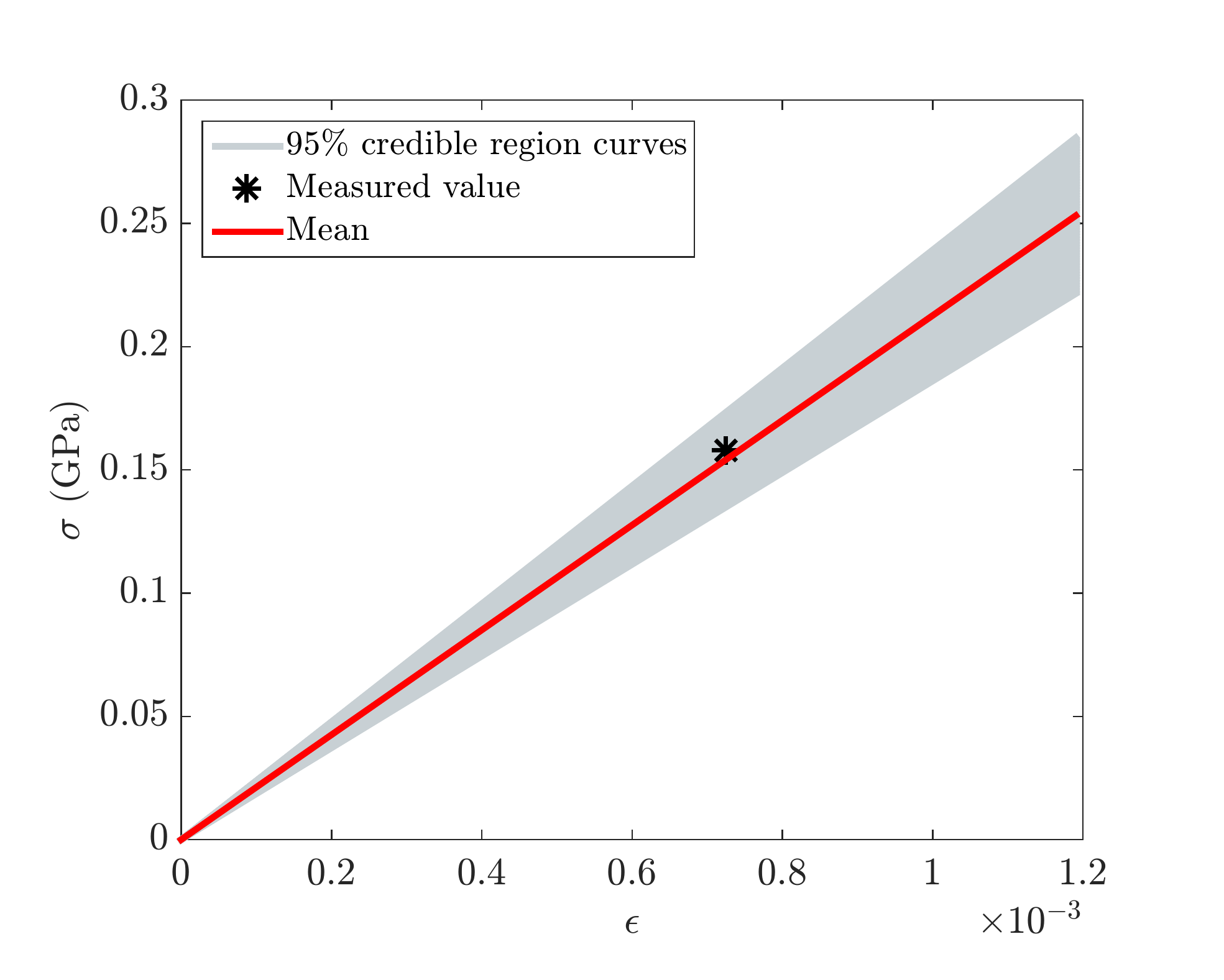}}
\end{minipage}
\begin{minipage}[t]{0.5\linewidth}
\centering
\subcaptionbox{Ten observations\label{fig:4_b}}{\includegraphics[width=\textwidth,trim={0.75cm 0.5cm 1cm 1cm},clip]{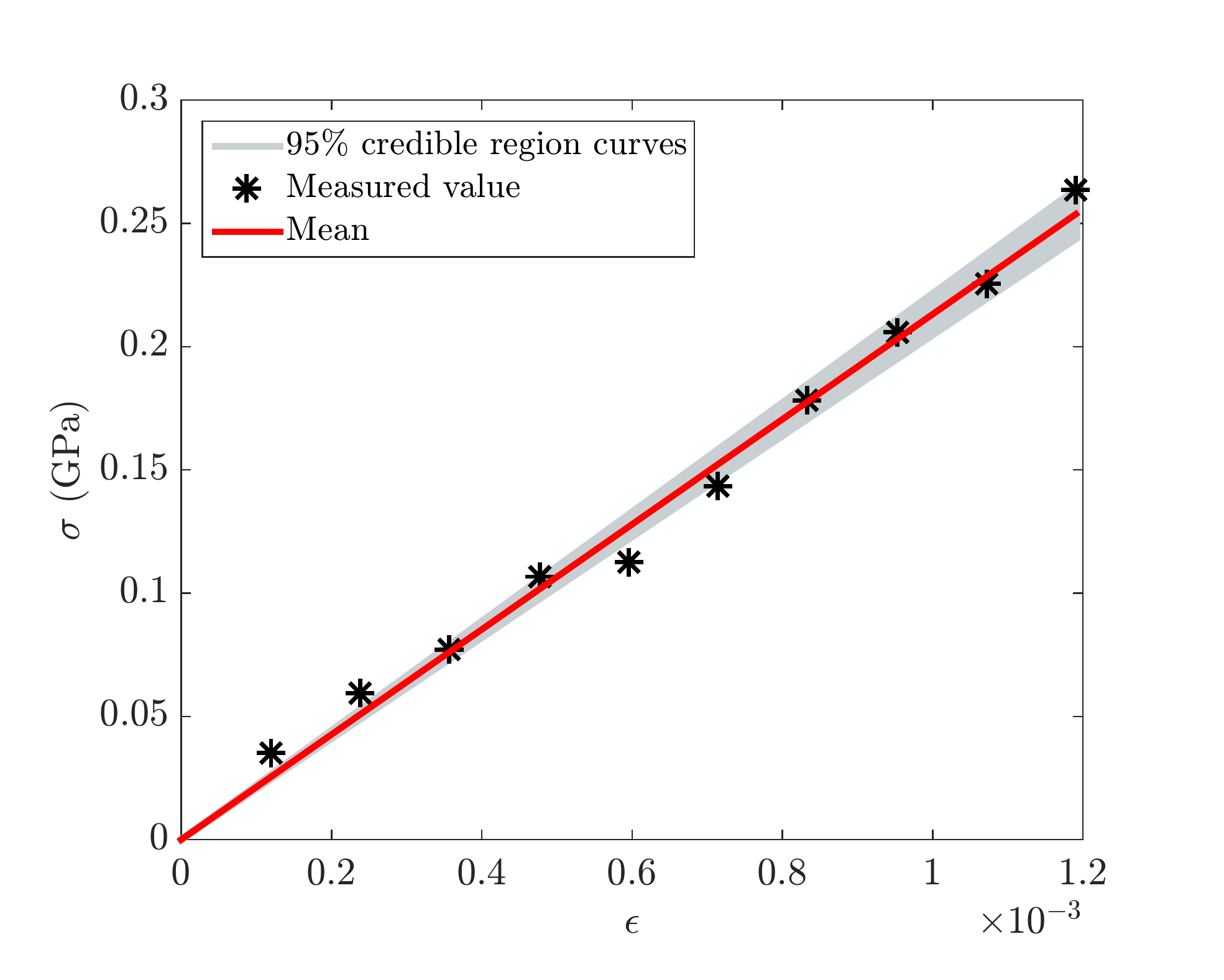}}
\end{minipage}
\caption{Linear elastic: The measurements, and the stress-strain curves created using the posterior for \subref{fig:4_a} one measurement and \subref{fig:4_b} ten measurements. Increasing the number of measurements leads to a smaller uncertainty, since the bandwidth of the responses associated with the $95\%$ credible region is much smaller in \subref{fig:4_b} than in \subref{fig:4_a}.}
\label{fig:4}
\end{figure}

Fig.~\ref{fig:3} shows this posterior distribution, as well as the prior distribution and the value predicted by the least squares method when one measurement is made and also when five measurements are made. Fig.~\ref{fig:4} presents the linear elastic responses for the case when one measurement is made and for the case when ten measurements are made. The figure also shows the stress-strain responses made with Young's moduli taken within the $95\%$ credible region (i.e.~the region that contains $95\%$ of the posterior) of the posterior.

Two points can be observed based on Fig.~\ref{fig:3}. First, the strain at which a measurement is made has a strong influence on the posterior. This can be observed when the posterior of Fig.~\ref{fig:3_a} is compared to the posterior of Fig.~\ref{fig:3_b} when only the first measurement is incorporated (the distribution in red, denoted by $\pi(E |\sigma^{m}_{1})$). The latter distribution is significantly wider and its MAP point is relatively distant from the specimen's Young's modulus. Hence, a measurement made at a comparatively large strain reduces the width of the posterior distribution (i.e.~reduces the uncertainty).

The second remark is that for an increasing number of measurements, the posterior becomes narrower and the MAP point moves closer to the specimen's Young's modulus.

When the MAP point for a single measurement in Fig.~\ref{fig:3_b} ($\mu_\text{post}=207.2821\ \textrm{GPa}$) is compared with the result of the least squares method for the same measurement ($E_\text{ls}=210.2216\ \textrm{GPa}$), one can notice the effect of the selected prior distribution. One interpretation of this is that the least squares method gives a more accurate result than BI (although this depends the selected prior), as the result of the least squares method is closer to the specimen's Young's modulus than the MAP point determined using BI. On the other hand, the result determined using the least squares method is not the actual Young's modulus of the specimen ($210\ \textrm{GPa}$), whereas the posterior distribution of BI does include this value. Furthermore, the MAP point and mean value of BI, comes with an uncertainty in terms of the parameter value itself. This can be considered an advantage when one wants to include this uncertainty, instead of including one deterministic value.

The main point is that BI cannot be directly compared to the least squares method, because contrary to the latter, it considers the Young's modulus to be stochastic in nature. This means that BI accounts for the fact that parameters originate from a distribution (for which the user is required to give an \emph{a-priori} belief, which is reflected in the prior). In the least squares method the Young's modulus is a deterministic value, which can only be determined using the current measurements. Hence, it cannot consider the effect that other measurements could have been made (due to different noise realisations).

\begin{figure}[H]
\begin{center}
\includegraphics[width=0.6\textwidth,trim={0.5cm 1cm 0.75cm 0.5cm},clip]{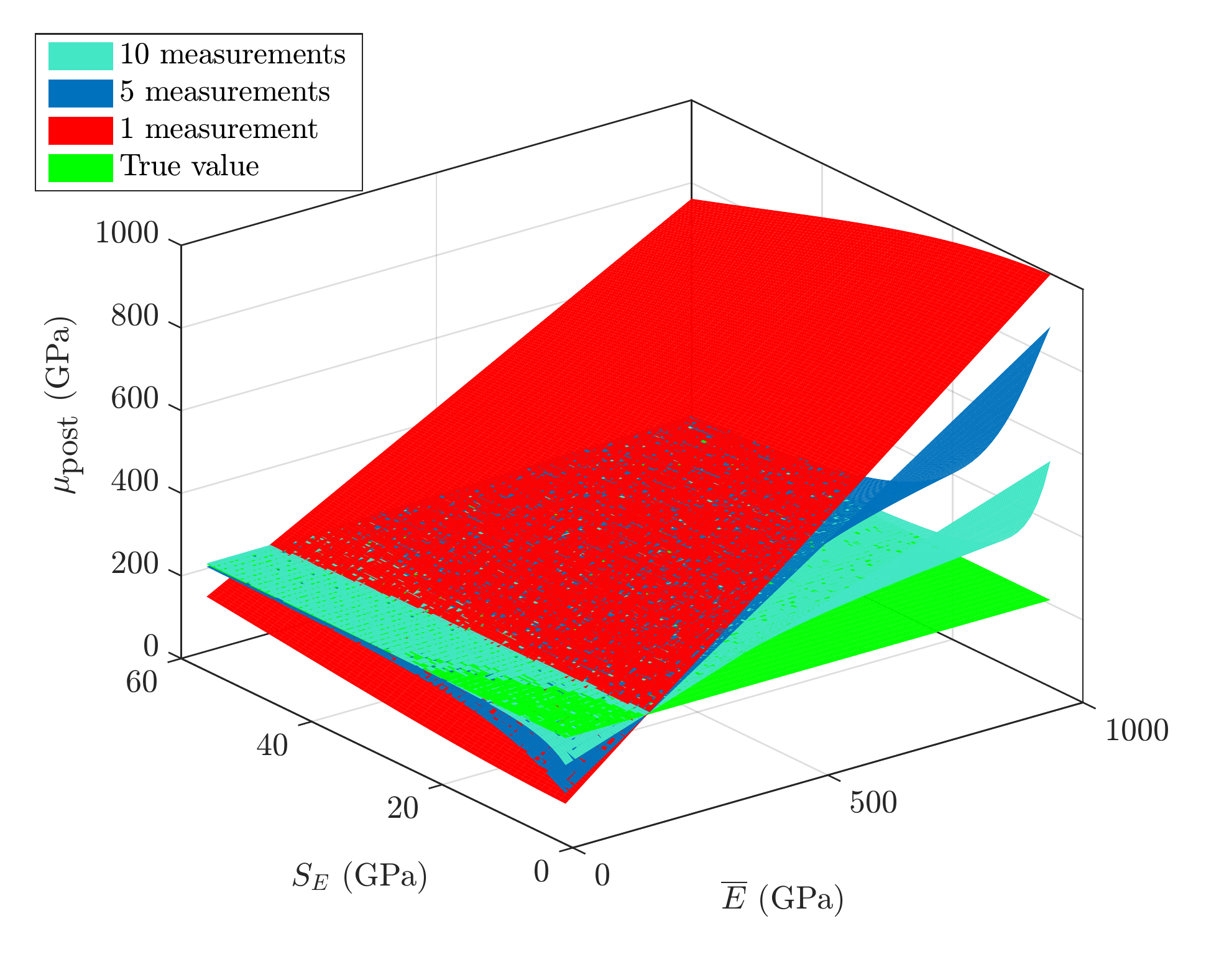}
\end{center}
\caption{Linear elastic: The influence of the prior (i.e.~the mean value and the standard deviation) on the resulting MAP point for different numbers of measurements. Increasing the number of the measurements leads to a flatter surface which indicates a decreasing influence of the prior distribution.}
\label{fig:5}
\end{figure}

\paragraph{The influence of the prior on the identified Young's modulus} It is nevertheless interesting to study the effect of the prior distribution on the MAP point (which is the same as the mean value for the normal posteriors in this subsection). In Fig.~\ref{fig:5} the MAP points are shown as a function of the mean and the standard deviation of the prior. The MAP points are presented for different numbers of measurements. As can be seen, an increase of the number of measurements leads to a flatter surface, which means that the influence of the prior distribution decreases.

\begin{figure}[H]
\begin{center}
\includegraphics[width=0.5\textwidth,trim={0.5cm 0.75cm 0.75cm 0.5cm},clip]{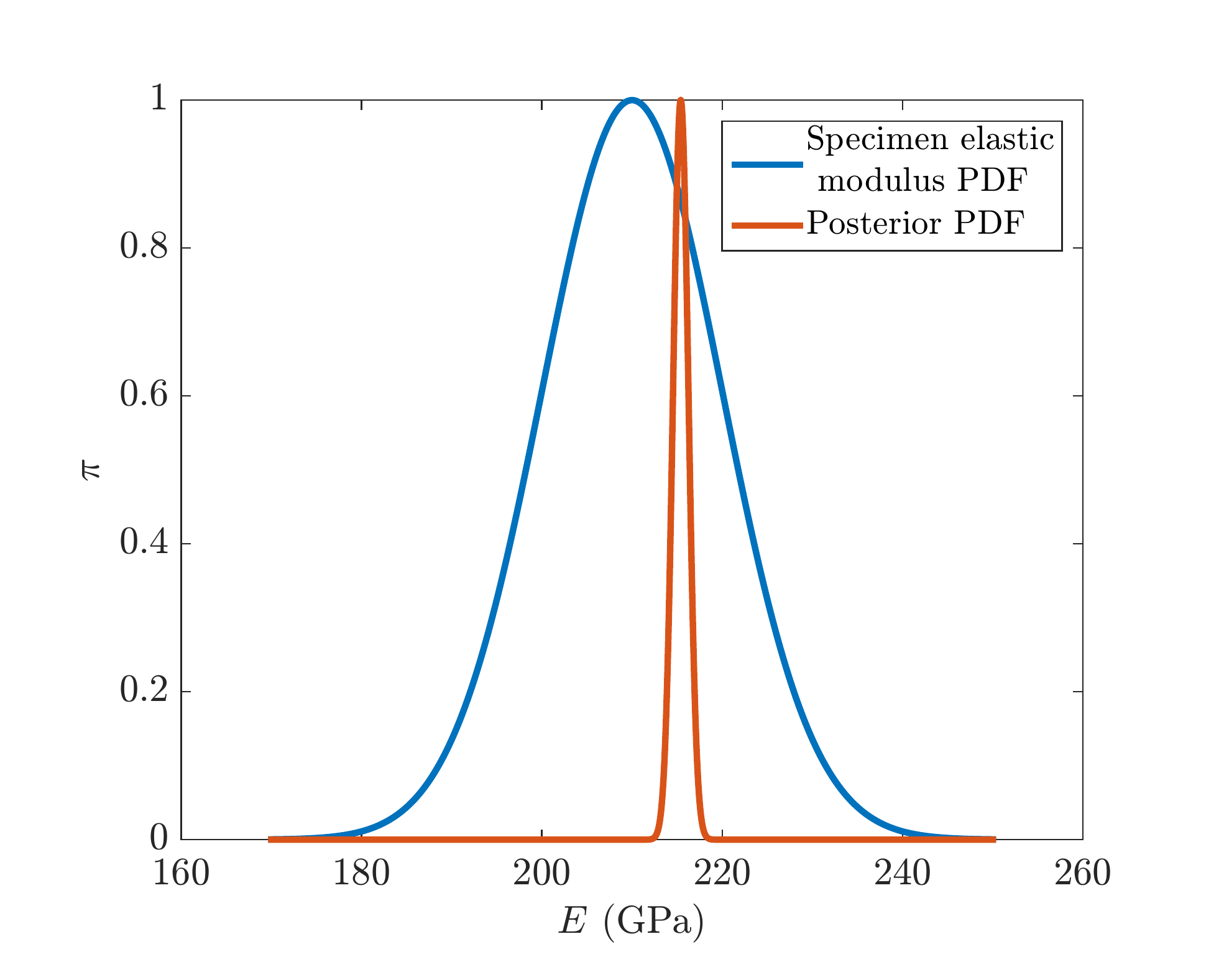}
\end{center}
\caption{Linear elastic: The distribution of the specimens' Young's moduli and the resulting posterior. The PDFs are not normalised. The current formulation is clearly not able to recover the material heterogeneity. To be able to recover the material heterogeneity one needs to consider both the inherent uncertainty in the material parameters as well as that in the measurements.}
\label{fig:6}
\end{figure}

\paragraph{Recovering the heterogeneity of the Young's modulus} A last important point to show using the linear elastic model is BI's ability (or inability for the current formulation) to capture the heterogeneity in the material parameters. The question here is thus if BI is able to recover the distribution of the Young's modulus when multiple specimens are tested and their Young's moduli are taken from a specific underlying distribution. To this end, $25$ specimens are considered of which the Young's moduli are taken from a normal distribution with a mean value of $210\ \textrm{GPa}$ and a standard deviation of $10\ \textrm{GPa}$ (blue curve in Fig.~\ref{fig:6}). For each specimen ten measurements are made. The same noise model and noise distribution are applied.

The resulting posterior is represented by the red curve in Fig.~\ref{fig:6}, which is a (modified) normal distribution with $\mu_\text{post}=215.3971\ \textrm{GPa}$ and $S_\text{post}=0.8561\ \textrm{GPa}$. The posterior is substantially narrower than the distribution of the specimens' Young's moduli and hence, using the BI formulations of this contribution, the material heterogeneity cannot be captured. This entails that the width of the posterior distributions (represented by $S_\text{post}$ in this subsection) is only a measure of the uncertainty of the MAP points and the mean value and not of the material heterogeneity.

To be able to recover the material heterogeneity one needs to consider both the inherent uncertainty in the material parameters as well as the uncertainty in the measurements. In that case, one does not only aim to determine the posterior of the material parameter(s), but also the posterior of the variance of the assumed material parameter distribution. Our future work will focus on this.

\subsubsection{Linear elastic-perfectly plastic (LE-PP)}

\label{subsubsection:Examples Linear elastic-perfectly plastic single noise}

\paragraph{Identification of the material parameters} In the first example of this subsection one linear elastic-perfectly plastically behaving specimen is considered with Young's modulus $E=210\ \textrm{GPa}$ and initial yield stress $\sigma_{y0}=0.25\ \textrm{GPa}$. Twelve measurements are generated by employing the same noise distribution as in the previous subsection. The prior distribution of Eq.~(\ref{eq:51}) is furthermore selected with the following mean vector and covariance matrix:

\begin{equation}
\label{eq:99}
\overline{\textbf{x}}=\begin{bmatrix}
200 \\
0.29
\end{bmatrix}\ \textrm{GPa},\ \boldsymbol{\Gamma_{\textbf{x}}}=\begin{bmatrix}
2500 & 0 \\
0 & 2.7778\times10^{-4}
\end{bmatrix} \textrm{GPa}^{2}.
\end{equation}

Consequently, the posterior for the LE-PP model of subsection \ref{subsubsection:Linear elastic-perfectly plastic} is of the form as presented in Eq.~(\ref{eq:52}), which is investigated by the MCMC approach given in subsection \ref{subsubsection:The standard and the adaptive Metropolis algorithm}. Running the chain for $10^{4}$ samples whilst burning the first 3000 samples (i.e.~the first $3000$ samples
are not used to determine the mean, the covariance matrix and the MAP estimate) yields:

\begin{equation}
\label{eq:100}
\widehat{\boldsymbol{\mu}}_\text{post}=\begin{bmatrix}
208.9859\\
0.2578
\end{bmatrix}\ \textrm{GPa},\ \widehat{\boldsymbol{\Gamma}}_\text{post}=\begin{bmatrix}
29.807  & 4.1064\times10^{-4}\\
4.1064\times10^{-4} & 1.5067\times{10}^{-5}
\end{bmatrix}\ \textrm{GPa}^{2},
\end{equation}

\noindent and

\begin{equation}
\label{eq:101}
\widehat{\textbf{MAP}}=\begin{bmatrix}
208.4475 \\
 0.2578 
\end{bmatrix}\ \textrm{GPa},
\end{equation}

\noindent where the hat sign ($\widehat{\cdot}$) denotes the numerical approximation.

\begin{figure}[H]
\begin{minipage}[t]{0.5\linewidth}
\centering
\subcaptionbox{Samples generated by the adaptive MCMC approach\label{fig:7_a}}{\includegraphics[width=\textwidth,trim={0.5cm 1cm 1cm 1cm},clip]{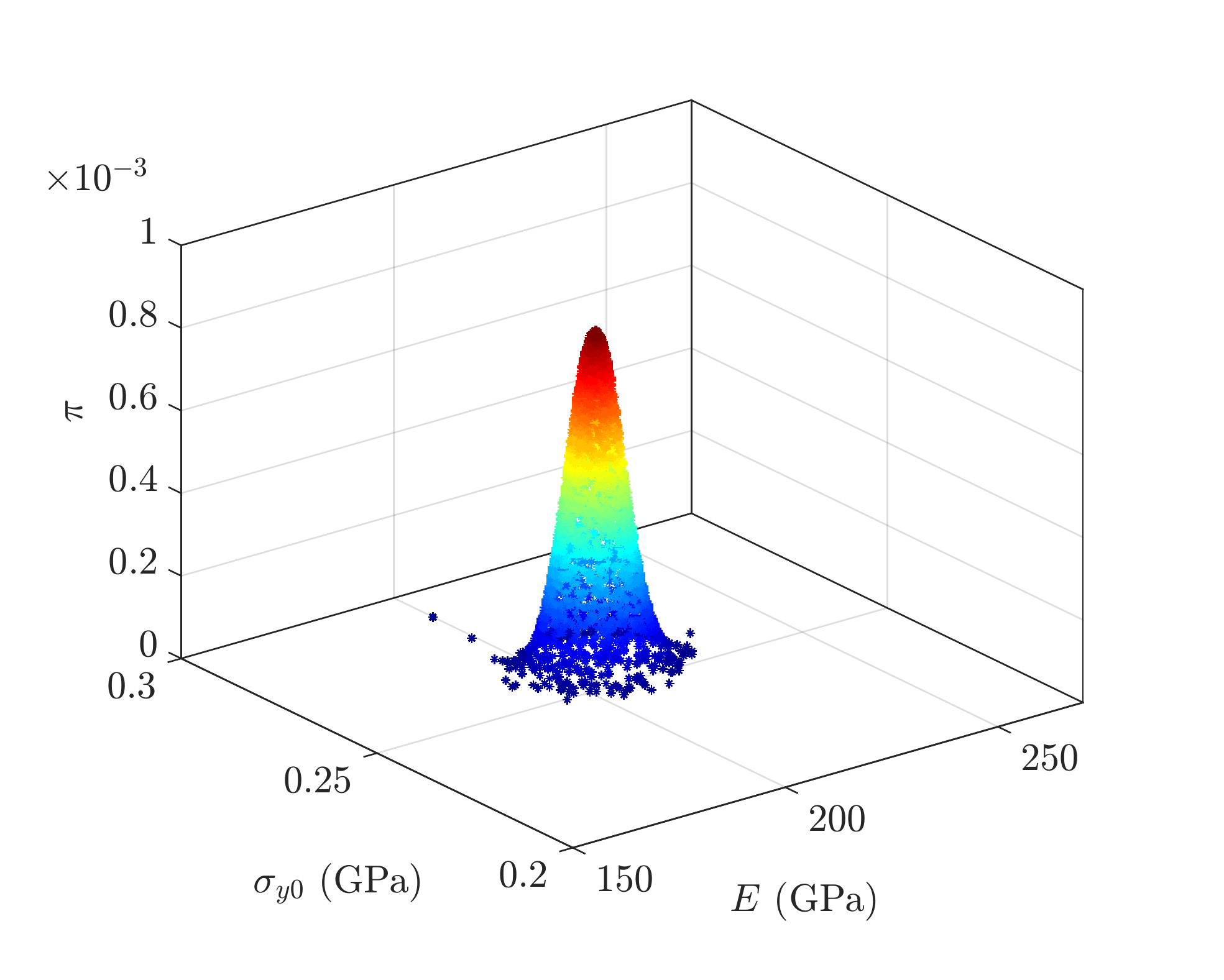}}
\end{minipage}
\begin{minipage}[t]{0.5\linewidth}
\centering
\subcaptionbox{Samples generated by the adaptive MCMC approach (top view) including different domains\label{fig:7_b}}{\includegraphics[width=\textwidth,trim={0.75cm 0.5cm 1cm 0.75cm},clip]{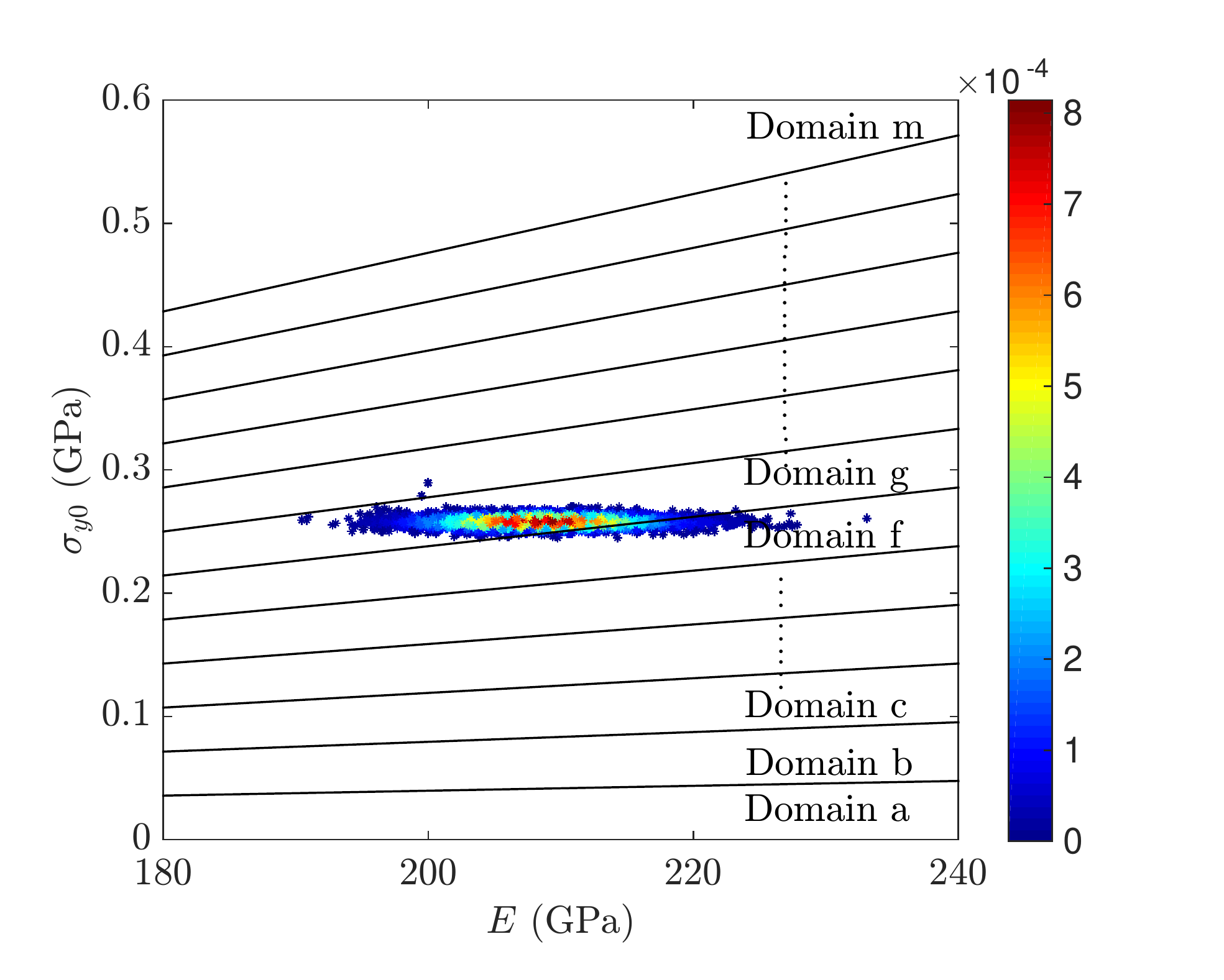}}
\end{minipage}
\caption{Linear elastic-perfectly plastic: Two different views of the samples generated by the adaptive MCMC approach to approximate the posterior. The colours represent the value of the posterior, which in the left image is also shown along the z-axis. In Fig.~\ref{fig:7_b} several domains are shown. For each of these domains, the measurements considered to be in the elastic part and the measurements considered to be in the plastic part remain the same.}
\label{fig:7}
\end{figure}

Fig.~\ref{fig:7_a} shows the samples generated by the adaptive MCMC approach which are used to approximate the posterior distribution. The domains presented in Fig.~\ref{fig:7_b} show which of the measurements are included in the elastic part and which fall within the plastic part. These discrete domains are a result of the $C_{0}$-continuity of Eq.~(\ref{eq:17}). In domain `a' (in which no samples are generated by the adaptive MCMC approach), all the measurements are considered to be in the plastic part. In domain `b' on the other hand, the first measurement (the one with the smallest strain) is considered to be in the elastic part and the others remain in the plastic part. Continuing like this, in domain `c' the second measurement is also considered to be in the elastic part. Finally, in domain `m' all measurements are considered to fall within the plastic domain. Based on Fig.~\ref{fig:7_b} the MAP point is clearly located in the domain in which the first six measurements are considered to be in the elastic part and the remaining in the plastic part.

\begin{figure}[H]
\begin{minipage}[t]{0.5\linewidth}
\centering
\subcaptionbox{ The $95 \%$ credible region\label{fig:8_a}}{\includegraphics[width=\textwidth,trim={0.6cm 0.5cm 1cm 0.75cm},clip]{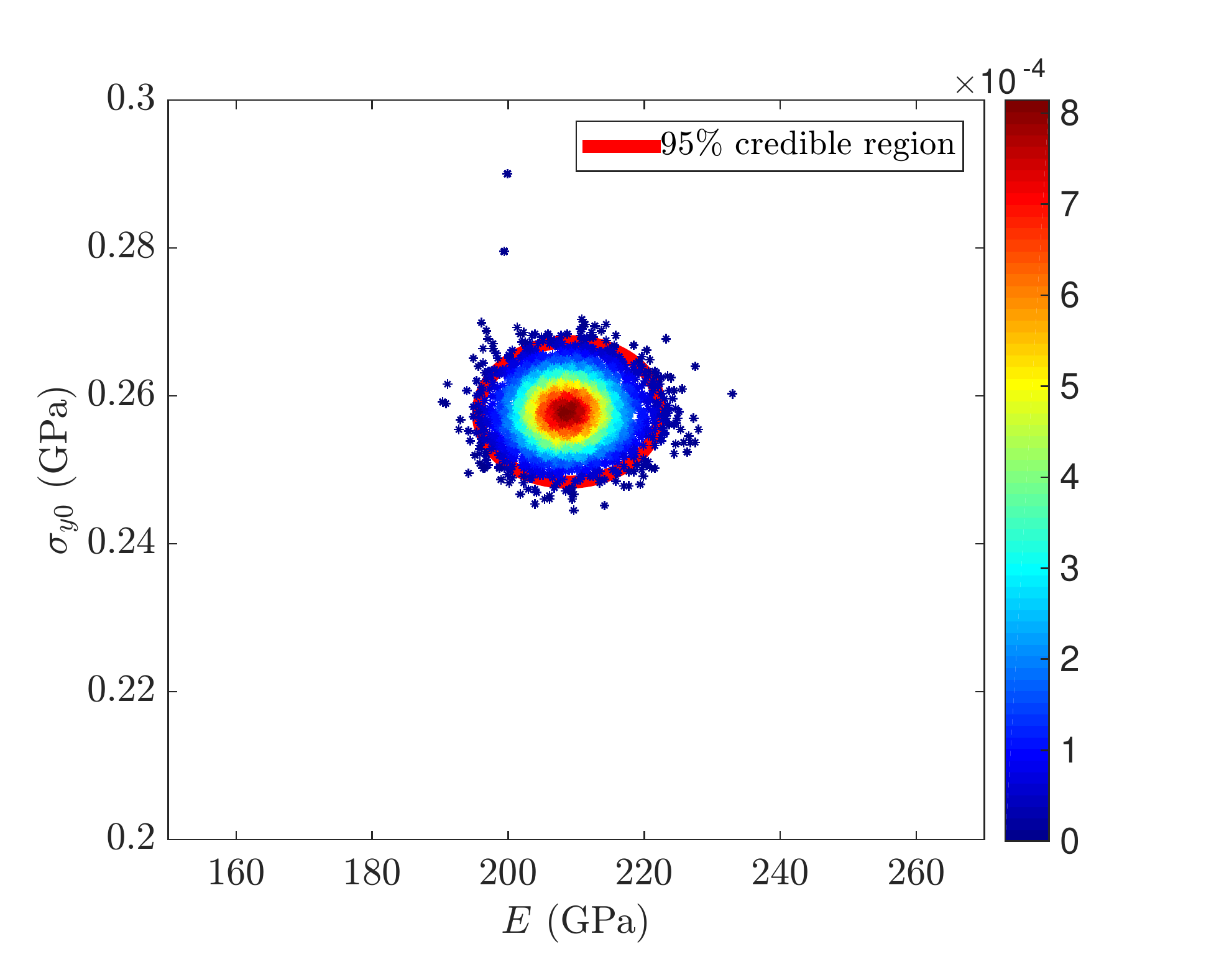}}
\end{minipage}
\begin{minipage}[t]{0.5\linewidth}
\centering
\subcaptionbox{The measurements and the stress-strain curves\label{fig:8_b}}{\includegraphics[width=\textwidth,trim={0.85cm 0.5cm 1cm 0.75cm},clip]{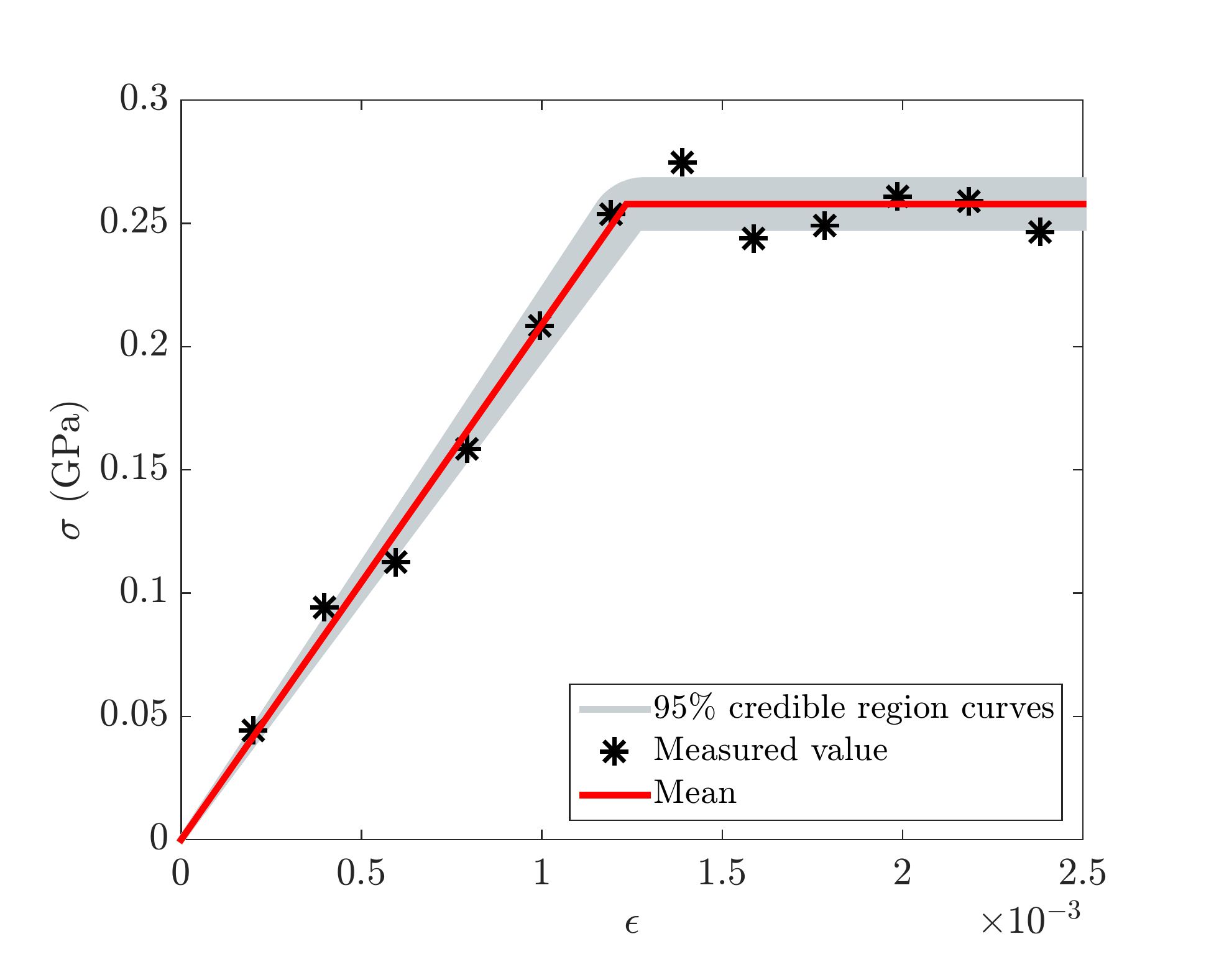}}
\end{minipage}
\caption{Linear elastic-perfectly plastic: The $95 \%$ credible region and the posterior distribution \subref{fig:8_a}, the measurements and the stress-strain curves created using the posterior \subref{fig:8_b}. The responses associated with the $95\%$ credible region in \subref{fig:8_b} are plotted using the points inside the $95\%$ credible region marked by the red ellipse in \subref{fig:8_a}.}
\label{fig:8}
\end{figure}

The $95\%$ credible region is shown together with the posterior distribution in Fig.~\ref{fig:8_a}. The possible stress-strain responses inside the credible region are presented in Fig.~\ref{fig:8_b}.

\begin{figure}
\begin{minipage}[t]{\columnwidth}
\centering
\subcaptionbox{Effect of $(\Gamma_{\textbf{x}}){12}$ on $(\widehat{\Gamma}_\text{post})_{11}$\label{fig:9_a}}{\includegraphics[width=0.32\textwidth,trim={0.5cm 0.5cm 1cm 0.5cm},clip]{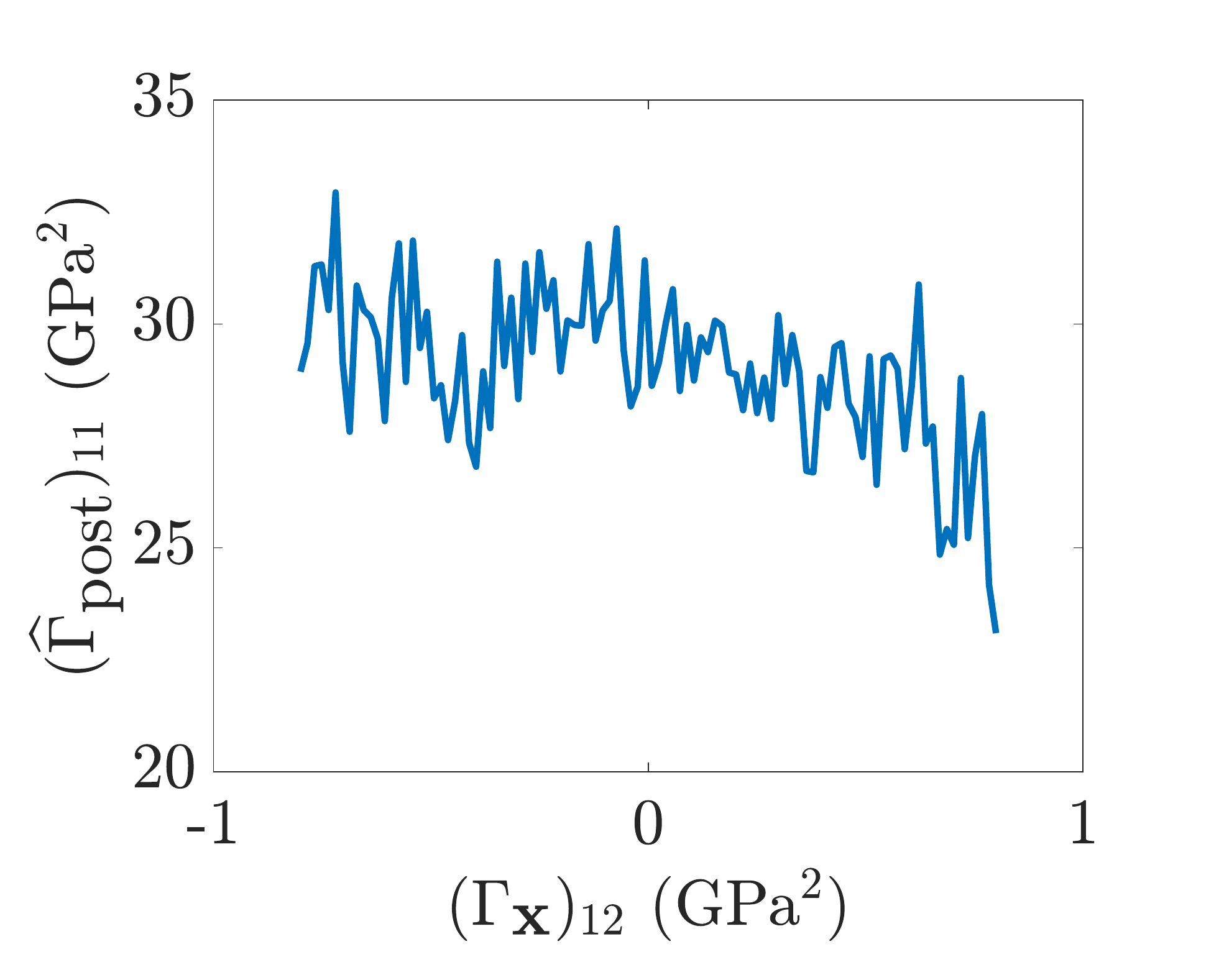}}
\subcaptionbox{Effect of $(\Gamma_{\textbf{x}})_{12}$ on $(\widehat{\Gamma}_\text{post})_{12}$\label{fig:9_b}}{\includegraphics[width=0.32\textwidth,trim={0.5cm 0.5cm 1cm 0.5cm},clip]{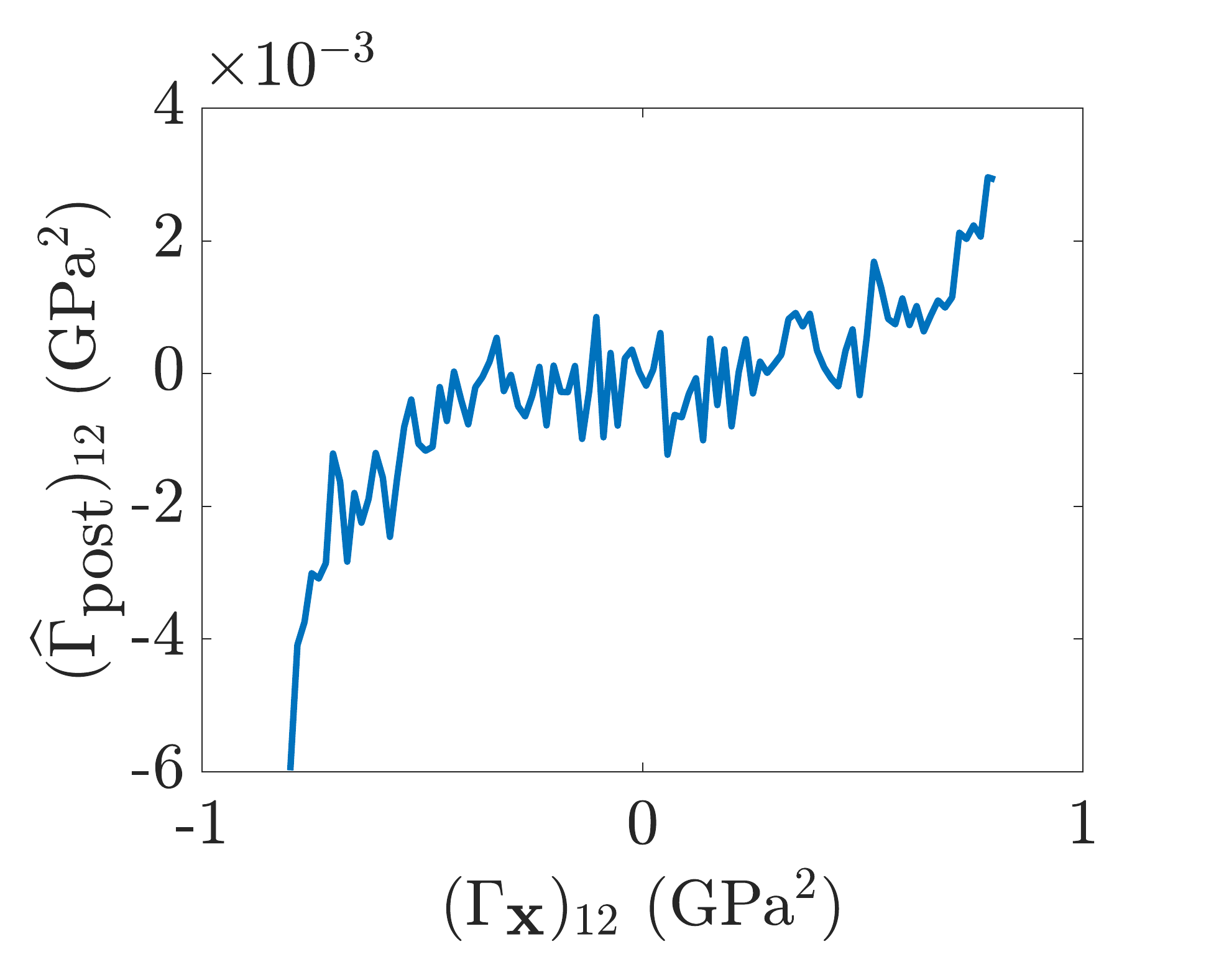}}
\end{minipage}
\begin{minipage}[t]{\columnwidth}
\centering
\subcaptionbox{Effect of $(\Gamma_{\textbf{x}})_{12}$ on $(\widehat{\Gamma}_\text{post})_{22}$\label{fig:9_c}}{\includegraphics[width=0.32\textwidth,trim={0.5cm 0.5cm 1cm 0.5cm},clip]{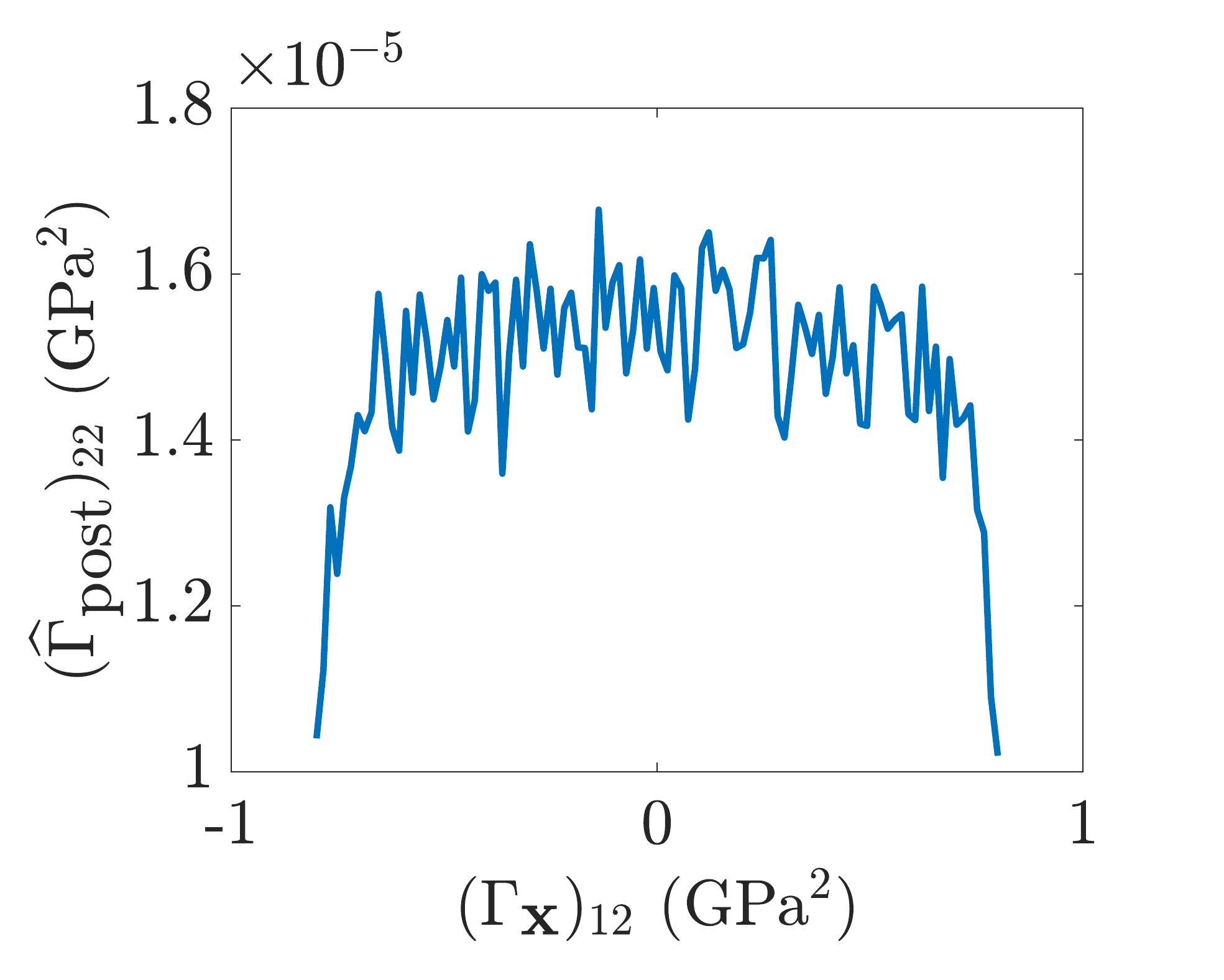}}
\end{minipage}
\caption{Linear elastic-perfectly plastic: Effect of the off-diagonal component of the prior covariance matrix on the posterior's covariance matrix. It seems that an increase of $(\Gamma_\textbf{x})_{12}$ leads to some decreasing trend of $(\widehat{\Gamma}_\text{post})_{11}$ and some increasing trend of $(\widehat{\Gamma}_\text{post})_{12}$. However, it is difficult to assess whether or not a true trend is present in these results. The hat sign ($\widehat{\cdot}$) indicates that the values are computed numerically using the adaptive MCMC approach.}
\label{fig:9}
\end{figure}

The posterior distribution seems be roughly of an an elliptical shape with the primary axes almost along the $E$-axis and $\sigma_{y0}$-axis. This entails that the correlation between the two material parameters is not significant. One has to notice though, that the assumed prior is uncorrelated. In other words, the prior covariance matrix ($\boldsymbol{\Gamma}_{\textbf{x}}$) is diagonal. It is therefore interesting to investigate the influence of the off-diagonal term of the prior covariance matrix on the posterior covariance matrix. In Fig.~\ref{fig:9} this influence is graphically presented for the three terms of the posterior covariance matrix (note that both the prior covariance matrix and the posterior covariance matrix are symmetric). It seems that an increase of $(\Gamma_\textbf{x})_{12}$ leads to some decreasing trend for $(\widehat{\Gamma}_\text{post})_{11}$ and some increasing trend for $(\widehat{\Gamma}_\text{post})_{12}$. However, it is difficult to assess whether or not these trends can be considered as meaningful.

\paragraph{The influence of the prior on the correlation between the material parameters} The next example focuses on the ability of the current formulation to \mbox{capture} a correlation between the Young's modulus and the initial yield stress when they are correlated. To this end, ten specimens are considered of which the material parameters are governed by a normal distribution with the following mean vector and covariance matrix:  

\begin{equation}
\label{eq:102}
\overline{\textbf{x}}_\text{spc}=\begin{bmatrix}
210 \\
0.25
\end{bmatrix}\ \textrm{GPa},\ \boldsymbol{\Gamma}_\text{spc}=\begin{bmatrix}
100 & 10^{-4} \\
10^{-4} & 1.1111\times10^{-4}
\end{bmatrix} \textrm{GPa}^{2}.
\end{equation}    

\noindent For each specimen twelve measurements are made. Using the same prior as in the previous example (see Eq.~(\ref{eq:99})) and the adaptive MCMC approach for $10^{4}$ samples whilst burning the first $3000$ samples yields:

\begin{equation}
\label{eq:103}
\widehat{\boldsymbol{\mu}}_\text{post}=\begin{bmatrix}
211.1077\\
0.2519
\end{bmatrix}\ \textrm{GPa},\ \widehat{\boldsymbol{\Gamma}}_\text{post}=\begin{bmatrix}
5.5373 &-8.396\times10^{-4}\\
-8.396\times10^{-4}& 1.8174\times{10}^{-6}
\end{bmatrix}\ \textrm{GPa}^{2}.
\end{equation}

\noindent The MAP estimate is given by:

\begin{equation}
\label{eq:104}
\widehat{\textbf{MAP}}=\begin{bmatrix}
210.5923 \\
0.2521 
\end{bmatrix}\ \textrm{GPa}.
\end{equation}

\begin{figure}
\begin{minipage}[t]{\columnwidth}
\centering
\subcaptionbox{Effect of $(\Gamma_{\textbf{x}})_{12}$ on $((\widehat{\Gamma}_\text{post})_{11}$\label{fig:10_a}}{\includegraphics[width=0.32\textwidth,trim={0.5cm 0.5cm 1cm 0.5cm},clip]{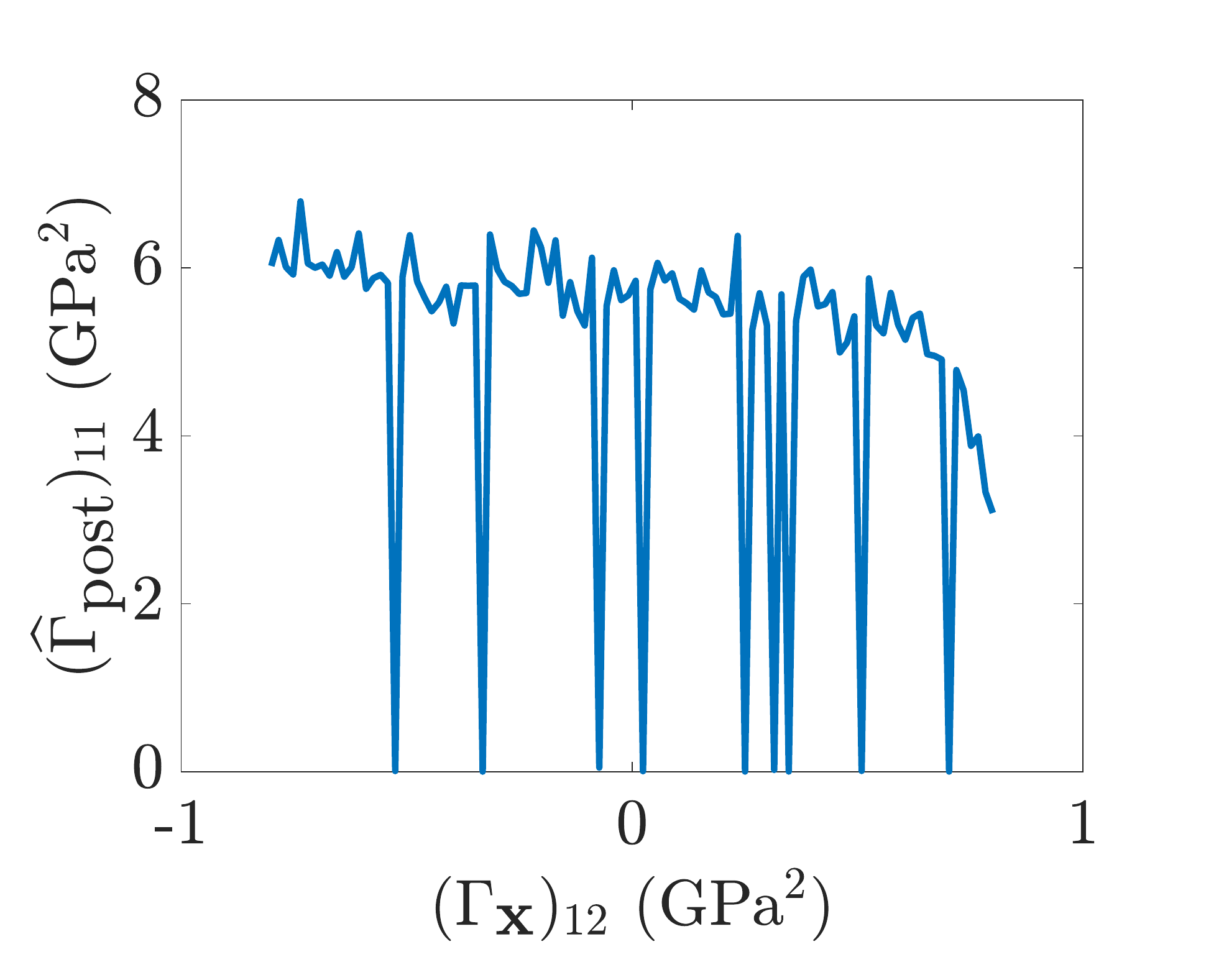}}
\subcaptionbox{Effect of $(\Gamma_{\textbf{x}})_{12}$ on $((\widehat{\Gamma}_\text{post})_{12}$\label{fig:10_b}}{\includegraphics[width=0.32\textwidth,trim={0.5cm 0.5cm 1cm 0.5cm},clip]{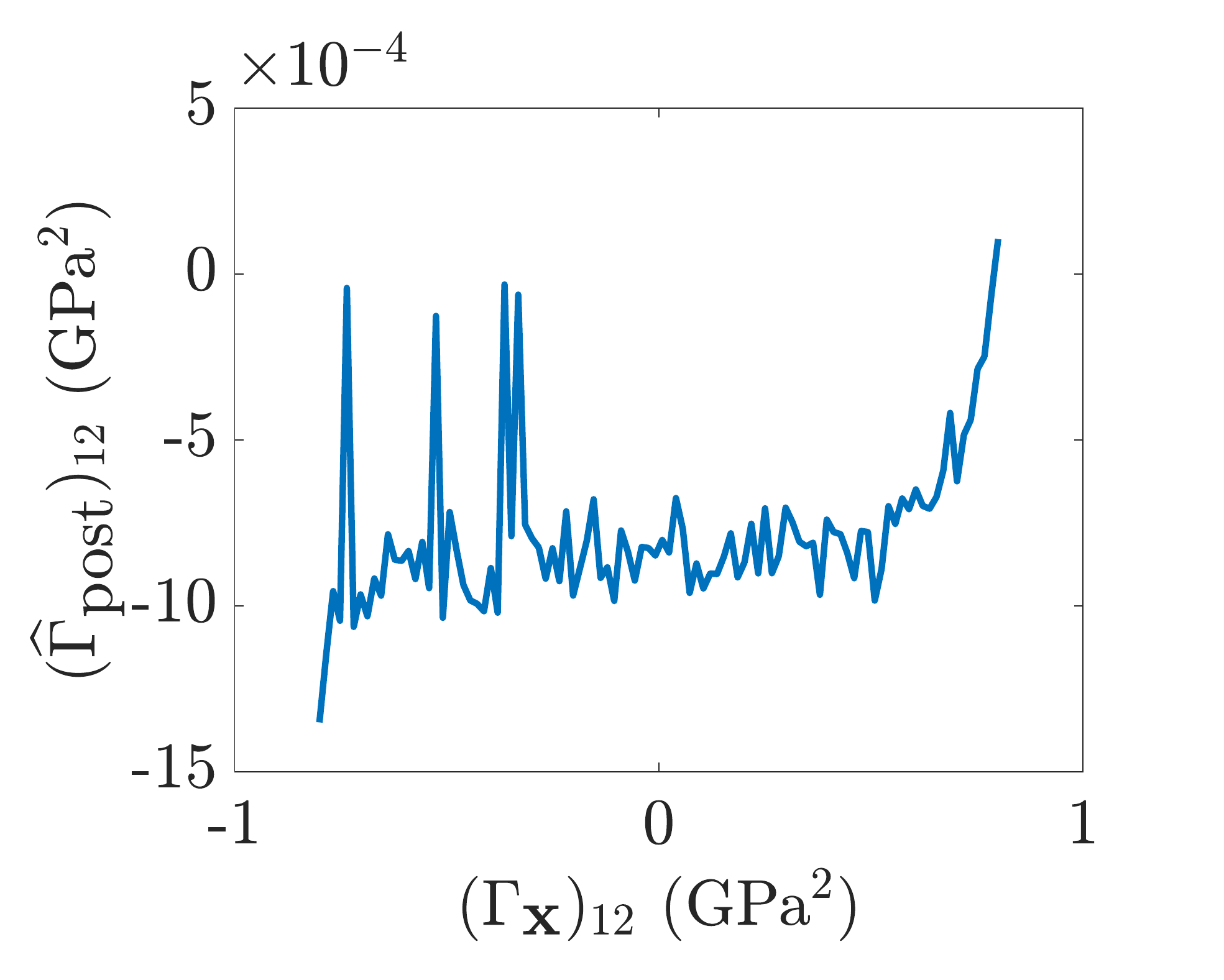}}
\end{minipage}
\begin{minipage}[t]{\columnwidth}
\centering
\subcaptionbox{Effect of $(\Gamma_{\textbf{x}})_{12}$ on $(\widehat{\Gamma}_\text{post})_{22}$\label{fig:10_c}}{\includegraphics[width=0.32\textwidth,trim={0.5cm 0.5cm 1cm 0.5cm},clip]{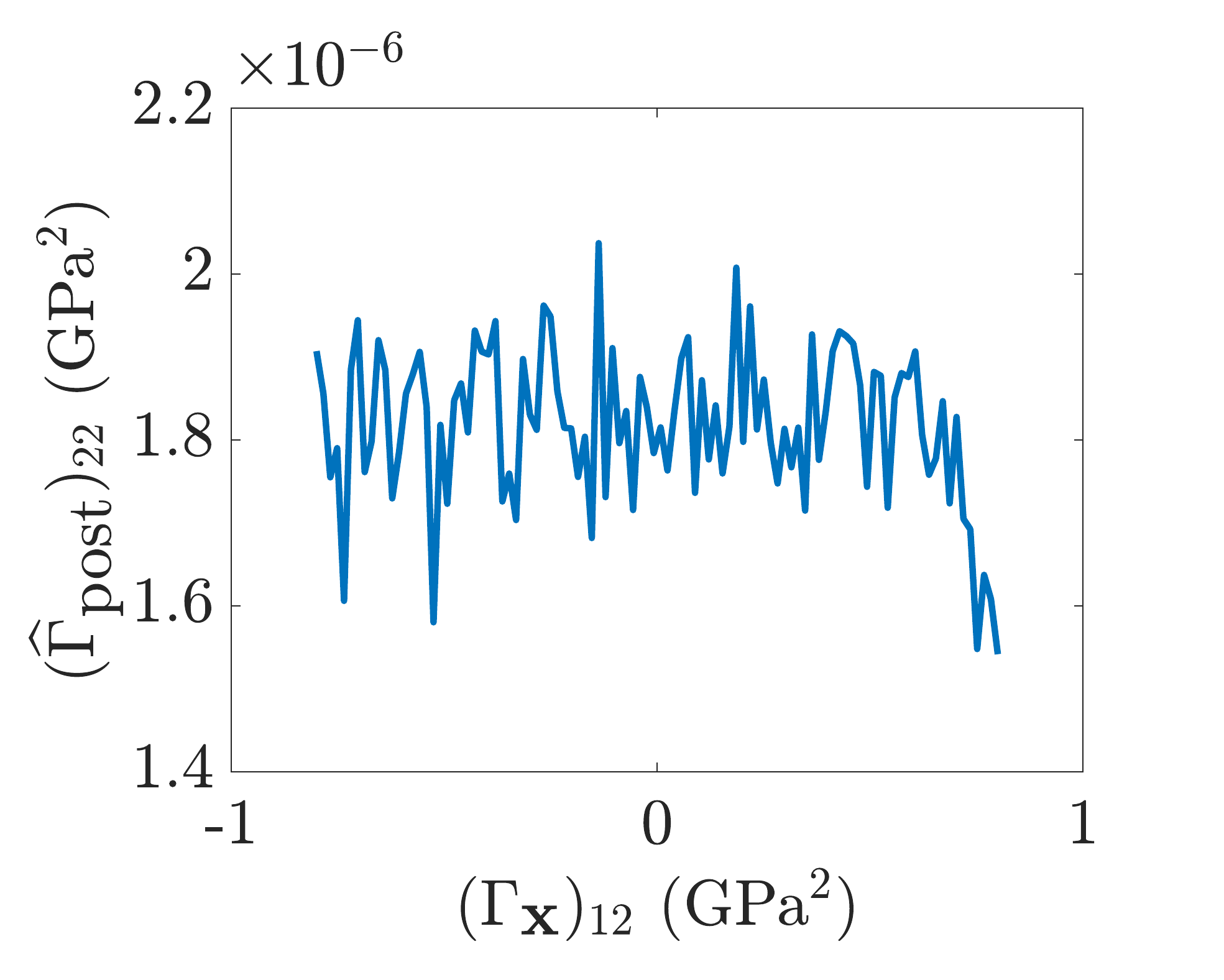}}
\end{minipage}
\caption{Linear elastic-perfect plastic: Effect of the off-diagonal components of the prior covariance matrix on the posterior's covariance matrix when the measurements are generated from specimens drawn from a normal distribution given in Eq.~(\ref{eq:102}). No real trends can be observed. The hat sign ($\widehat{\cdot}$) indicates that the values are computed numerically using the adaptive MCMC approach.}
\label{fig:10}
\end{figure}

These results show that the correlation of the posterior is not same as that of the distribution of the actual material. This corresponds closely with the observation that the formulations in this contribution are not able to capture any of the intrinsic uncertainty of the material parameters. Fig.~\ref{fig:10} shows the effect of the off-diagonal component of the prior covariance matrix ($\boldsymbol{\Gamma}_{\textbf{x}}$) on the components of the posterior's covariance matrix ($\boldsymbol{\Gamma}_\text{post}$). Again, no specific trends can be observed.

\paragraph{Application of the perfect plasticity formulation for a material that hardens nonlinearly} In the last example of this subsection, we will regard how the formulation for the linear elastic-perfectly plastic model behaves when it is used for a material that hardens nonlinearly. To this end, fifteen measurements are generated using the linear elastic-nonlinear hardening model with $E=210\ \textrm{GPa}$, $\sigma_{y0}=0.25\ \textrm{GPa}$, $H=2\ \textrm{GPa}$ and $n=0.5$ and the same noise model and noise distribution as in the previous examples. Employing the prior distribution in the form of Eq.~(\ref{eq:51}) with the mean and covariance matrix of Eq.~(\ref{eq:99}) and running the adaptive MCMC approach for $10^{4}$ samples whilst burning the first 3000 samples leads to:

\begin{equation}
\label{eq:105}
\widehat{\boldsymbol{\mu}}_\text{post}=\begin{bmatrix}
206.4357\\
0.2941
\end{bmatrix}\ \textrm{GPa},\ \widehat{\boldsymbol{\Gamma}}_\text{post}=\begin{bmatrix}
31.2818 &-3.0214\times10^{-3}\\
-3.0214\times10^{-3}& 1.1279\times{10}^{-5}
\end{bmatrix}\ \textrm{GPa}^{2},
\end{equation}

\noindent and

\begin{equation}
\label{eq:106}
\widehat{\textbf{MAP}}=\begin{bmatrix}
204.8997\\
0.2944
\end{bmatrix}\ \textrm{GPa}.
\end{equation}

\begin{figure}[H]
\begin{center}
\includegraphics[width=0.5\textwidth,trim={0.5cm 0.75cm 0.75cm 0.5cm},clip]{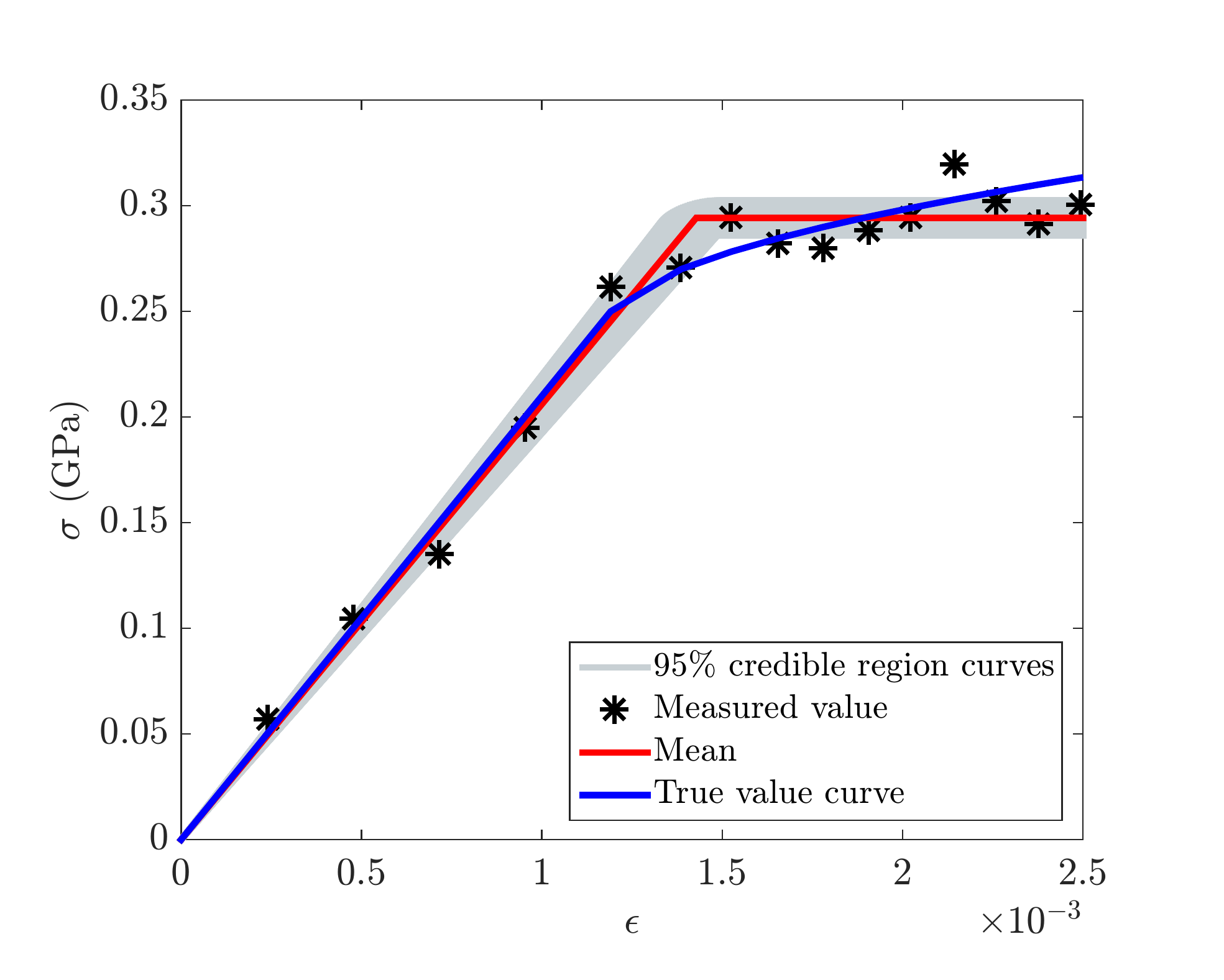}
\end{center}
\caption{Linear elastic-perfectly plastic: The measurements, the stress-strain curves using $\boldsymbol{\mu}_\text{post}$ including the 95\% credible region and the linear elastic-nonlinear hardening curve used to create the measurements.}
\label{fig:11}
\end{figure}

The stress-strain curves associated with the $95\%$ credible region in Fig.~\ref{fig:11} show that six measurements belong to the elastic part of the response for the mean.

\subsubsection{Linear elastic-linear hardening (LE-LH)}

\label{subsubsection:Examples Linear elastic-linear hardening single noise}

\paragraph{Identification of the material parameters} This subsection deals with the Bayesian formulation for the linear elastic-linear hardening model. A specimen with Young's modulus $E=210\ \textrm{GPa}$, initial yield stress $\sigma_{y0}=0.25\ \textrm{GPa}$ and plastic modulus $H=50\ \textrm{GPa}$ is regarded. Twelve measurements are created  by employing the same noise distribution as in the previous subsection. The prior distribution is given by Eq.~(\ref{eq:54}) with the following properties:

\begin{equation}
\label{eq:107}
\overline{\textbf{x}}=\begin{bmatrix}
200 \\
0.29\\
60
\end{bmatrix}\ \textrm{GPa},\ \boldsymbol{\Gamma_{\textbf{x}}}=\begin{bmatrix}
2500 & 0 & 0 \\
0 & 2.7778\times10^{-4} & 0\\
0 & 0 & 100
\end{bmatrix} \textrm{GPa}^{2}.
\end{equation}

The adaptive MCMC algorithm for $10^{4}$ samples whilst burning the first $3000$ samples gives:

\begin{equation}
\label{eq:108}
\widehat{\boldsymbol{\mu}}_\text{post}=\begin{bmatrix}
207.4586\\     
0.2533\\
55.9187
\end{bmatrix}\ \textrm{GPa}, \widehat{\boldsymbol{\Gamma}}_\text{post}=\begin{bmatrix}
36.5642 & -1.2746\times10^{-2} & -3.7886\\
-1.2746\times10^{-2} & 4.0359\times{10}^{-5} &-2.6218\times{10}^{-2}\\
-3.7886 & -2.6218\times{10}^{-2} & 66.8214
\end{bmatrix}\ \textrm{GPa}^{2},
\end{equation}

\noindent and

\begin{equation}
\label{eq:109}
\widehat{\textbf{MAP}}=\begin{bmatrix}
206.9528\\
0.2548\\
55.2838 
\end{bmatrix}\ \textrm{GPa}.
\end{equation}

\begin{figure}[H]
\begin{center}
\includegraphics[width=0.6\textwidth,trim={0.5cm 0.75cm 0.75cm 0.5cm},clip]{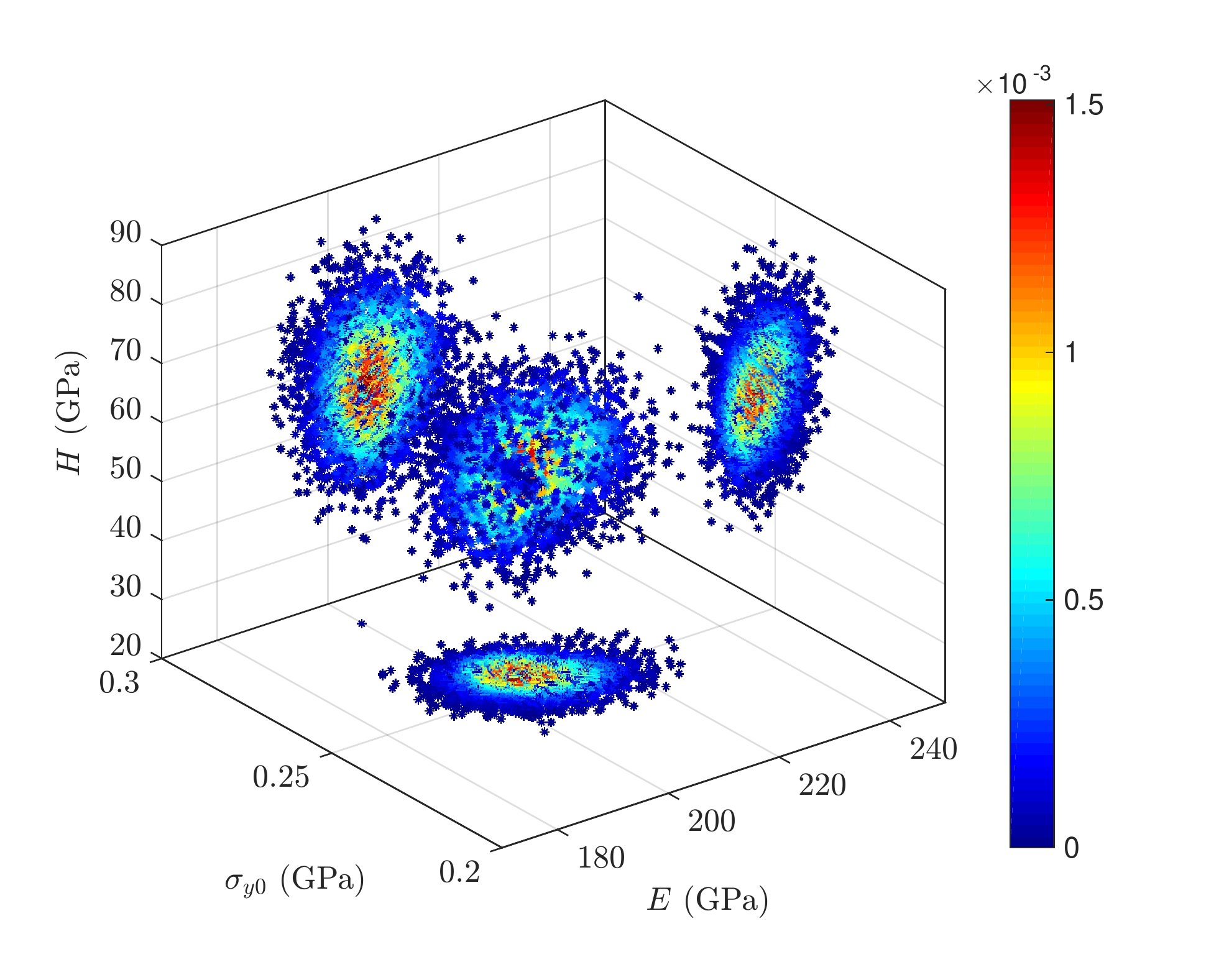}
\end{center}
\caption{Linear elastic-linear hardening: Samples generated by the adaptive MCMC approach to approximate the posterior distribution and its projection on the three planes.}
\label{fig:12}
\end{figure}

\noindent Fig.~\ref{fig:12} shows the generated samples by the adaptive MCMC approach in  the $E-\sigma_{y0}-H$ space, including the projections on the $E-\sigma_{y0}$, $E-H$ and $\sigma_{y0}-H$ planes.

\begin{figure}[H]
\begin{minipage}[t]{0.5\linewidth}
\centering
\subcaptionbox{ The $95 \%$ credible region\label{fig:13_a}}{\includegraphics[width=\textwidth,trim={0.6cm 0.5cm 1cm 0.75cm},clip]{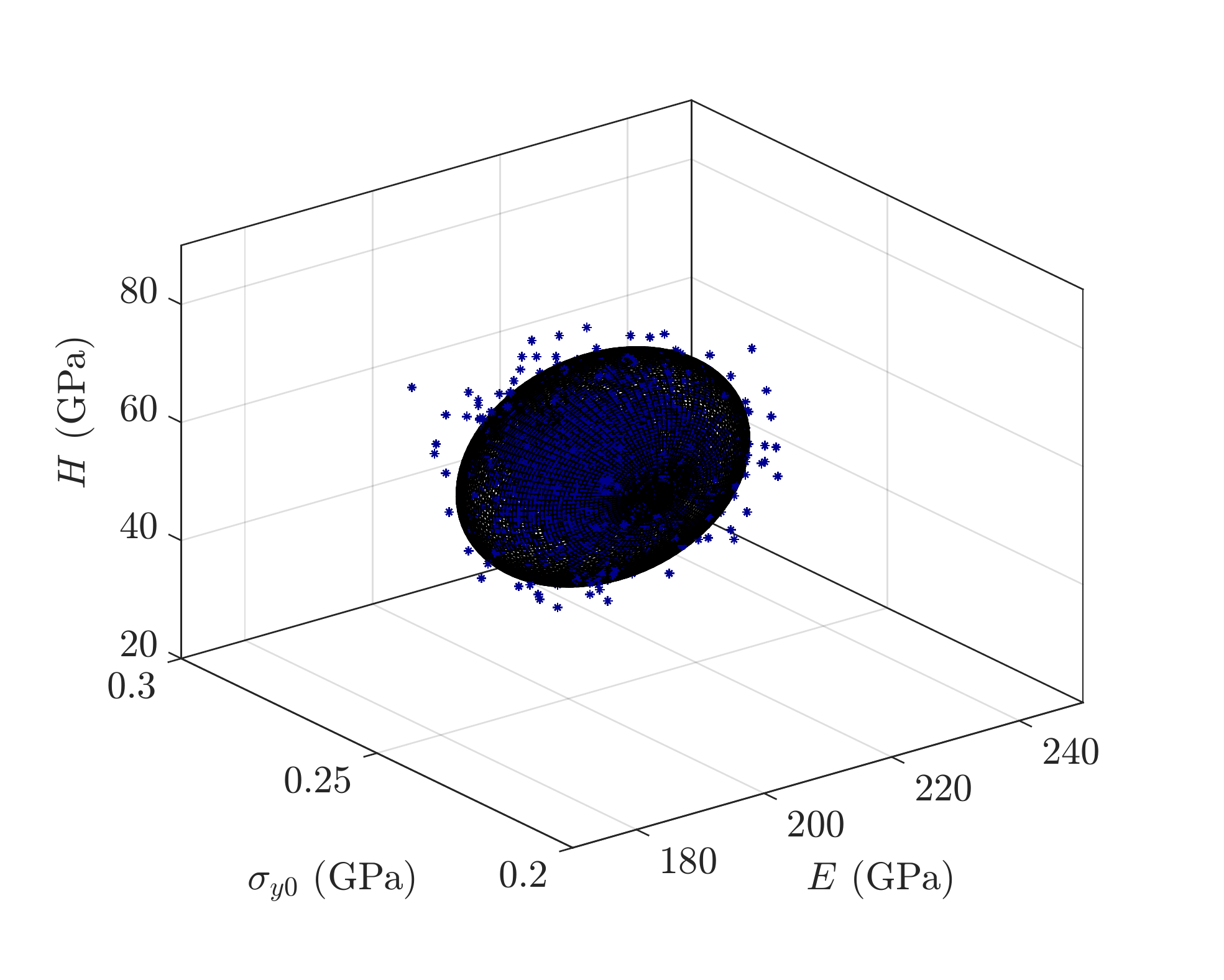}}
\end{minipage}
\begin{minipage}[t]{0.5\linewidth}
\centering
\subcaptionbox{The measurements and the stress-strain curves\label{fig:13_b}}{\includegraphics[width=\textwidth,trim={0.6cm 0.5cm 1cm 0.75cm},clip]{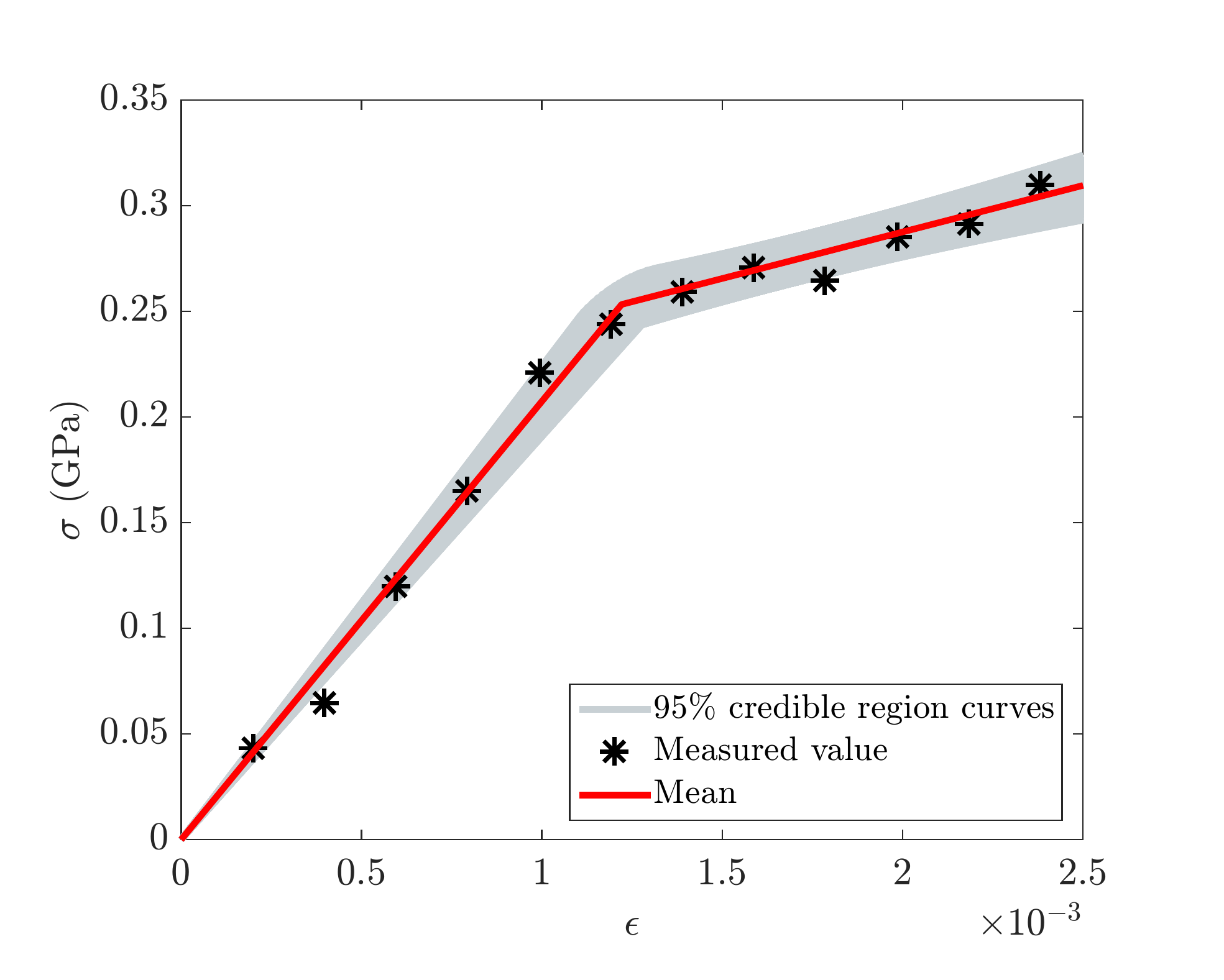}}
\end{minipage}
\caption{Linear elastic-linear hardening: The $95\%$ credible region and the posterior distribution \subref{fig:13_a} and the measurements and the stress-strain associated with the $95\%$ credible region \subref{fig:13_b}}
\label{fig:13}
\end{figure}

The $95\%$ credible region is presented in Fig.~\ref{fig:13_a} and the possible stress-strain responses associated with it are shown in Fig.~\ref{fig:13_b}.

\subsubsection{Linear elastic-nonlinear hardening (LE-NH)} 

\label{subsubsection:Examples Linear elastic-nonlinear hardening single noise}

\paragraph{Identification of the material parameters} For this subsection twelve measurements are generated based on a specimen with $E=210\ \textrm{GPa}$, $\sigma_{y0}=0.25\ \textrm{GPa}$, $H=2\ \textrm{GPa}$, $n=0.57$ and the same noise distribution as in the previous subsections. The prior distribution is selected in the form of Eq.~(\ref{eq:64}) with the following mean vector and covariance matrix:

\begin{equation}
\label{eq:110}
\overline{\textbf{x}}=\begin{bmatrix}
200 \\
0.29\\
2.5\\
0.57\ \textrm{GPa}^{-1}
\end{bmatrix}\ \textrm{GPa},\ \boldsymbol{\Gamma_{\textbf{x}}}=\begin{bmatrix}
2500 & 0 & 0 & 0\\
0 & 2.7778\times10^{-4} & 0 & 0\\
0 & 0 &  0.1111 & 0\\
0 & 0 & 0 & 0.0025\ \textrm{GPa}^{-2}
\end{bmatrix} \textrm{GPa}^{2}.
\end{equation}

\noindent Running the adaptive MCMC approach for $10^{4}$ samples after burning the first $3000$ samples yields:

\begin{equation}
\label{eq:111}
\begin{gathered}
\widehat{\boldsymbol{\mu}}_\text{post}=\begin{bmatrix}
209.8035 \\     
0.2566\\
2.1602\\
0.6034\ \textrm{GPa}^{-1}
\end{bmatrix}\ \textrm{GPa},\\
\widehat{\boldsymbol{\Gamma}}_\text{post}=\begin{bmatrix}
25.6975 &-3.9287\times10^{-3} & -9.0948\times10^{-2} & -6.8572\times10^{-3}\ \textrm{GPa}^{-1}\\
-3.9287\times10^{-3} & 7.7856\times{10}^{-5} &-5.2198\times{10}^{-4} & 1.4961\times10^{-4}\ \textrm{GPa}^{-1}\\
-9.0948\times10^{-2} & -5.2198\times10^{-4} & 9.6433\times10^{-2} & 5.7512\times10^{-3}\ \textrm{GPa}^{-1}\\
-6.8572\times10^{-3}\ \textrm{GPa}^{-1} & 1.4961\times10^{-4}\ \textrm{GPa}^{-1} & 5.7512\times10^{-3}\ \textrm{GPa}^{-1} & 9.6817\times10^{-4}\ \textrm{GPa}^{-2}
\end{bmatrix}\ \textrm{GPa}^{2},
\end{gathered}
\end{equation}

\noindent and

\begin{equation}
\label{eq:112}
\widehat{\textbf{MAP}}=\begin{bmatrix}
209.4125\\
0.2551\\
2.1266\\
0.597\ \textrm{GPa}^{-1} 
\end{bmatrix}\ \textrm{GPa}.
\end{equation}

\begin{figure}[H]
\begin{center}
\includegraphics[width=0.5\textwidth,trim={0.5cm 0.75cm 0.75cm 0.5cm},clip]{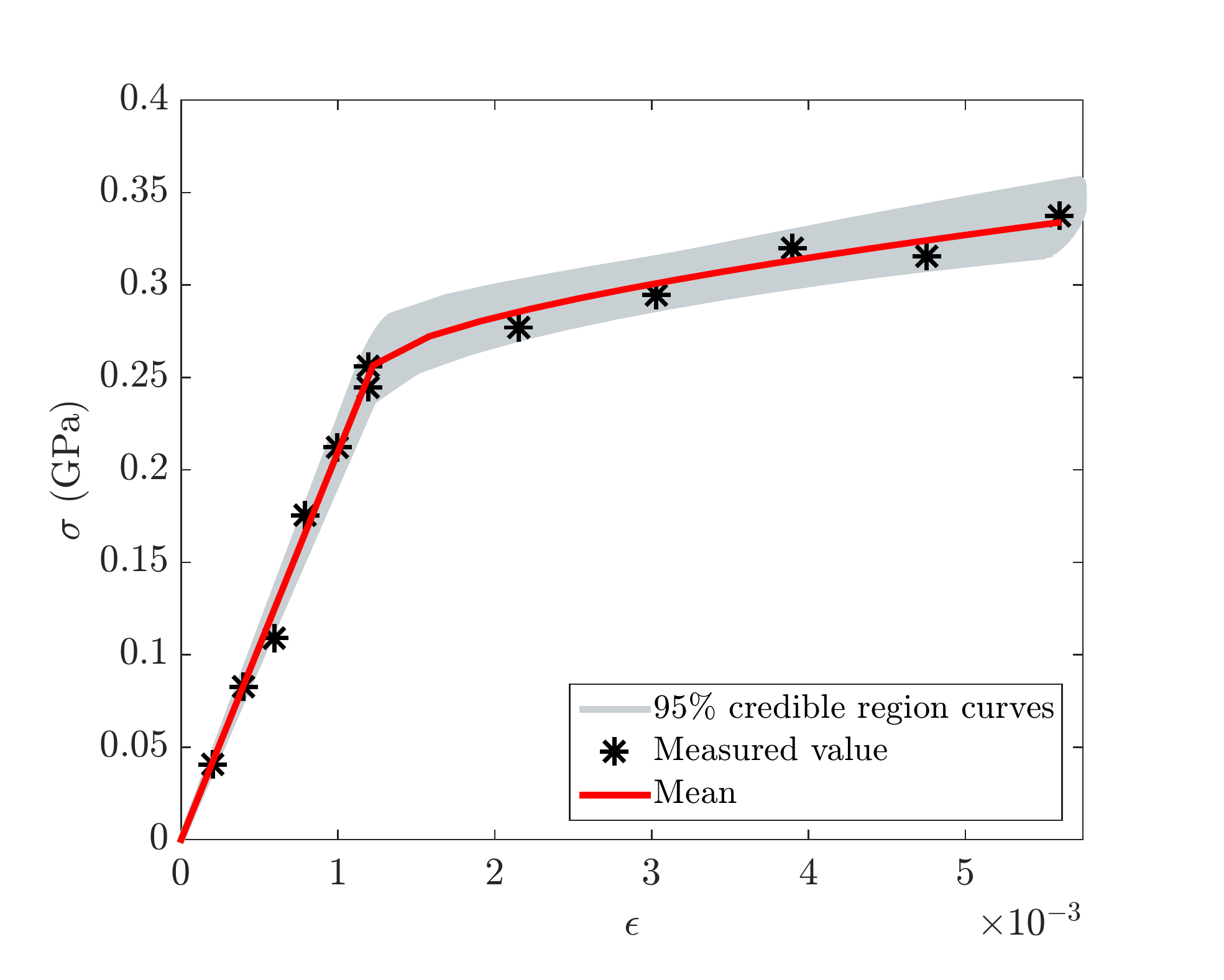}
\end{center}
\caption{Liner elastic-nonlinear hardening: The measurements and the stress-strain curves associated with the $95\%$ credible region.}
\label{fig:14}
\end{figure}

\noindent The possible stress-strain responses using the $95\%$ credible region for the posterior are presented in Fig.~\ref{fig:14}. 

\subsection{Bayesian inference with noise in stress and strain}

\label{subsection: Examples Bayesian inference with noise in stress and strain}

All the measurements used in subsection \ref{subsection: Examples Bayesian inference with noise in stress} are assumed to be exact in the strain. The aim of current subsection is to show how the previous results change when the strain is also contaminated by noise. To accurately investigate the influence of this, exactly the same measurements are used as in the previous subsection.

\subsubsection{Linear elastic (LE)}

\label{subsubsection:Examples linear elastic double noise}

\paragraph{Identification of the Young's modulus} In the first example the same specimen and the same measurements as in subsection \ref{subsubsection:Examples linear elastic single noise} ($E=210\ \textrm{GPa}$) are considered. The noise distribution obtained from the `calibration experiments' is a normal distribution in the form of Eq.~(\ref{eq:72}) with $S_{\sigma}=0.01\ \textrm{GPa}$ and $S_{\epsilon}=0.0001$. For the same measurement pair with $\sigma^{m}=0.1576\ \textrm{GPa}$ and $\epsilon^{m}=7.25\times10^{-4}$, the posterior distribution can be calculated using Eq.~(\ref{eq:78}). The same prior is selected as in the first example of subsection \ref{subsubsection:Examples linear elastic single noise} ($\overline{E}=150 \textrm{GPa}$ and $S_{E}=50\ \textrm{GPa}$) and the upper limit for the strain is infinity. Running the adaptive MCMC approach for $10^{4}$ samples whilst burning the first $3000$ samples leads to a posterior with $\widehat{\mu}_\text{post}=202.7767\ \textrm{GPa}$, $\widehat{S}_\text{post}=24.1867\ \textrm{GPa}$ and $\widehat{\textrm{MAP}}=197.5282\ \textrm{GPa}$. 

\begin{figure}[H]
\begin{center}
\includegraphics[width=0.6\textwidth,trim={0.5cm 0.75cm 0.75cm 0.5cm},clip]{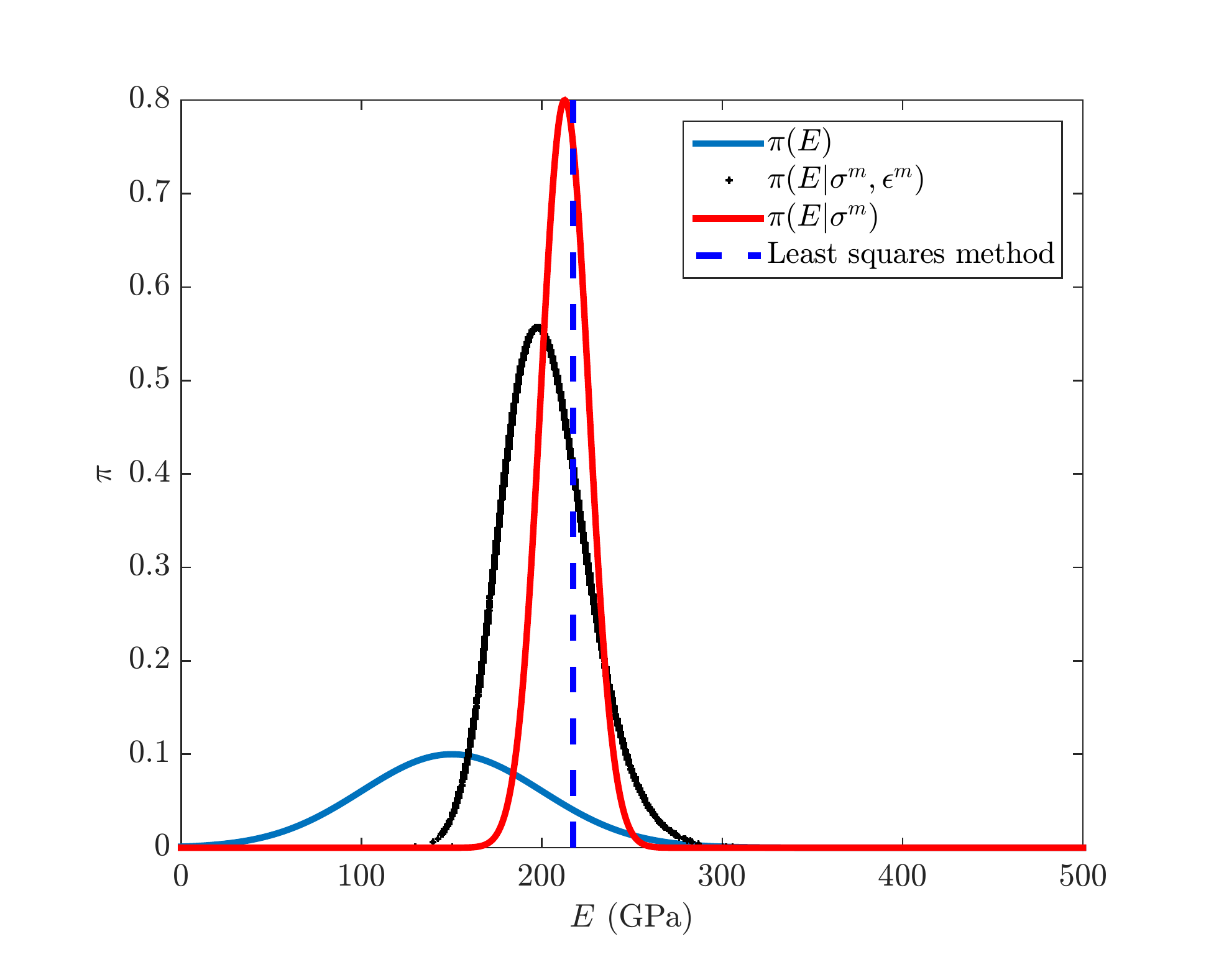}
\end{center}
\caption{Linear elastic: The prior and posterior when both the stress and the strain are corrupted by noise (black dots), the posterior when only the stress is contaminated (red) and the value predicted by the least squares method (blue dashed). The posterior for the case with noise in the stress and strain is wider than the posterior for the case with noise in the stress only. It furthermore does not have the form of a (modified) normal. The distributions are not normalised.}
\label{fig:15}
\end{figure}

Fig.~\ref{fig:15} shows this posterior, together with the posterior of subsection \ref{subsubsection:Examples linear elastic single noise} when only the noise in the stress is present. It can clearly be seen that the newly established posterior is wider and it does not have the form of a (modified) normal distribution, as the posterior when only the noise in the stress is considered.

\begin{figure}[H]
\begin{minipage}[t]{0.5\linewidth}
\centering
\subcaptionbox{Uncertainty in both the stress and the strain \mbox{measurements}\label{fig:16_a}}{\includegraphics[width=\textwidth,trim={0.6cm 0.5cm 1cm 0.75cm},clip]{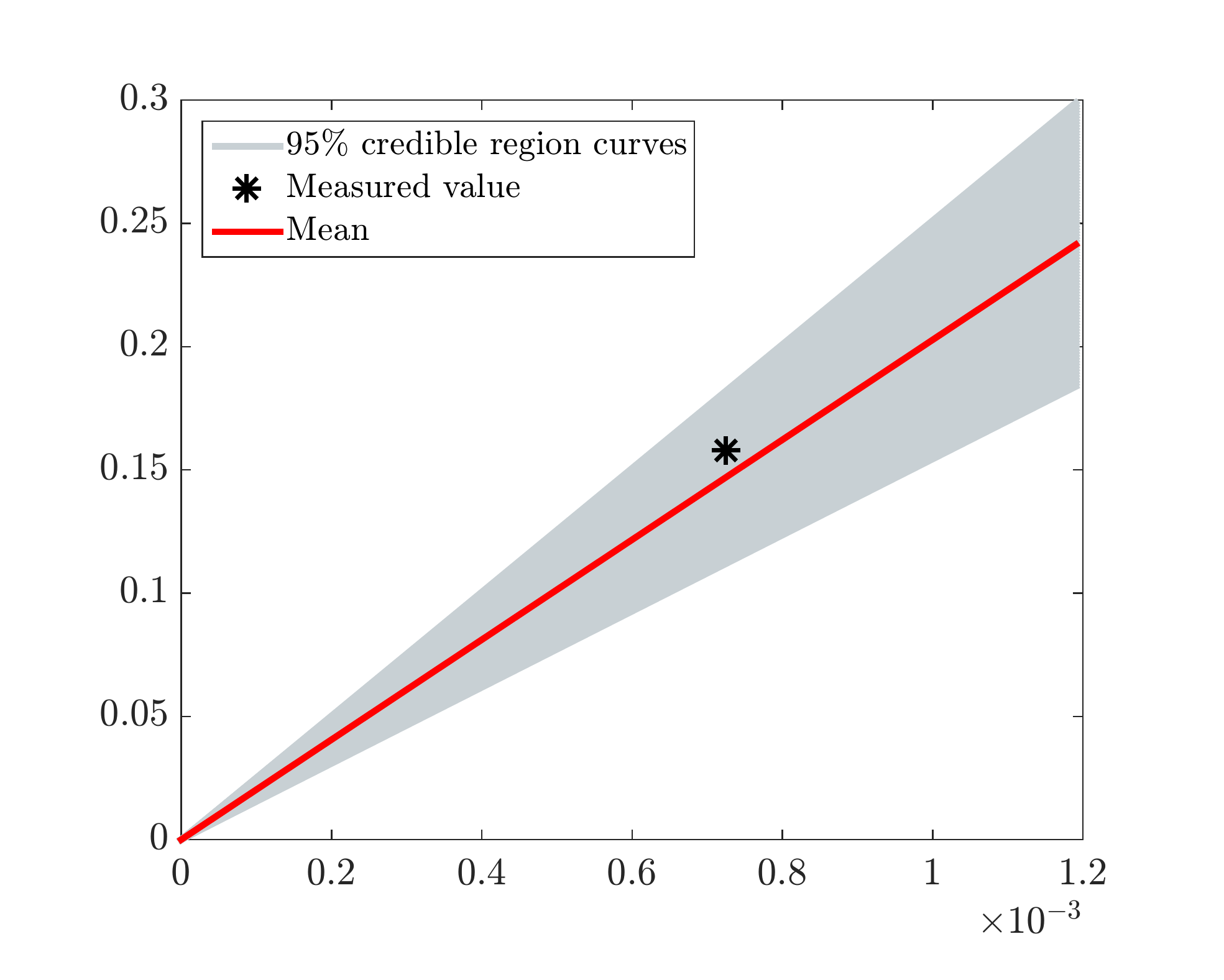}}
\end{minipage}
\begin{minipage}[t]{0.5\linewidth}
\centering
\subcaptionbox{Uncertainty in the stress only\label{fig:16_b}}{\includegraphics[width=\textwidth,trim={0.6cm 0.5cm 1cm 0.75cm},clip]{Fig_4_a.pdf}}
\end{minipage}
\caption{Linear elastic: The measurements, the stress-strain curves associated with the mean and the $95\%$ credible region for \subref{fig:16_a} noise in the stress and strain and \subref{fig:16_b} noise in the stress only. The uncertainty is larger for the case with noise in the stress and strain than  that for the case with noise in the stress only.}
\label{fig:16}
\end{figure}

The stress-strain responses created using the $95\%$ credible region are shown in Fig.~\ref{fig:16_a}. For comparison, the same responses are shown in Fig.~\ref{fig:16_b} when only the stress is polluted by noise. The estimated values for this case were $\mu_\text{post}=209.1364\ \textrm{GPa}$, $S_\text{post}=9.6642\ \textrm{GPa}$ and $\textrm{MAP}=207.9963\ \textrm{GPa}$.

\begin{figure}[H]
\begin{minipage}[t]{0.5\linewidth}
\centering
\subcaptionbox{Uncertainty in both the stress and the strain measurements\label{fig:17_a}}{\includegraphics[width=\textwidth,trim={0.6cm 0.5cm 1cm 0.75cm},clip]{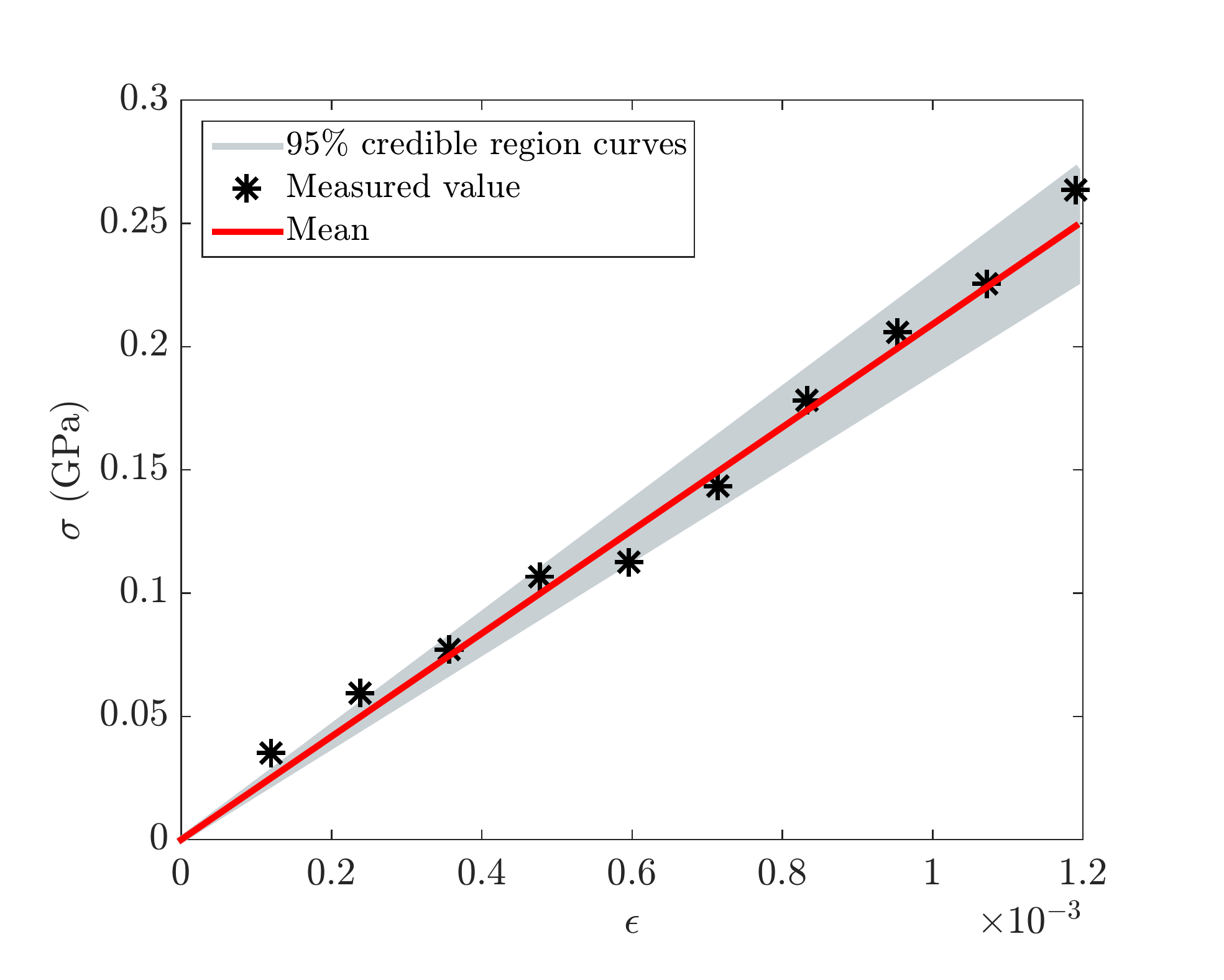}}
\end{minipage}
\begin{minipage}[t]{0.5\linewidth}
\centering
\subcaptionbox{Uncertainty in the stress only\label{fig:17_b}}{\includegraphics[width=\textwidth,trim={0.6cm 0.5cm 1cm 0.75cm},clip]{Fig_4_b.pdf}}
\end{minipage}
\caption{Linear elastic: The measurements, and the stress-strain curves associated with the mean and the $95\%$ credible region for ten measurements \subref{fig:17_a} noise in the stress and the strain and \subref{fig:17_b} noise in the stress only. The uncertainty is larger for the case with noise in the stress and strain than  that for the case with noise in the stress only.}
\label{fig:17}
\end{figure} 

\subsubsection{Linear elastic-perfectly plastic (LE-PP)}

\label{subsubsection:Examples Linear elastic-perfectly plastic double noise}

\paragraph{Identification of the material parameters} Similar to the example of subsection \ref{subsubsection:Examples Linear elastic-perfectly plastic single noise}, in this subsection a specimen is considered  with a Young's modulus $E=210\ \textrm{GPa}$ and an initial yield stress $\sigma_{y0}=0.25\ \textrm{GPa}$. Furthermore, to be able to compare the current results with those of subsection \ref{subsubsection:Examples Linear elastic-perfectly plastic single noise} the same measurements are considered again, together with the same noise distribution as in the previous subsection. The prior distribution is furthermore selected as in Eq.~(\ref{eq:51}) with its properties given in Eq.~(\ref{eq:99}). The resulting posterior is of the form of Eq.~(\ref{eq:82}) with Eq.~(\ref{eq:81}) as the likelihood function. Similar to subsection \ref{subsubsection:Examples Linear elastic-perfectly plastic single noise}, the upper limit for the theoretical strain ($a$ in Eqs.~(\ref{eq:70}) and (\ref{eq:81})) is infinite. Employing the adaptive MCMC approach for $10^{4}$ samples and burning the first $3000$ samples results in: 

\begin{equation}
\label{eq:113}
\widehat{\boldsymbol{\mu}}_\text{post}=\begin{bmatrix}
 204.0458\\
0.2572
\end{bmatrix}\ \textrm{GPa},\ \widehat{\boldsymbol{\Gamma}}_\text{post}=\begin{bmatrix}
215.3918 & -6.751\times10^{-4}\\
-6.751\times10^{-4} & 1.4978\times{10}^{-5}
\end{bmatrix}\ \textrm{GPa}^{2},
\end{equation}

\noindent and

\begin{equation}
\label{eq:114}
\widehat{\textbf{MAP}}=\begin{bmatrix}
202.5858 \\
0.2573 
\end{bmatrix}\ \textrm{GPa}.
\end{equation}

\begin{figure}[H]
\begin{minipage}[t]{0.5\linewidth}
\centering
\subcaptionbox{Samples generated by the adaptive MCMC approach\label{fig:18_a}}{\includegraphics[width=\textwidth,trim={0.6cm 0.5cm 1cm 0.75cm},clip]{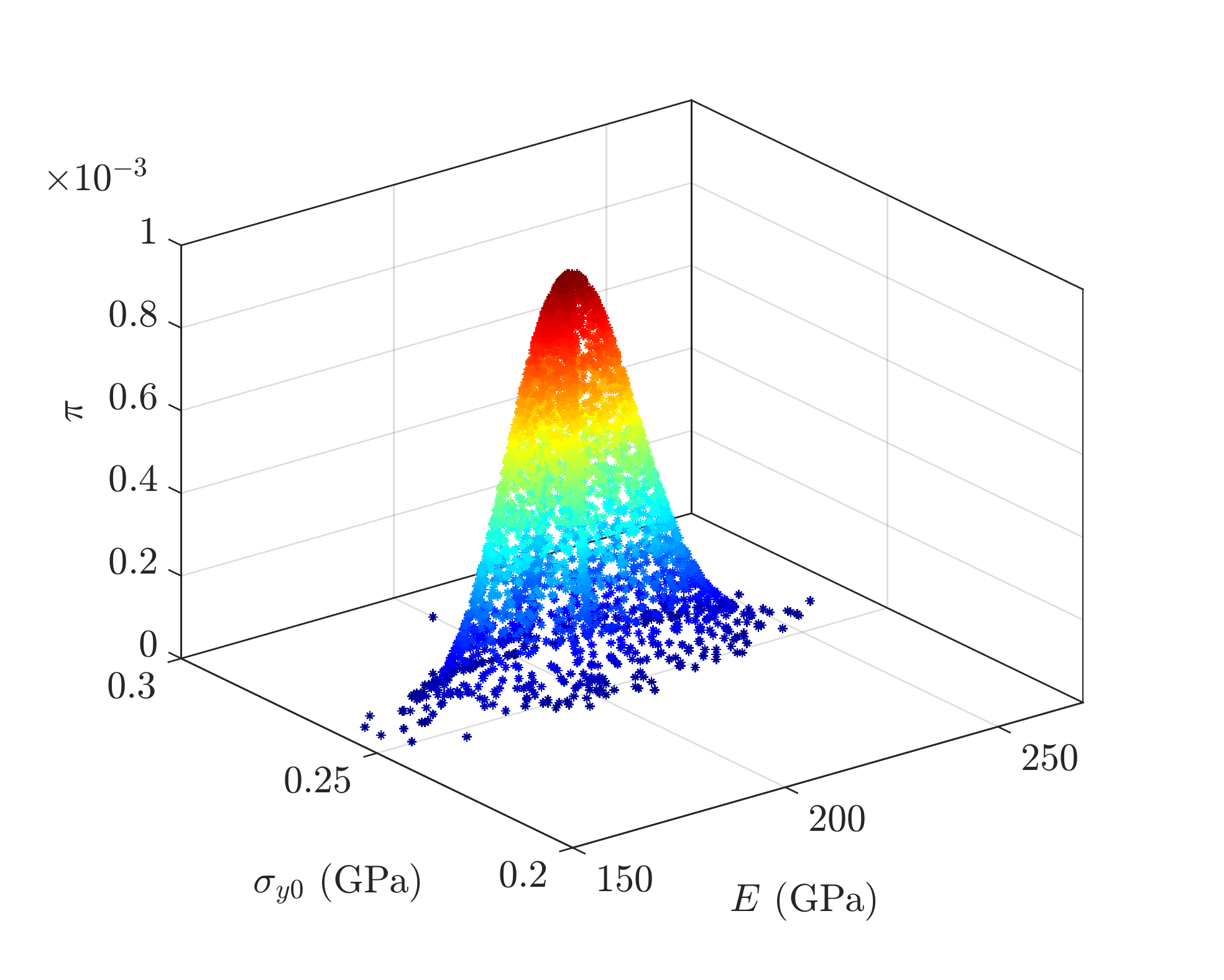}}
\end{minipage}
\begin{minipage}[t]{0.5\linewidth}
\centering
\subcaptionbox{$95\%$ credible region\label{fig:18_b}}{\includegraphics[width=\textwidth,trim={0.6cm 0.5cm 1cm 0.75cm},clip]{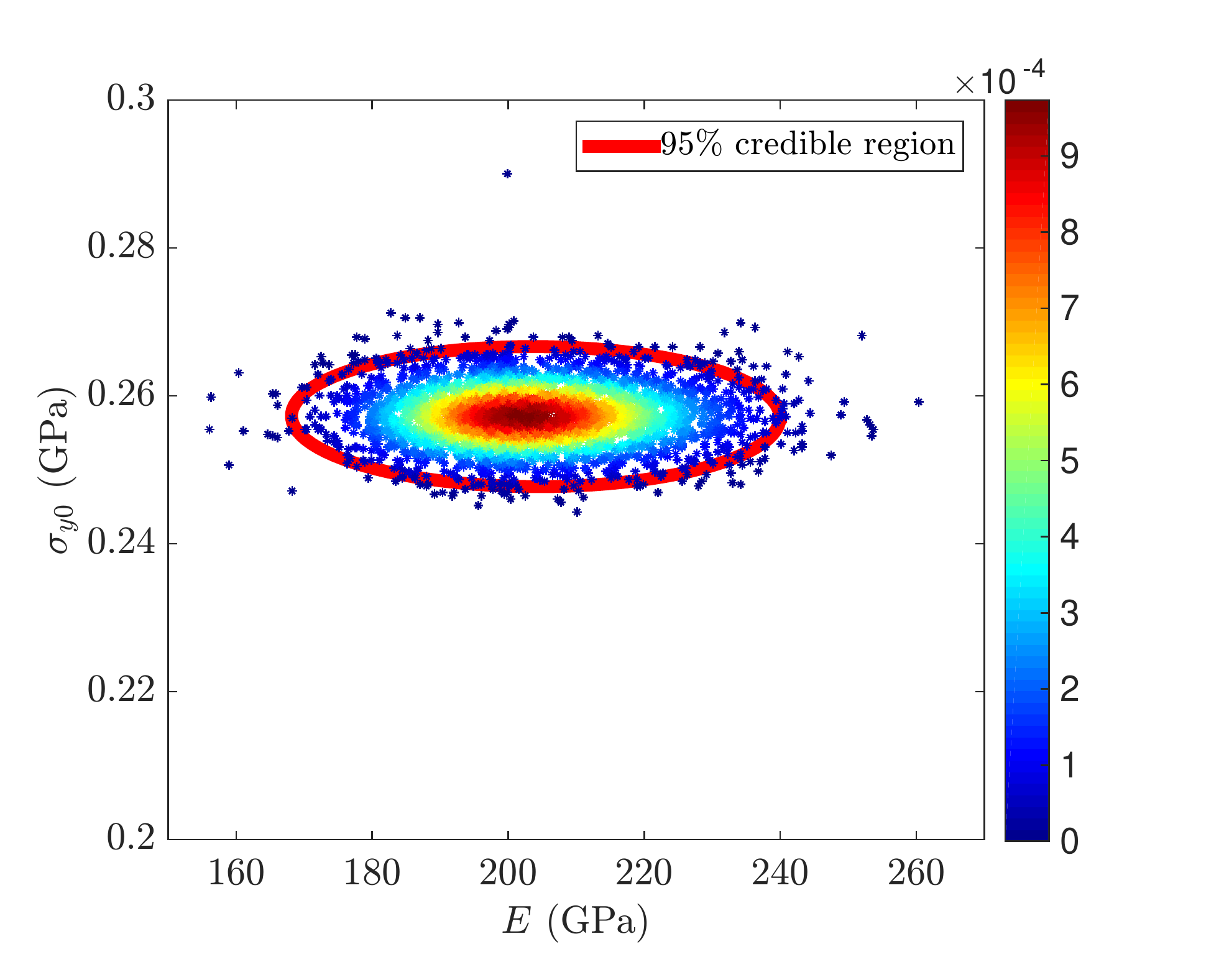}}
\end{minipage}
\caption{Linear elastic-perfectly plastic: The samples generated by the adaptive MCMC approach to approximate the posterior distribution \subref{fig:18_a} and the $95\%$ credible region \subref{fig:18_b}.}
\label{fig:18}
\end{figure}

\noindent By comparing the posterior's covariance matrices (Eqs.~\ref{eq:100} and \ref{eq:113}), one can see that the additional uncertainty in the strain measurement has a significant effect on $(\widehat{\Gamma}_\text{post})_{11}$. Fig.~\ref{fig:18} shows the samples generated by the adaptive MCMC approach to approximate the posterior distribution and the associated $95\%$ credible region.

\begin{figure}[H]
\begin{minipage}[t]{0.5\linewidth}
\centering
\subcaptionbox{Uncertainty in both the stress and the strain \mbox{measurements}\label{fig:19_a}}{\includegraphics[width=\textwidth,trim={0.6cm 0.5cm 1cm 0.75cm},clip]{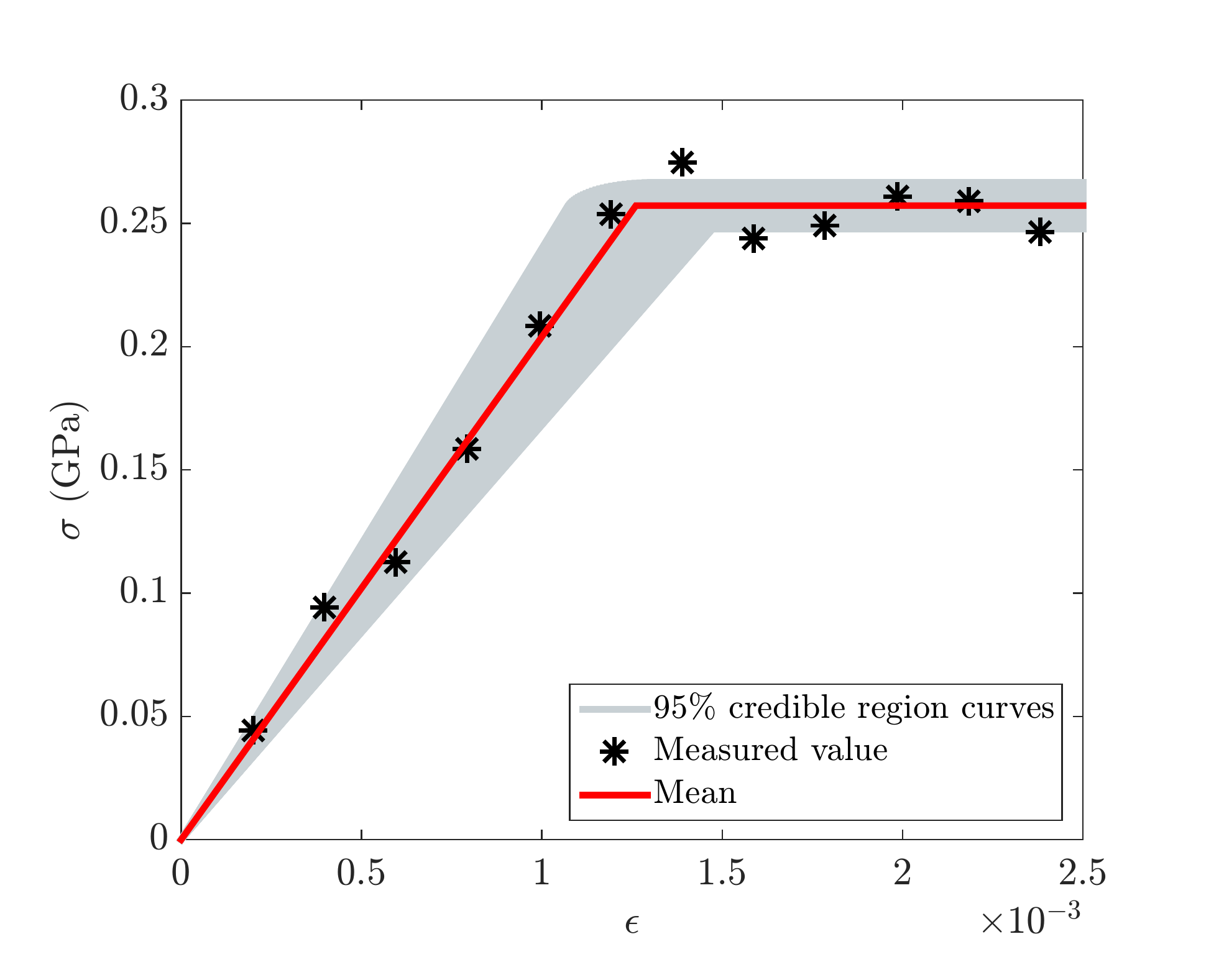}}
\end{minipage}
\begin{minipage}[t]{0.5\linewidth}
\centering
\subcaptionbox{Uncertainty in the stress measurements only\label{fig:19_b}}{\includegraphics[width=\textwidth,trim={0.6cm 0.5cm 1cm 0.75cm},clip]{Fig_8_b.pdf}}
\end{minipage}
\caption{Linear elastic-perfectly plastic: The measurements, and the stress-strain responses associated with the mean and the $95\%$ credible region for twelve measurements: \subref{fig:19_a}  noise in the stress and strain measurements and \subref{fig:19_b} noise in the stress measurements only. The additional uncertainty in the strain has a considerable effect on the elastic response, but not on the plastic response. This is caused by the fact that the plastic part of the response does not depend on the strain.}
\label{fig:19}
\end{figure}

The stress-strain curves associated with the credible region are shown in Fig.~\ref{fig:19_a}. The same curves for the case with noise in the stress measurements only are presented in Fig.~\ref{fig:19_b}. Comparing Figs.~\ref{fig:19_a} and \ref{fig:19_b}, one can see that the additional uncertainty in the strain has a considerable effect on the elastic response, but not on the plastic response. This is caused by the fact that the plastic part of the response does not depend on the strain. Practically no influence on the yield stress can be observed, because this parameter is independent of the strain. This can also be observed by comparing $(\widehat{\Gamma}_\text{post})_{22}$ of Eq.~\ref{eq:100} and that of Eq.~\ref{eq:113}.

\subsubsection{Linear elastic-linear hardening (LE-LH)}

\label{subsubsection:Examples Linear elastic-linear hardening double noise}

\paragraph{Identification of the material parameters} In this subsection we will regard the results of the BI formulation for the linear elastic-linear hardening model, when both the stress and the strain measurements are polluted by statistical noise. The same noise model and noise distribution is employed as in the previous subsections. The same measurements as in subsection \ref{subsubsection:Examples Linear elastic-linear hardening single noise} are furthermore considered, which are created with $E=210\ \textrm{GPa}$, $\sigma_{y0}=0.25\ \textrm{GPa}$ and $H=50\ \textrm{GPa}$. Similar to in the previous subsections, it is assumed that the theoretical strain ranges from zero to infinity. Selecting the prior distribution in the form of Eq.~(\ref{eq:54}) with the mean vector and covariance matrix given by Eq.~(\ref{eq:107}), the resulting posterior is in the form of Eq.~(\ref{eq:85}), where the likelihood function is given by Eq.~(\ref{eq:83}). Employing the adaptive MCMC approach for $10^{4}$ sample generations and burning the first $3000$ samples yields:

\begin{equation}
\label{eq:115}
\widehat{\boldsymbol{\mu}}_\text{post}=\begin{bmatrix}
204.269\\     
0.2553\\
56.08
\end{bmatrix}\ \textrm{GPa},\widehat{\boldsymbol{\Gamma}}_\text{post}=\begin{bmatrix}
148.2602 & -5.1006\times10^{-2} &  -14.8463\\
-5.1006\times10^{-2} & 6.0511\times{10}^{-5} & -2.7164\times{10}^{-2}\\
 -14.8463 & -2.7164\times{10}^{-2} & 76.7817
\end{bmatrix}\ \textrm{GPa}^{2},
\end{equation}

\noindent and

\begin{equation}
\label{eq:116}
\widehat{\textbf{MAP}}=\begin{bmatrix}
201.8193\\
55.0417\\
0.256
\end{bmatrix}\ \textrm{GPa}.
\end{equation}

\begin{figure}[H]
\begin{minipage}[t]{0.5\linewidth}
\centering
\subcaptionbox{Samples generated by the adaptive MCMC approach\label{fig:20_a}}{\includegraphics[width=\textwidth,trim={0.6cm 0.5cm 1cm 0.75cm},clip]{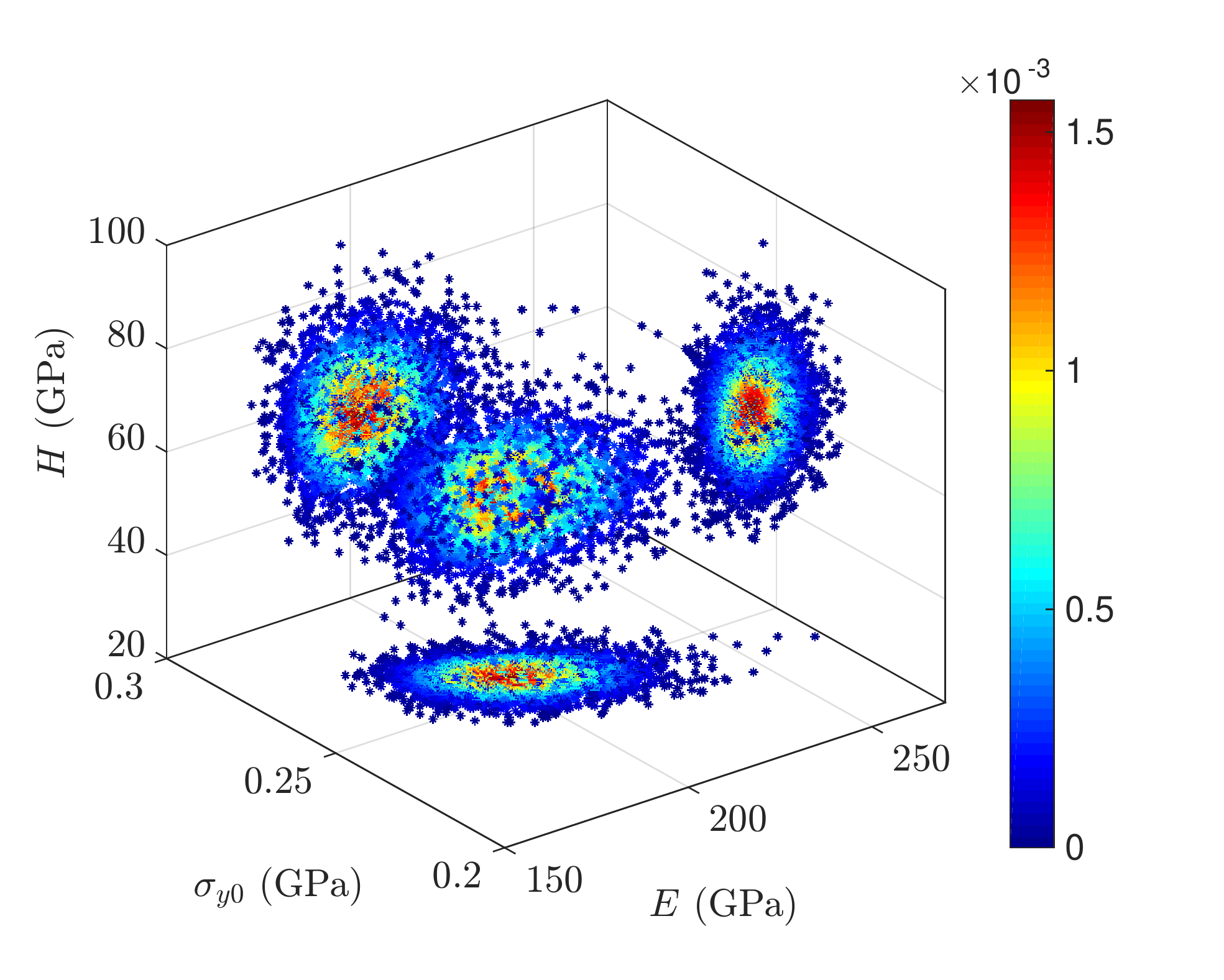}}
\end{minipage}
\begin{minipage}[t]{0.5\linewidth}
\centering
\subcaptionbox{$95\%$ credible region\label{fig:20_b}}{\includegraphics[width=\textwidth,trim={0.6cm 0.5cm 1cm 0.75cm},clip]{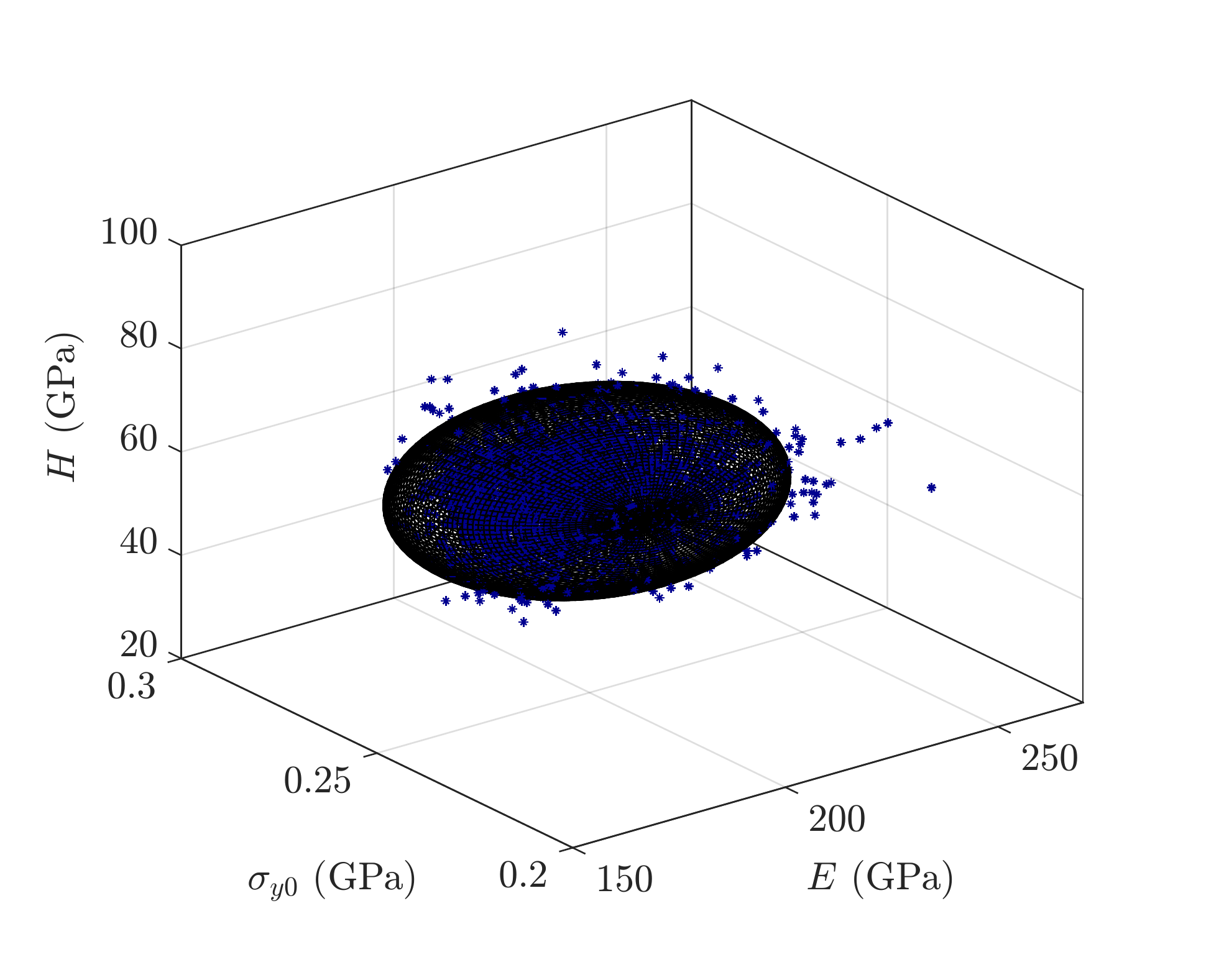}}
\end{minipage}
\caption{Linear elastic-linear hardening: Samples generated by the adaptive MCMC approach to approximate the posterior distribution, its projection on the three planes \subref{fig:20_a} and the $95\%$ credible region \subref{fig:20_b}.}
\label{fig:20}
\end{figure}

\begin{figure}[H]
\begin{minipage}[t]{0.5\linewidth}
\centering
\subcaptionbox{Uncertainty in both the stress and the strain \mbox{measurements}\label{fig:21_a}}{\includegraphics[width=\textwidth,trim={0.6cm 0.5cm 1cm 0.75cm},clip]{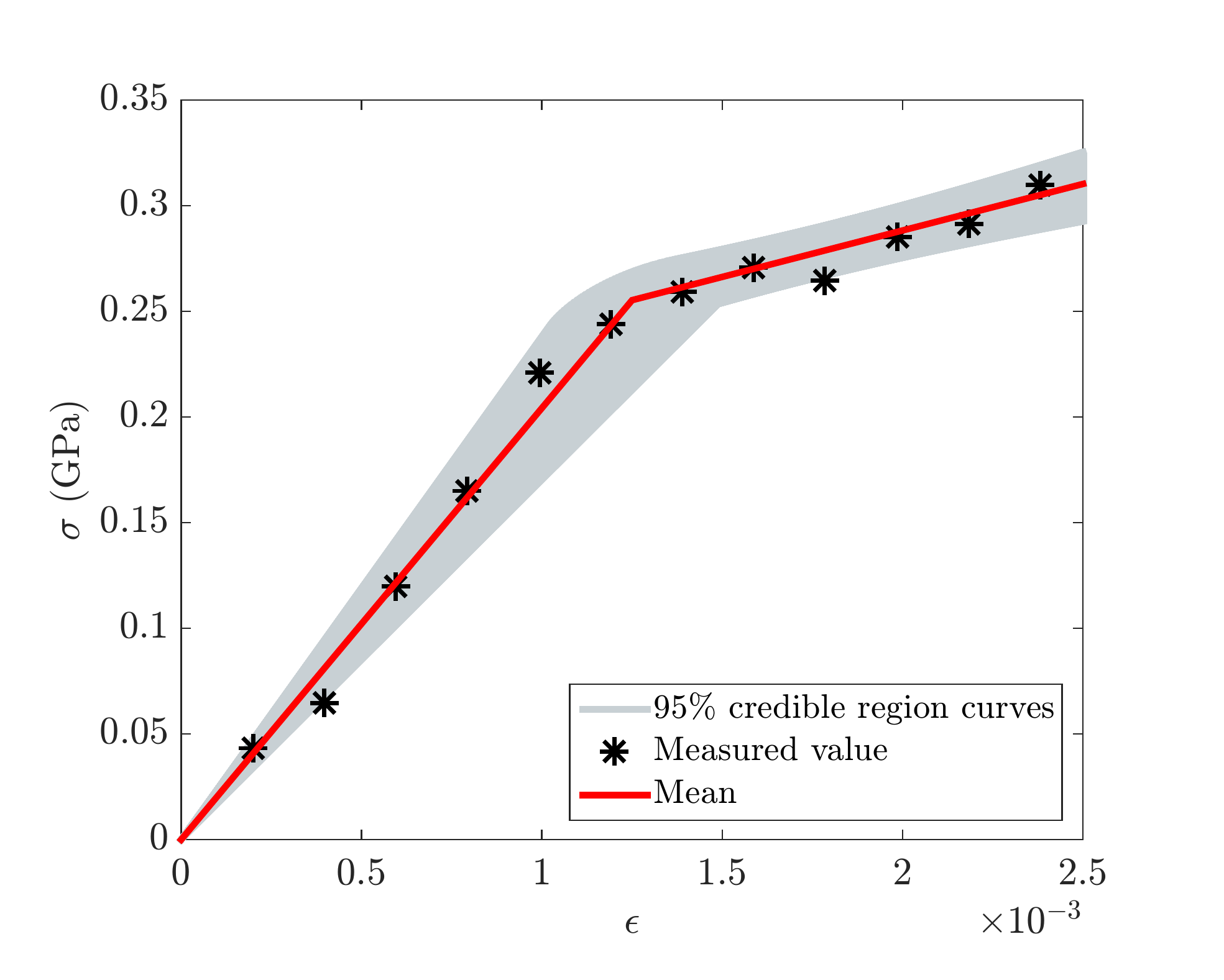}}
\end{minipage}
\begin{minipage}[t]{0.5\linewidth}
\centering
\subcaptionbox{Uncertainty in the stress measurements only\label{fig:21_b}}{\includegraphics[width=\textwidth,trim={0.6cm 0.5cm 1cm 0.75cm},clip]{Fig_13_b.pdf}}
\end{minipage}
\caption{Linear elastic-linear hardening: The measurements, and the stress-strain curves associated with the mean and the $95\%$ credible region for twelve measurements: \subref{fig:21_a} noise in the stress and strain measurements and \subref{fig:21_b} noise in the stress measurements only. For the case with noise in both the stress and strain measurements, the uncertainty of the Young's modulus is larger than for the case with noise in the stress measurements only. Hardly any difference can be distinguished for the plastic parameters however.}
\label{fig:21}
\end{figure}

Comparing the results above with those in Eqs.~(\ref{eq:108}) and (\ref{eq:109}) shows that the uncertainty in the strain has a larger effect on the Young's modulus and its corresponding components in the posterior's covariance matrix than on the initial yield stress and the plastic modulus. Fig.~\ref{fig:20} shows the samples generated by the adaptive MCMC approach and the associated $95\%$ credible region. The associated stress-strain responses are presented in Fig.~\ref{fig:21}. The large influence of the additional noise in the strain measurements on the Young's modulus compared to its influence on the plastic parameters can clearly be distinguished.

\subsubsection{Linear elastic-nonlinear hardening (LE-NH)}

\label{subsubsection:Examples Linear elastic-nonlinear hardening double noise}

\paragraph{Identification of the material parameters} The specimen in this subsection is the same as in subsection \ref{subsubsection:Examples Linear elastic-nonlinear hardening single noise} and has a Young's modulus $E=210\ \textrm{GPa}$, an initial yield stress $\sigma_{y0}=0.25\ \textrm{GPa}$, plastic parameters $H=2\ \textrm{GPa}$ and $n=0.57$. The same measurements as in subsection \ref{subsubsection:Examples Linear elastic-nonlinear hardening single noise} are considered, as well as the same noise distribution. The prior distribution is in the form of Eq.~(\ref{eq:64}) with its values given in Eq.~(\ref{eq:110}). Consequently, the posterior distribution is in the form of Eq.~(\ref{eq:97}) with conditional probability $\pi(\sigma^{m}|\textbf{x},\epsilon^{m})$ given by Eq.~(\ref{eq:91}). Assuming again that the theoretical strain ranges from zero to infinity and running the adaptive MCMC approach for $10^{4}$ samples whilst burning the first 3000 samples leads to:

\begin{equation}
\label{eq:117}
\begin{gathered}
\widehat{\boldsymbol{\mu}}_\text{post}=\begin{bmatrix}
209.2\\     
0.2557\\
2.1918\\
0.6033\ \textrm{GPa}^{-1}
\end{bmatrix}\ \textrm{GPa},\\ \widehat{\boldsymbol{\Gamma}}_\text{post}=\begin{bmatrix}
131.9068 &-2.7875\times10^{-2} & 0.4345 & 4.3498\times10^{-3}\ \textrm{GPa}^{-1}\\
-2.7875\times10^{-2} & 8.4839\times{10}^{-5} &-4.9175\times{10}^{-4} & 1.7498\times10^{-4}\ \textrm{GPa}^{-1}\\
0.4345 & -4.9175\times10^{-4} & 9.925\times10^{-2} & 6.1877\times10^{-3}\ \textrm{GPa}^{-1}\\
4.3498\times10^{-3}\ \textrm{GPa}^{-1} & 1.7498\times10^{-4}\ \textrm{GPa}^{-1} & 6.1873\times10^{-3}\ \textrm{GPa}^{-1} & 1.1031\times10^{-3}\ \textrm{GPa}^{-2}
\end{bmatrix}\ \textrm{GPa}^{2},
\end{gathered}
\end{equation}

\noindent and

\begin{equation}
\label{eq:118}
\widehat{\textbf{MAP}}=\begin{bmatrix}
 208.6431\\
0.2555\\
2.2619\\
0.6055\ \textrm{GPa}^{-1} 
\end{bmatrix}\ \textrm{GPa}.
\end{equation} 

\begin{figure}[H]
\begin{minipage}[t]{0.5\linewidth}
\centering
\subcaptionbox{Uncertainty in both the stress and the strain measurements\label{fig:22_a}}{\includegraphics[width=\textwidth,trim={0.6cm 0.5cm 1cm 0.75cm},clip]{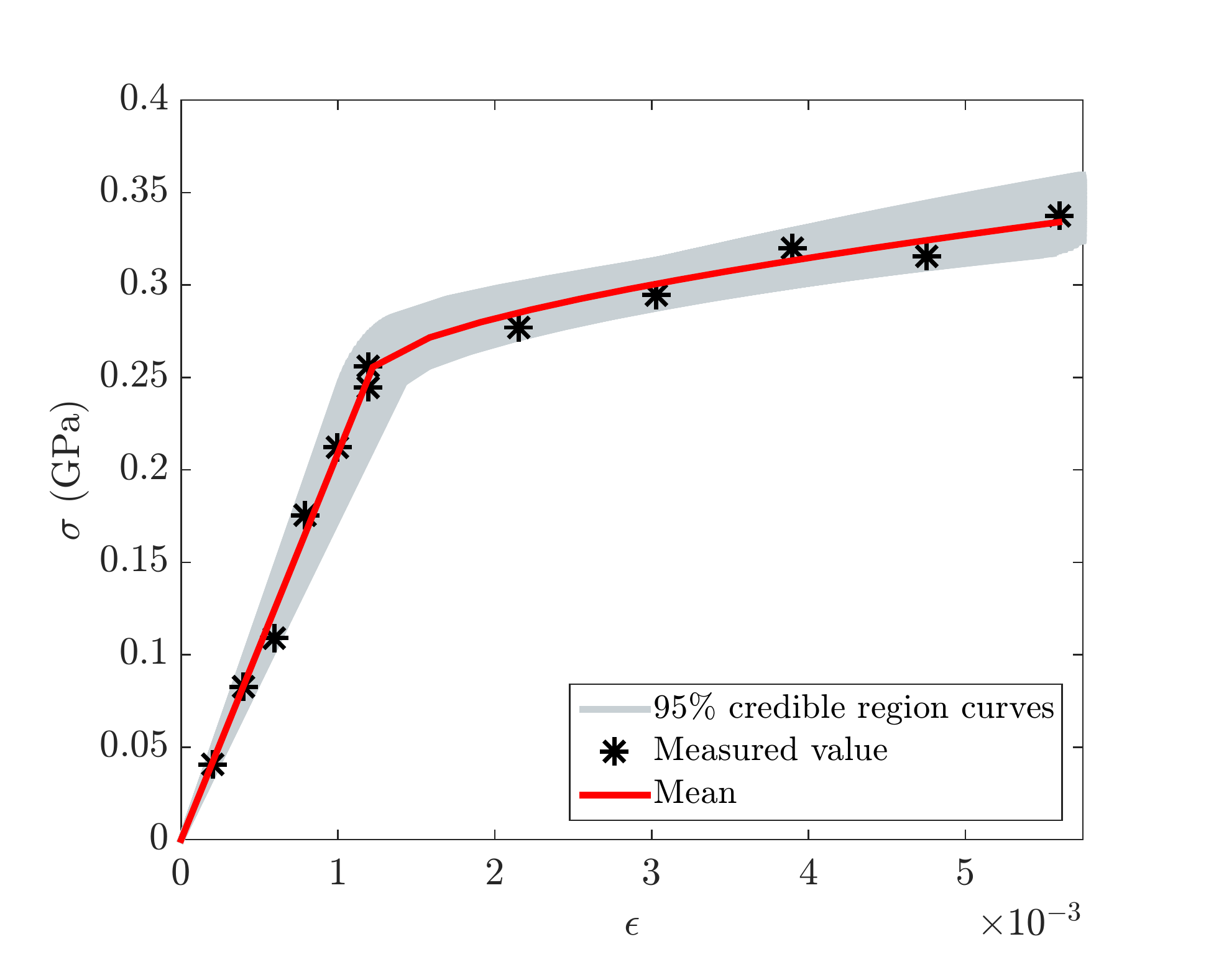}}
\end{minipage}
\begin{minipage}[t]{0.5\linewidth}
\centering
\subcaptionbox{Uncertainty in the stress measurements only\label{fig:22_b}}{\includegraphics[width=\textwidth,trim={0.6cm 0.5cm 1cm 0.75cm},clip]{Fig_14.pdf}}
\end{minipage}
\caption{Linear elastic-nonlinear hardening: The measurements, and the stress-strain curves associated with the mean and the $95\%$ credible region for twelve measurements: \subref{fig:22_a} noise in the stress and strain measurements and \subref{fig:22_b} noise in the stress measurements only. The uncertainty in the strain largely influences the Young's modulus because the multiplied gain to the strain ($E$) in the elastic part is much larger than the gain in plastic part.}
\label{fig:22}
\end{figure}

The associated stress-strain curves are shown in Fig.~\ref{fig:22} for both the single uncertainty and the double uncertainty cases. As can be seen, the additional uncertainty in the strain largely influences the Young's modulus because the multiplied gain to the strain ($E$) in the elastic part is substantially larger than the gain in plastic part.

\section{Conclusions}

\label{section:Conclusion}

In this contribution, we proposed a number of Bayesian inference formulations to identify elastoplastic material parameters of a single spring based on uniaxial tensile results. The main objectives of the paper were:

\begin{enumerate}[(1)]
\item to explain and propose BI formulations for the identification of elasto-plastic material parameters, when only the stress measurements are contaminated by statistical noise,
\item to propose BI formulations for the identification of elasto-plastic material parameters, when the stress measurements as well as the strain measurements are contaminated by statistical noise,
\item expose possible misconceptions of BI.
\end{enumerate}

In order to apply or develop a Bayesian approach, a noise model and noise distribution need to be formulated, based on calibration results. Our calibration results were artificially created, but this has enabled us to to make a direct comparison with the values used to create the measurements. We have furthermore made comparisons with the results of the least squares method. 

The examples given in section \ref{section:Examples} allow us to draw four conclusions:

\begin{enumerate}[(1)]
\item The results of BI cannot directly be compared to those of the least squares method. BI assumes that other measurements could have been made and hence, the actual measurements are only (a limited number of) realisations of statistical distributions. BI aims to take this awareness into account (via the prior distribution).
\item If one however wants to compare the results of both approaches, it is shown in Fig.~\ref{fig:3_a} that the selected prior distribution has a significant effect on the results. Fig.~\ref{fig:3_a} also shows that the influence of the prior decreases significantly if the number of measurements increases.
\item The standard deviations and correlations of the material parameters established using the `standard' BI formulations presented in this contribution, do \textit{not} reflect the heterogeneity of the material parameters. In other words, they are \textit{not} representative for the standard deviations and correlations of the material parameter distributions. Hence, they are only a measure for the uncertainty of the mean values and maximum-a-posteriori-probability (MAP) points. If one wants to determine the actual distributions of the material parameters, the `standard' BI formulations presented here need to be extended by assuming a probability density function (PDF) for the parameter vector, which is also a source of uncertainty in addition to measurement error. The objective is then to estimate the representative parameters of the probability density function (e.g.~the mean and variance). This is the focus of our future work.
\item Comparing the results when only the stress measurements are polluted by statistical noise to those when also the strain measurements are polluted by statistical noise, has shown that the uncertainty of the Young's modulus is substantially influenced by the additional noise in the strain measurements. The influence of the additional noise in the strain measurements on the uncertainty of the plastic parameters is limited however. It may therefore be worthwhile to consider noise in the strains (next to noise in the stresses) when one is interested in elastic parameters. If one is mostly interested in the plastic parameters however, it seems questionable if the substantially more complex BI formulations for double error sources are worth the extra effort.
\end{enumerate}

\section*{Acknowledgements}

The authors would like to acknowledge the financial support from the University of Luxembourg and the European Research Council Starting Independent Research Grant (ERC Stg grant agreement No.~279578) entitled ``Towards real time multiscale simulation of cutting in nonlinear materials with applications to surgical simulation and computer guided surgery". We also acknowledge the helpful discussion with Prof. J.T.~Oden on the philosophy of Bayesian inference. 


\bibliography{mybibfile.bib}

\end{document}